\documentclass[12pt]{article}
\pdfoutput=1

\usepackage{hyperref}

\usepackage{newclude}

\usepackage{array} 
\usepackage{amssymb}
\usepackage{graphics,graphpap}
\usepackage{graphicx}
\usepackage{color}
\usepackage{graphicx}
\usepackage{dcolumn}
\usepackage{epsfig}
\usepackage{epstopdf}
\usepackage{bbm}
\usepackage{amsmath}
\usepackage{amsfonts}
\usepackage{textcomp}
\usepackage{setspace}

\usepackage[all]{hypcap}
\usepackage[titletoc,title]{appendix}
\usepackage{cite}

\newcommand{\beq}{\begin{eqnarray}}
\newcommand{\eeq}{\end{eqnarray}}

\newcommand{\bmp}{\noindent\begin{minipage}{16cm}}
\newcommand{\emp}{\end{minipage}\vskip 7mm} 

\def\drawbox#1#2{\hrule height#2pt
        \hbox{\vrule width#2pt height#1pt \kern#1pt
              \vrule width#2pt}
              \hrule height#2pt}

\def\Asym#1#2{\vcenter{\vbox{\drawbox{#1}{#2}
              \kern-#2pt 
              \drawbox{#1}{#2}}}}

\def\simge{\mathrel{%
   \rlap{\raise 0.511ex \hbox{$>$}}{\lower 0.511ex \hbox{$\sim$}}}}

\def\simle{\mathrel{
   \rlap{\raise 0.511ex \hbox{$<$}}{\lower 0.511ex \hbox{$\sim$}}}}

\def\s#1{\setbox0=\hbox{$#1$}%
\rlap{\ifdim\wd0>.7em\kern.22\wd0\else\kern.1\wd0\fi /}#1}

\usepackage{slashed}

\newcommand{\equref}[1]{Eq.~\eqref{#1}}
\newcommand{\Equref}[1]{Eq.~\eqref{#1}}

\newcommand{\Figref}[1]{Fig.~\ref{#1}}

\newcommand{\tabref}[1]{Tab.~\ref{#1}}

\newcommand{\sub}[2]{#1_{\mathrm{#2}}} 						

\newcommand{\diffd}{\mathrm{d}}													
\newcommand{\dd}[1]{\frac{\mathrm{d}}{\mathrm{d} #1}}							
\newcommand{\DD}[2]{\frac{\mathrm{d} #1}{\mathrm{d} #2}}						

\newcommand{\Lag}{\mathcal{L}}											
\newcommand{\LagArg}[1]{\mathcal{L}_{\mathrm{#1}}}											

\newcommand{\orderof}[1]{\mathcal{O} \left(#1 \right)}					

\newcommand{\av}[1]{\ensuremath{\left\langle #1 \right\rangle}}

\newcommand{\CTabs}[3]{\ensuremath{\mathcal{C}^{#1}_{#2\leftrightarrow#3}}}
	
\newcommand{\hc}{\mathrm{h.c.}} 						

\newcommand{\delslashed}{\slashed{\partial}}

\newcommand{\dmf}[1]{dark matter\footnote{}}

\newcommand{\unit}[2]{#1 \, \mathrm{#2}}
\newcommand{\unitonly}[1]{\mathrm{#1}}

\newcommand{\code}[1]{\textalltt{#1}}
\makeatletter
\newcommand*{\textalltt}{}
\DeclareRobustCommand*{\textalltt}{%
	\begingroup
	\let\do\@makeother
	\dospecials
	\catcode`\\=\z@
	\catcode`\{=\@ne
	\catcode`\}=\tw@
	\verbatim@font\@noligs
	\@vobeyspaces
	\frenchspacing
	\@textalltt
}
\newcommand*{\@textalltt}[1]{%
	#1%
	\endgroup
}
\makeatother

\usepackage{xcolor}
\definecolor{red}{rgb}{1,0,0}
\definecolor{purple}{rgb}{0.5,0,0.5}
\definecolor{blue}{rgb}{0,0,1}

\begin{document}

\begin{titlepage}
\title{\vspace*{-2.0cm}
\hfill {\small MPP-2016-263}\\[20mm]
\bf\Large
 keV Sterile Neutrino Dark Matter from Singlet Scalar Decays: The Most General Case \\[5mm]\ }

\author{
Johannes K\"onig\thanks{email: \tt jkoenig@mpp.mpg.de}~,~~~Alexander Merle\thanks{email: \tt amerle@mpp.mpg.de}~,~~~and~~Maximilian Totzauer\thanks{email: \tt totzauer@mpp.mpg.de}
\\ \\
{\normalsize \it Max-Planck-Institut f\"ur Physik (Werner-Heisenberg-Institut),}\\
{\normalsize \it F\"ohringer Ring 6, 80805 M\"unchen, Germany}\\
}
\date{\today}
\maketitle
\thispagestyle{empty}

\begin{abstract}
\noindent
We investigate the early Universe production of sterile neutrino Dark Matter by the decays of singlet scalars. All previous studies applied simplifying assumptions and/or studied the process only on the level of number densities, which makes it impossible to give statements about cosmic structure formation. We overcome these issues by dropping all simplifying assumptions (except for one we showed earlier to work perfectly) and by computing the full course of Dark Matter production on the level of non-thermal momentum distribution functions. We are thus in the position to study all aspects of the resulting settings and apply all relevant bounds in a reliable manner. We have a particular focus on how to incorporate bounds from structure formation on the level of the linear power spectrum, since the simplistic estimate using the free-streaming horizon clearly fails for highly non-thermal distributions. Our work comprises the most detailed and comprehensive study of sterile neutrino Dark Matter production by scalar decays presented so far. 
\end{abstract}
\end{titlepage}

\section{\label{sec:Introduction}Introduction}

Invoking something invisible that even our most sensitive detectors cannot detect does not sound like science. Yet, in contemporary cosmology, we have so much indirect evidence for Dark Matter (DM) that hardly any scientist doubts its existence. DM is a non-luminous form of matter, i.e., it does not interact with light -- unlike everyday objects around us. Nevertheless, having entered an era of precision cosmology, we have been able to determine that DM outweighs ordinary matter by a factor of about five in the energy balance of the Universe~\cite{Planck:2015xua}. Further observational evidence such as galaxy~\cite{Begeman:1991iy} or cluster dynamics~\cite{COMA} and the Bullet Cluster~\cite{Clowe:2006eq} allow us to constrain the properties of DM: we are sure that it is a form of matter (and not, e.g., modified gravity~\cite{Lage:2014yxa}), and our best guess for its identity is a new elementary particle that is electrically neutral and massive~\cite{Bertone:2004pz}. Importantly, it must have been produced in the early Universe in the right amounts and with a momentum spectrum suitable not to spoil the formation of cosmic structures.

It is generally accepted that DM was responsible for the emergence of structures in the Universe~\cite{Primack:1997av}, as seen in elaborate $N$-body simulations on high performance computers~\cite{Springel:2005nw}. Historically, the most generic DM candidates were WIMPs (Weakly Interacting Massive Particles), which appear in popular scenarios like supersymmetry. Such particles typically form \emph{cold} DM (CDM), i.e., particles produced by thermal freeze-out~\cite{Lee:1977ua,Bernstein:1985th} which have non-relativistic velocities. The velocity of the particles strongly impacts structure formation. For example, \emph{hot} DM (HDM), which is highly relativistic, is excluded as it would have wiped out all small structures in the Universe~\cite{Abazajian:2004zh,dePutter:2012sh}. But CDM may have problems, too: it possibly forms ``too many'' small halos, which might even have the ``wrong'' structure (known small scale issues include the missing satellite problem~\cite{Klypin:1999uc,Moore:1999nt}, the abundance of isolated small halos~\cite{Klypin:2014ira}, the too-big-to-fail problem~\cite{BoylanKolchin:2011de,Papastergis:2014aba}, and the cusp-core problem~\cite{Dubinski:1991bm,Moore:1994yx}). While these may be cured once baryons are correctly included in the simulations~(see, e.g.,~\cite{Schaller:2014uwa,Chan:2015tna,Henson:2016eip}), an alternative attempt is to consider non-cold DM settings. In the literature, these ideas have triggered the slightly unfortunate terminology of \emph{warm} DM (WDM), i.e., DM particles with a \emph{thermal} spectrum (i.e., Bose-Einstein or Fermi-Dirac) of a temperature close to their mass. This is also assumed in many astrophysical studies, from Lyman-$\alpha$ bounds~\cite{Narayanan:2000tp,Viel:2005qj,Boyarsky:2008xj,Viel:2013apy} over galaxy formation~\cite{Menci:2013ght,Menci:2016eww,Menci:2016eui} to $N$-body simulations~\cite{Yoshida:2003rm,Lovell:2013ola,Bose:2016irl}. However, looking at realistic settings, many DM-models feature a \emph{non-thermal} spectrum, which one cannot associate any temperature with.

Among the most popular non-cold DM candidates is a sterile neutrino with a mass of a few keV, see Ref.~\cite{Adhikari:2016bei} for a recent collection of information on this topic. Sterile neutrinos are motivated from a particle theory point of view, as they are related to light (active) neutrinos and possibly even involved in their mass generation, see e.g.\ Refs.~\cite{Ky:2005yq,Dias:2005yh,Shaposhnikov:2006nn,Cogollo:2009yi,Dias:2010vt,Lindner:2010wr,Kusenko:2010ik,Merle:2011yv,Araki:2011zg,Adulpravitchai:2011rq,Zhang:2011vh,Robinson:2012wu,Mavromatos:2012cc,Dev:2012sg,Takahashi:2013eva,Borah:2013waa,Merle:2013gea,Robinson:2014bma} for concrete models. Furthermore, the topic has gained some attention in recent years, because a sterile neutrino $N$ would -- with a very small rate -- decay like $N\to \nu \gamma$, thereby producing a nearly monoenergetic X-ray photon. A detection of the corresponding line signal has been claimed by two groups in 2014~\cite{Bulbul:2014sua,Boyarsky:2014jta}, but it was very actively disputed -- see Ref.~\cite{Adhikari:2016bei} for a detailed discussion and Ref.~\cite{Aharonian:2016gzq} for the (still ambiguous) data that the Hitomi satellite could take before enduring its unfortunate fate. It may also be worth mentioning that there could be a ``dip'' in the cluster data~\cite{Conlon:2016lxl}, tending to make a non-observation of the line in stacked data more consistent.

Sterile neutrinos with a sufficiently large lifetime are excellent DM candidates provided that 1.)~an efficient production mechanism exists in the early Universe which 2.)~produces a spectrum in accordance with cosmic structure formation. The most generic idea is to produce sterile neutrinos by their small admixtures to active neutrinos, by non-resonant active-sterile transitions, a mechanism first proposed by Langacker~\cite{Langacker:1989sv} and related to DM by Dodelson and Widrow (DW)~\cite{Dodelson:1993je}. In modern terms, one would refer to these sterile neutrinos as FIMPs (Feebly Interacting Massive Particles~\cite{Hall:2009bx}), which do not thermalise but are instead produced gradually in the early Universe via freeze-in by their feeble couplings to Standard Model (SM) particles. While this mechanism is rather simple, it is excluded by data, due to the particles being too close to the HDM limit~\cite{Seljak:2006qw,Viel:2013apy}; still, it is an unavoidable addendum to the spectrum produced by any other mechanism, which can modify the DM distribution function at least for sterile neutrino masses below $3$~keV~\cite{Merle:2015vzu}.

A popular alternative is based on a resonant enhancement of the active-sterile neutrino transitions by the presence of a primordial lepton number asymmetry. First proposed by Enqvist and collaborators~\cite{Enqvist:1990ek} and related to DM by Shi and Fuller (SF)~\cite{Shi:1998km}, this mechanism has been studied actively~\cite{Abazajian:2001nj,Laine:2008pg,Kishimoto:2008ic,Canetti:2012kh,Abazajian:2014gza,Ghiglieri:2015jua,Venumadhav:2015pla}, and it does indeed yield a spectrum colder than that produced by DW. However, while there is no perfect agreement between the results of different groups, at least for the results that are publicly available~\cite{Abazajian:2001nj,Abazajian:2014gza,Venumadhav:2015pla} it seems unclear whether they are in full agreement with cosmic structure formation~\cite{Merle:2014xpa,Horiuchi:2015qri,Schneider:2016uqi}.

On the other hand, one could produce sterile neutrinos thermally via freeze-out, if they had non-trivial charges beyond the SM, as long as the resulting overabundance is diluted by a sufficiently efficient production of additional entropy~\cite{Bezrukov:2009th,Nemevsek:2012cd,Patwardhan:2015kga}; this, however, is constrained by big bang nucleosynthesis~\cite{King:2012wg}.

A completely different direction is to produce sterile neutrinos by the decays of other particles. A generic possibility is a decaying singlet scalar $S$, which can easily couple to sterile neutrinos after having been produced itself in the first place. Apart from this scalar being the inflaton~\cite{Shaposhnikov:2006xi,Bezrukov:2009yw,Bezrukov:2014nza}, in which case it had been present all along in the early Universe, one can tune the Higgs portal coupling of $S$ in such a way that it is either produced like a WIMP via freeze-out~\cite{Kusenko:2006rh,Petraki:2007gq,Kusenko:2009up} or like a FIMP via freeze-in~\cite{Merle:2013wta,Merle:2015oja}. Variants have been presented in~\cite{Klasen:2013ypa,Kang:2014cia,Frigerio:2014ifa,Adulpravitchai:2014xna,Humbert:2015epa,Ayazi:2015jij,McDonald:2015ljz,Adulpravitchai:2015mna,Shakya:2015xnx,Kaneta:2016vkq}, some also featuring other parent particles such as vectors~\cite{Boyanovsky:2008nc,Shuve:2014doa,Biswas:2016bfo}, Dirac fermions~\cite{Abada:2014zra}, or pions~\cite{Lello:2014yha,Lello:2015uma}; related aspects such as influences from inflation~\cite{Nurmi:2015ema} or thermal corrections~\cite{Drewes:2015eoa} have been discussed, too. 

Our main goal is to close a gap in the treatment of sterile neutrino production from scalar decays. In earlier works~\cite{Kusenko:2006rh,Petraki:2007gq,Merle:2013wta,Merle:2015oja}, simplifying assumptions have been applied to at all arrive at a result, such as neglecting active-sterile mixing, taking the number of relativistic degrees of freedom to be constant, and assuming a heavy scalar. This is also true for a previous paper by two of us (AM \& MT)~\cite{Merle:2015oja}, which was the first to show how to numerically compute momentum distribution functions of sterile neutrinos from scalar decay. While it has been proven that neglecting active-sterile mixing is in fact a very good assumption~\cite{Merle:2015vzu}, going to the low-mass region of the singlet scalar -- where the number of degrees of freedom is not constant -- poses new technical challenges. The only treatment available for this regime was put forward in~\cite{Adulpravitchai:2014xna}. However, the authors only used rate equations, such that an actual computation of the momentum distribution function and the confrontation with structure formation data cannot be reliably performed. We will in this work present the full numerical computation on the level of momentum distribution functions instead, including a detailed treatment of all subtleties related to the many new technical aspects that arise in this regime. We will give an a-posteriori justification of some of the assumptions made in~\cite{Adulpravitchai:2014xna} (such as the Higgs staying equilibrated during $S$-production), while we will also reveal that others (like the Higgs degrees of freedom in the unbroken phase) are maybe less justified. We will furthermore show in great detail how to constrain the resulting spectra by cosmic structure formation.

Our study presents the first complete treatment of scalar decay production in the whole parameter space. It involves a fully general solution of the production equations on the level of distribution functions. Apart from our techniques being transferable to virtually any type of decay production, our study will guide the particle physics community towards using limits from cosmic structure formation without having to leave their comfort zone completely. We thus contribute to closing the gap between particle physics models, early DM-production, and their phenomenological consequences.

This text is structured as follows. We first present a qualitative discussion of decay production of sterile neutrinos in Sec.~\ref{sec:QualitativeDiscussion}, before explaining how to solve the evolution equations and to apply the relevant bounds in Sec.~\ref{sec:Technicalities}. Our main results are presented in Sec.~\ref{sec:Results}, which features a thorough discussion of all relevant aspects from the production process over the distribution functions to structure formation. We conclude in Sec.~\ref{sec:CandO}. Technical aspects on the Boltzmann equation, on the evolution of the Higgs distribution, on the failure of the free-streaming horizon as a reliable estimator, and on the robustness of our half-mode analysis are discussed in Appendices~\ref{app:A:DetailsComputation}, \ref{app:B:Higgs-FO}, \ref{app:C:FSvsHalfMode}, and~\ref{app:D:HalfmodeThreshold}, respectively.


\section{\label{sec:QualitativeDiscussion}The basic idea: Qualitative discussion}

This section is intended to introduce the particle physics model used in our analysis and to give a qualitative understanding of its different regimes. Furthermore, we will justify the assumptions that go into the numerical computations.

Our setting introduces one real scalar singlet $S$ (with mass $m_S$) and one right-handed neutrino $N$ (with mass $m_N$) beyond the particle content of the SM.\footnote{In general, any number of right-handed neutrinos can be assumed in this model. In the case of more than one generation, the scalar will then decay into all kinematically accessible right-handed states $N_i$ with branching fractions given by $y_i^2/\sum_k y_k^2$. If the mixing between the different right-handed states is large enough, all right-handed neutrinos will decay into the lightest state ($N_1$ by convention) quickly and the results of this paper stay unaltered under the substitution $y^2 \rightarrow \sum_k y_k^2$. If, however, the mixing inside the sterile sector is small, there might be additional complications due to late injection of highly energetic DM particles by the decay of the heavier states into the lighter ones.} The right-handed neutrino is coupled to the singlet scalar via a Yukawa-type interaction with coupling $y$,
\begin{align}
 \Lag \supset - \frac{y}{2} S \overline{N^c}N + \hc \,,
 \label{eq:YukawaCouplingLagrangia}
\end{align}
while the new scalar singlet is coupled to the Higgs doublet $\Phi$ via the most generic potential symmetric under a global $\mathbb{Z}_4$-symmetry:\footnote{Suitable charge assignments are given by $S \to -S$ and $N\to \pm i N$. For further comments on this assumption, see~\cite{Merle:2013wta,Merle:2015oja} and references therein. The (rather mild) consequences of giving up this simplification are discussed in~\cite{Petraki:2007gq}.}
\begin{align}
  \sub{V}{scalar} =   \frac{1}{2} m_S^2 S^2 + \frac{\lambda_S}{4} S^4 +2\lambda \left(\Phi^\dagger \Phi\right)S^2 \,.
 \label{eq:ScalarPotentialLagrangian}
\end{align}
Accordingly, the complete Lagrangian of the model reads
\begin{align}
  \Lag = \LagArg{SM} + \left[\frac{i}{2}\overline{N} \delslashed N + \frac{1}{2}\left(\partial_\mu S\right)\left(\partial^\mu S\right) - \frac{y}{2}S\overline{N^c}N +\hc \right] - \sub{V}{scalar} + \Lag_\nu \,,
 \label{eq:ModelLagrangian}
\end{align}
where $\Lag_\nu$ is the part of the Lagrangian that can give mass to the active neutrinos. We do \emph{not} assume any vacuum expectation value (VEV) for $S$ in our analysis, however, we will include the constraints which would arise of the VEV of $S$ gave the sterile neutrinos their mass, so that the case of $\langle S \rangle \neq 0$ can be qualitatively recovered from our analysis.

Note that we assume active-sterile mixing to be so small that the contribution from the DW mechanism to the production of sterile neutrinos is negligible compared to the part produced by scalar decay, based on taking into account X-ray limits on the decay of sterile neutrino DM. In Ref.~\cite{Merle:2015vzu} it has been shown that the DW modification to any previously produced population of sterile neutrinos (from whichever main production mechanism) is of a few percent at most, and completely negligible for masses $m_N$ larger than about $\unit{4}{keV}$.

Our setup mainly allows us to produce scalars from their coupling to the SM with interactions of the type $\sub{\overline{X}}{SM}\sub{X}{SM} \leftrightarrow SS$. Subsequently, the scalars decay into sterile neutrinos via $S \rightarrow N N$. We have to distinguish three different regimes:
\begin{enumerate}
 \item[I] Production before the electroweak phase transition (EWPT): all four degrees of freedom of the $SU(2)_L$-doublet Higgs $\Phi$ contribute equally to the production/depletion of scalars from/into the thermal bath.
 \item[II] Production after EWPT with $m_S > m_h/2$, where $m_h$ is the mass of the Higgs after electroweak symmetry breaking. Now the Higgs and the massive gauge bosons interact with the scalar different ways.
 \item[III] Production after EWPT with $m_S < m_h/2$. This is similar to case~II, the difference being that the Higgses present in the thermal plasma are now kinematically allowed to decay into pairs of scalars.
\end{enumerate}

\begin{table}[t!]
 \centering
 \begin{tabular}{|c|cccccc|}
  \hline
  \rotatebox{90}{regime~} & \multicolumn{6}{|c|}{production channels} \\ 
  \hline 
  I & 
  \includegraphics[scale=0.32]{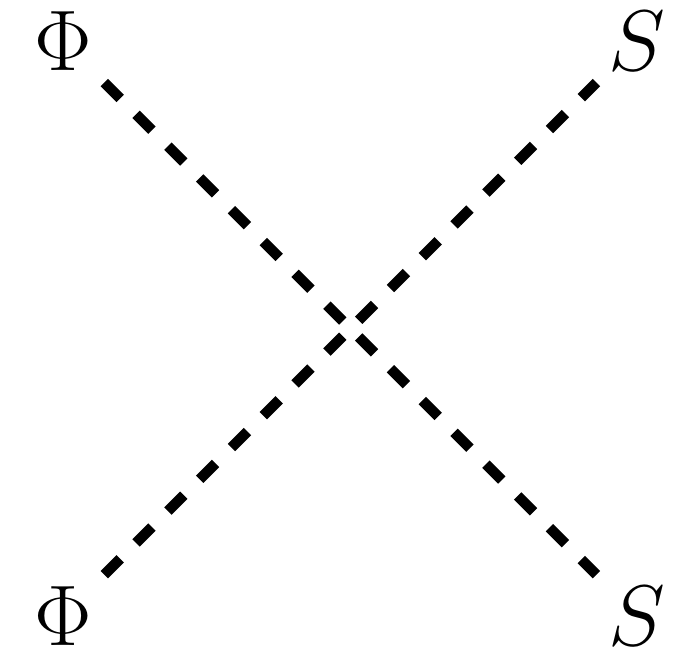} & & & & &\\
  \hline
  II &
  \includegraphics[scale=0.32]{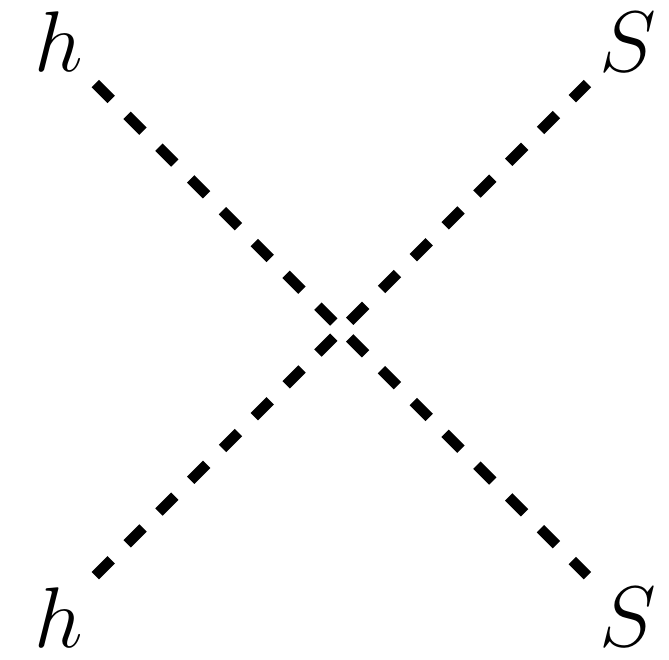} &
  \includegraphics[scale=0.32]{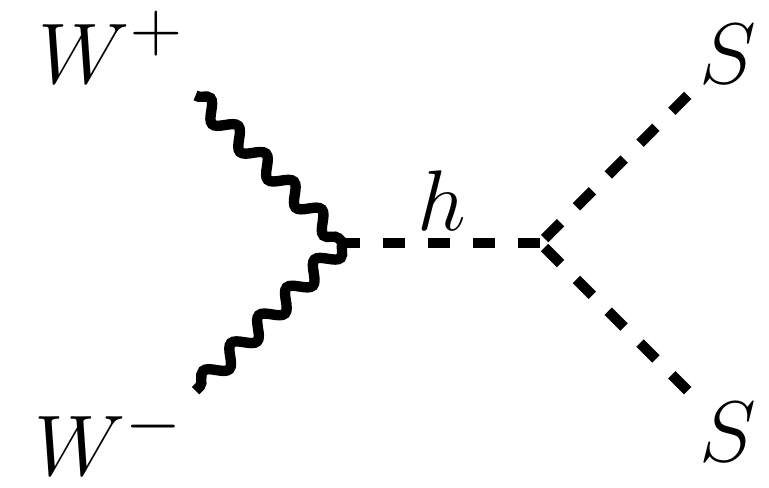} &
  \includegraphics[scale=0.32]{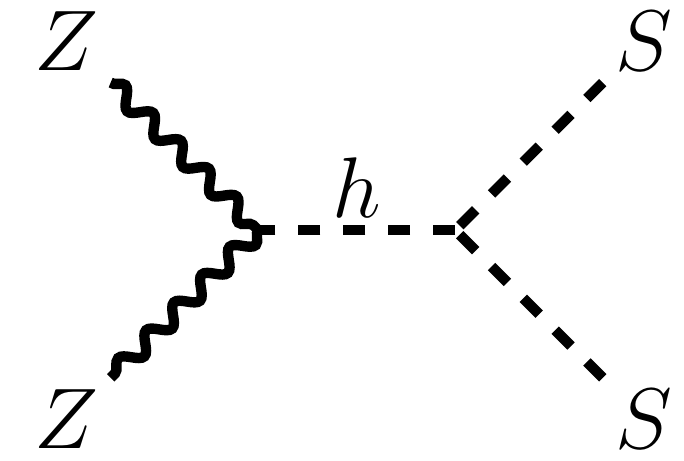} &
  \includegraphics[scale=0.32]{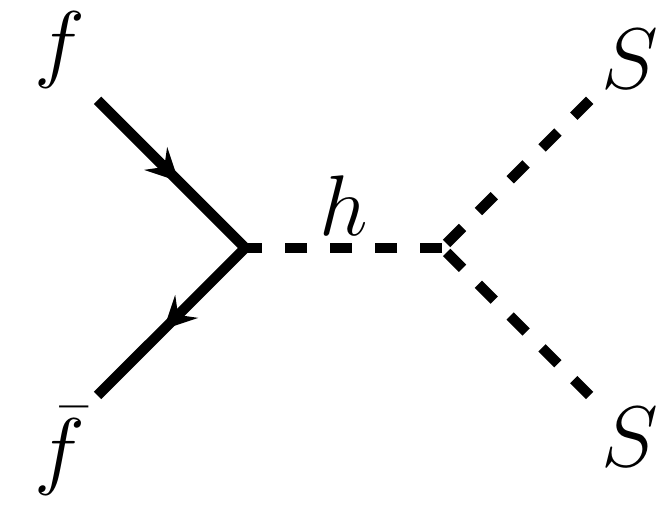} &
  \includegraphics[scale=0.32]{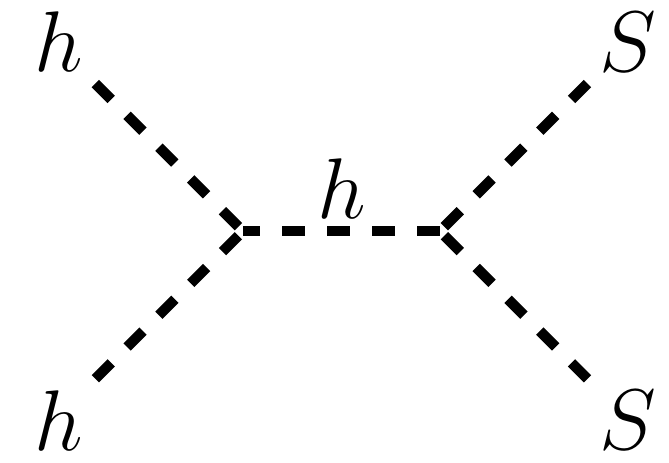} & \\
  \hline
  III & 
  \includegraphics[scale=0.32]{figures/feynman_diagrams/hhssfeynman.pdf} &
  \includegraphics[scale=0.32]{figures/feynman_diagrams/wwssfeynman.pdf} &
  \includegraphics[scale=0.32]{figures/feynman_diagrams/zzssfeynman.pdf} &
  \includegraphics[scale=0.32]{figures/feynman_diagrams/ffssfeynman.pdf} &
  \includegraphics[scale=0.32]{figures/feynman_diagrams/hhssfeynman_s-channel.pdf} & 
  \includegraphics[scale=0.32]{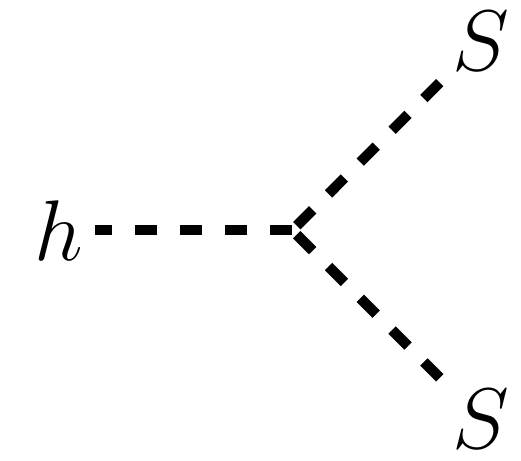} \\
  \hline
 \end{tabular}
  \caption{\label{tab:RegimesFeynman}Relevant production channels in regimes~I--III.}
\end{table}

The channels corresponding to regimes I--III are listed in \tabref{tab:RegimesFeynman}. Note that we have omitted the diagrams with a scalar $S$ in the $t$- or the $u$-channels, since they will contribute to the production only at order $\orderof{\lambda^3}$, i.e., suppressed by another small factor of $\lambda$ compared to the leading order. In regimes~II and~III, the fermionic initial state $f$ only receives a contribution from the top quark in practice, since all other channels are suppressed by their small Yukawa couplings. Couplings to lighter fermions may appear relevant at first sight, since in that case the Higgs could potentially be on-shell, which may enhance the cross section by orders of magnitude. However, this case of an on-shell Higgs needs to be subtracted adequately to avoid double-counting of decay events of thermal Higgses~\cite{Frigerio:2011in}, and they ultimately do not contribute much. Note that, at precision level, even thermal corrections to the parameters can play a role~\cite{Drewes:2015eoa}.

In general, production will always start in regime~I and then, unless finished in that regime, proceed either in regime~II or in regime~III after EWPT, depending on the value of $m_S$. In the case of small Higgs portal couplings $\lambda$, the scalar will freeze in. This production mechanism is most efficient when $T \sim m_S$. In the case of larger $\lambda$, the scalar first equilibrates and then freezes out. In this scenario, the freeze-out time will also be linked to $m_S$. Accordingly, there is a finite span in cosmic time (or in the temperature $T$, equivalently) in which the production is effectively taking place. 

\begin{figure}
 \centering
 \includegraphics[width=0.8 \textwidth]{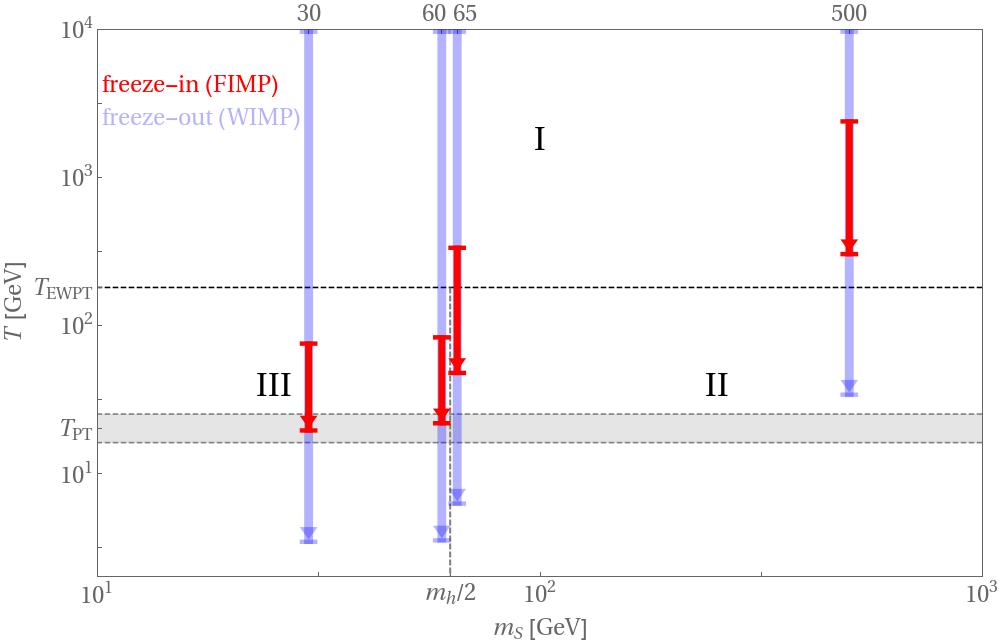}
 \caption{\label{fig:RegimeExampleCases}Temperature ranges of the plasma temperature $T \equiv T_\gamma$ at which the scalar singlet $S$ is efficiently produced in the freeze-in scenario (red) and in the freeze-out scenario (light blue). We present four different scalar masses (30, 60, 65, and 500~GeV) that span the variety of possible production times. $\sub{T}{EWPT}$ indicates the temperature when electroweak phase transition happens, while $\sub{T}{PT}$ is supposed to give a rough idea of the temperature when the Higgs and gauge bosons become strongly Boltzmann-suppressed, even if still equilibrated with the remaining SM degrees of freedom.}
\end{figure}

This time span is illustrated in \Figref{fig:RegimeExampleCases} for four different masses $m_S$. Red arrows correspond to small Higgs portal couplings $\lambda$, implying that the scalar never equilibrates, but freezes in instead. In this case, we have defined the time span of production starting when 10\% of the final yield $Y$ of a would-be stable scalar is produced and ending when 90\% are produced. Blue arrows indicate the time spans for scalars freezing out before decaying into sterile neutrinos. In this case, there is no physically preferred initial time. Note that, even when assuming a vanishing initial abundance, thermalisation is fast compared to the time scale of the decay. The late-end boundary (i.e., at low temperatures) of the time span is, however, defined by $Y/\sub{Y}{eq}=10$, where $\sub{Y}{eq}$ denotes the equilibrium yield.

One can clearly see that, in cases where $m_S<m_h/2$, the freeze-in of scalars occurs only after EWPT. This can be explained by the fact that the decay of thermal Higgses after EWPT completely dominates the production as soon as this channel is kinematically accessible~\cite{Adulpravitchai:2014xna}. For $m_S=\unit{65}{GeV}$, production sets in before EWPT and stops at temperatures of some tens of $\unitonly{GeV}$. In the case of a much heavier scalar (say, $m_S=\unit{500}{GeV}$), production even ends before EWPT, since the kinetic energies of the Higgses after EWPT are insufficient to produce the heavy scalar. 

Note that we need the particle distribution function of all sourcing species in the plasma, like gauge bosons or Higgses, to account for the production of scalars induced by the channels shown in \tabref{tab:RegimesFeynman}. In general, the dynamics of each of these species is determined by their own Boltzmann equation. Fortunately, it is not necessary to incorporate more than two equations into the system, since we can safely assume these particles to be in thermal equilibrium, an assumption that has been readily made in~\cite{Adulpravitchai:2014xna} even for small $T$, but not proven to work. We have explicitly checked that the Higgs indeed stays in equilibrium at least until $T \approx \unit{1}{GeV}$. Note that the usual estimate for WIMP-like particles of mass $m$ is that freeze-out occurs at $T_\text{freeze-out}\sim m/20$. The reason that the Higgs stays in equilibrium longer are inverse decay processes, and similar arguments apply to the gauge bosons and to the top-quark, cf.\ App.~\ref{app:B:Higgs-FO}. Still, the number densities of all SM particles become strongly Boltzmann-suppressed for temperatures of, say, $10$--$\unit{20}{GeV}$. Hence, production \emph{always} ceases at these temperatures, simply because there are hardly any particle left in the bath, even if they are still in equilibrium.

\section{\label{sec:Technicalities}Technicalities}

\subsection{\label{sec:Technicalities:BoltzmannEquation}The Boltzmann equation and its solution}

In order to compute the distribution functions of the scalar and the sterile neutrino, we employ the Boltzmann equation, $\hat{L}[f]=\mathcal{C}[f]$, where $f$ is the distribution function we want to determine, $\hat{L} = \frac{\partial}{\partial t} - H p \frac{\partial}{\partial p}$ (with the Hubble function $H$) is the Liouville operator in a homogeneous and isotropic Friedman-Robertson-Walker universe, and $\mathcal{C}$ contains the collision terms describing interactions of the particles.\footnote{This is already an approximation in itself because, by using the Boltzmann equation -- which is a classical equation -- we do neglect quantum corrections. To include those, Kadanoff-Baym equations must be used, which are way more difficult to solve. The error caused by the approximation we make here is of $\mathcal{O}(10\%)$ in the abundance~\cite{Hamaguchi:2011jy}.} In our case, we actually need to solve two coupled Boltzmann equations, for the two distribution functions $f_N$ and $f_S$ of the sterile neutrino $N$ and the scalar $S$, respectively:
\begin{eqnarray}
 \hat{L}f_S &=& \mathcal{C}^S \label{eq:Boltzmann_Scalar}\ \ \ {\rm and}\\
 \hat{L} f_N &=& \mathcal{C}^N. \label{eq:Boltzmann_Neutrino}
\end{eqnarray}
The collision terms are themselves sums of several individual terms which encode the details of the respective processes as well as information on whether they contribute to production or depletion of the species under consideration. Depending on the regime, various diagrams listed in Tab.~\ref{tab:RegimesFeynman} can contribute to the production of scalars and hence, ultimately sterile neutrinos. Explicitly, the scalar collision term $\mathcal{C}^S$ consists of the following:
\begin{itemize}
\item Regime~I ($T>T_\text{EWPT}$):
\begin{align}\label{eq:ct_scalar_above_pt}
\mathcal{C}^S_\text{I} = \mathcal{C}^S_{\phi\phi \leftrightarrow SS} + \mathcal{C}^S_{S \rightarrow NN}.
\end{align}
\item Regime~II ($T<T_\text{EWPT}$  and  $m_S> m_h/2$):
\begin{align}\label{eq:ct_scalar_below_pt_large_m}
\mathcal{C}^S_\text{II} = \mathcal{C}^S_{hh \leftrightarrow SS} + \mathcal{C}^S_{t\bar{t} \leftrightarrow SS} + \mathcal{C}^S_{W^+W^- \leftrightarrow SS} + \mathcal{C}^S_{ZZ \leftrightarrow SS} + \mathcal{C}^S_{S \rightarrow NN}.
\end{align}
\item Regime~III ($T<T_\text{EWPT}$ and  $m_S < m_h/2$):
\begin{align}\label{eq:ct_scalar_below_pt_small_m}
\mathcal{C}^S_\text{III} =\mathcal{C}^S_\text{II} + \mathcal{C}^S_{h \leftrightarrow SS}.
\end{align}
\end{itemize}
The labels should be quite intuitive. For example, in regime~I, the term $\mathcal{C}^S_{\phi\phi \leftrightarrow SS}$ encodes both the annihilation of two Higgs doublets $\Phi$ into two scalars $S$ and vice versa, with opposite signs for the two cases. On the other hand, $\mathcal{C}^S_{S \rightarrow NN}$ will always come with a negative sign, since it only describes the decay of the scalar into sterile neutrinos. Similarly, the only difference between regimes~II and~III is the appearance of the Higgs decays into two scalars, $\mathcal{C}^S_{h \leftrightarrow SS}$, for sufficiently small scalar masses. 

The sterile neutrino collision term is comparatively simple, and it is given by:
\begin{align}
 \mathcal{C}^N = \mathcal{C}^N_{S \rightarrow NN},
 \label{eq:ct_sterile_neurtrino}
\end{align}
but in this case with a positive sign as it creates sterile neutrinos. Here we neglected the inverse decay, i.e.\ the process $NN \rightarrow S$, because the sterile neutrino is a FIMP, such that its abundance is always far below its would-be equilibrium abundance for large enough temperatures $T$. Hence the process $NN \rightarrow S$ is suppressed by the small number of sterile neutrinos present in the Universe.

Note that the collision terms will in general depend on the distribution functions $f_i$ of \emph{all} particle species $i$. However, as argued, we only need to compute the distributions $f_S$ and $f_N$, since all SM species can be assumed to be in thermal equilibrium such that their distributions are well approximated by simple Boltzmann factors.\footnote{As already anticipated, this is a non-trivial assumption for the SM Higgs boson. We have shown explicitly that the Higgs only freezes out at relatively small temperatures, $T\sim 3$~GeV, which is due to its decay into SM particles keeping it in equilibrium for a longer time than naively expected for a WIMP-like particle. This also provides an a posteriori justification of the assumption made by Adulpravitchai and Schmidt in Ref.~\cite{Adulpravitchai:2014xna}. See App.~\ref{app:B:Higgs-FO} for details.} To anticipate one particular example collision term, we quote from Eq.~\eqref{eq:CT_N_App} the one describing the production of sterile neutrinos of momentum $p$ at temperature $T$ by the decaying scalars,
\begin{align}
 \mathcal{C}^N_{S \rightarrow NN}[f_S](p , T) = \frac{y^2 m_S^2}{16 \pi p^2}\int\limits_{p'_\text{min,N}}^\infty \mathrm{d}p'\ \frac{p' \; f_S(p',T)}{\sqrt{m_S^2 + p'^2}},
 \label{eq:CT_N}
\end{align}
where the lower boundary of the integral over scalar momenta $p'$ is given by $p'_\text{min,N} = \left| p - \frac{m_S^2}{4 p} \right|$. This term is indeed positive, since it produces sterile neutrinos. As expected, this equation contains the distribution function $f_S$ of the scalar at temperature $T$, making it explicit that Eqs.~\eqref{eq:Boltzmann_Scalar} and~\eqref{eq:Boltzmann_Neutrino} are coupled and have to be solved as a system of equations. For a detailed discussion of all collision terms we refer the reader to App.~\ref{app:coll_terms}.

\subsubsection{\label{sec:Technicalities:BoltzmannEquation-solution}How to crack a coupled system of non-linear partial integro-differential equations in two variables}

Glancing at the explicit forms of the collision terms in App.~\ref{app:coll_terms}, we can see that the task of computing sterile neutrino production from scalar decay is a highly non-trivial one: it necessitates the solution of a coupled system of partial integro-differential equations in two variables. In an abstract, yet intuitive manner, the system of equations we have to solve turns out to have the following form:\footnote{Note that, at this point, we could also have taken an alternative route. The part of the integral that only contains Boltzmann distributions does \emph{not} depend on $f_S$ and could hence, in principle, be integrated numerically. This would result in a numerical function of the form $\mathcal{F}(p,T)$ on the right-hand side of the first Eq.~\eqref{eq:Boltzmann_system}. However, this strategy would have at least two drawbacks: 1.) First, the structure of the equation would be much less clear, since several crucial dependencies would just be ``hidden'' inside a numerical function $\mathcal{F}(p,T)$. 2.) Furthermore, this procedure would ultimately force us to subtract two numerically computed integrals from each other, which is numerically less accurate than first computing the difference of the integrand functions before evaluating the integral over their difference. We thus stick to the strategy outlined in the main text.} 
\begin{eqnarray}
 \hat{L} f_S\left(p,T\right) &=& \mathcal{D}\left(p,T\right) f_S\left(p,T\right) + \int\limits_{0}^{\infty}{\diffd p'\mathcal{K}^{S}\left(p,p',T\right)\left[f_S\left(p,T\right) f_S\left(p',T\right) - f_S^{\mathrm{eq}}\left(p,T\right) f_S^{\mathrm{eq}}\left(p',T\right) \right]}  \,, \nonumber\\
 \hat{L} f_N\left(p,T\right) &=& \int\limits_{0}^{\infty}{\diffd p'\mathcal{K}^N\left(p,p',T\right) f_S\left(p',T\right)} \,,  \label{eq:Boltzmann_system}
\end{eqnarray}
where $\mathcal{K}^{S/N}$ and $\mathcal{D}$ are known functions directly related to the collision terms $\mathcal{C}$, and $f_i^{\mathrm{eq}}$ denotes the (possibly hypothetical, in case the interactions are too feeble) equilibrium distribution of particle $i$.

The question is now how to tackle such a problem. Before starting the computation we note that, while the sterile neutrino distribution function $f_N$ does depend on the distribution function $f_S$ of the scalar, the converse is not true. The reason is that, as we had argued, we can neglect any inverse processes involving sterile neutrinos. This already yields to a considerable simplification, allowing us to solve the first Eq.~\eqref{eq:Boltzmann_system} independently and to then insert the result obtained into the second equation. However, even then, we are still left with a non-linear partial integro-differential equation in $f_S$ which is highly non-trivial to solve numerically (and impossible to solve analytically). We will thus need to play some tricks, in order to tame this equation.

The first step is to perform a transformation of variables, $(t,p) \to (r,\xi)$, such that the left-hand side of the first Eq.~\eqref{eq:Boltzmann_system} only involves a derivative with respect to a single variable. As shown in App.~\ref{app:transf_variables}, this is possible for $r=f(t)$ being an arbitrary function $f$ independent of the scalar momentum $p$, and $\xi(p,t) = g \left( \frac{a(t)}{a(t_0)} p \right)$ simultaneously being an arbitrary function $g$ of a specific combination of time $t$ and momentum $p$, which involves the scale factor $a(t)$ and an arbitrary reference time $t_0$. Thus, in fact, there exists a whole family of functions $f$ and $g$ which would simplify our complicated equation, however, choosing them in a smart manner may even lead to further simplifications. In our view, the following choice of variables is particularly convenient:\footnote{We exploit the one-to-one correspondence between cosmic time $t$ and plasma temperature $T$.}
\begin{align}\nonumber\label{eq:xi_and_r_definition}
r &= \frac{m_0}{T},\\
\xi &= \frac{1}{T_0}\frac{a(t)}{a(t(T_0))} \; p = \left( \frac{g_{s}(T_0)}{g_{s}(T)} \right)^{1/3}\;\frac{p}{T},
\end{align}
where $g_s(T)$ is the number of effective entropy degrees of freedom (for which we have used the numerical fit developed in Ref.~\cite{Wantz:2009it}). The arbitrary reference mass $m_0$ and temperature $T_0$ have been chosen by us to both equal the Higgs mass, $m_0 = T_0 = m_h$, which will later prevent us from working with extremely small or large numbers in our numerical computation. This choice of variables indeed implies certain simplifications, e.g., the Liouville operator now reads:
\begin{align}
\hat{L} = \frac{\partial r}{\partial t}\frac{\partial}{\partial r}= r H\left(r\right) \left( \frac{T g_s'}{3 g_s} + 1 \right)^{-1}\frac{\partial}{\partial r},\label{eq:liouville_final_form}
\end{align}
where $'$ denotes a derivative with respect to the temperature $T$. 

A valid interpretation of the new variables is to view $r$ as ``time'' variable and $\xi$ as ``momentum'' variable. Indeed, $r$ increases as the temperature decreases, just as the cosmic time $t$, so it can really be thought of as a rescaled or stretched time. In turn, $\xi$ can be thought of as something like a comoving rescaled momentum. Alternatively, one could think of it as a rescaled momentum that is red- or blue-shifted with respect to the reference temperature $T_0$. Indeed, the big advantage of the variable $\xi$ is that the distributions remain constant in time $r$ as soon as the production of particles is finished. For example, if the scalar $S$ was stable, its distribution for late time would be given by $f_S(\xi,r) \equiv f_S(\xi, r_\text{prod}),~~~\forall r \geq r_\text{prod}$, where $r_\text{prod}$ marks the time when no $S$-particles are produced anymore (e.g.\ the freeze-out time in case $S$ was a stable WIMP). This effect translates into the sterile neutrino distribution, i.e., $f_N$ will also be constant in $r$ once the scalar decays cease to be efficient. All redshifts of the momenta are automatically included in the definition of $\xi$.

The resulting form of the Boltzmann equations used in our implementation is
\begin{align}
 \frac{\partial f_i}{\partial r}( \xi, r ) = \frac{1}{r H(r)} \left( 1 - \frac{r }{3} \frac{\partial}{\partial r} \ln [g_s(r)] \right) \mathcal{C}^i[f_i,f_{j\neq i}],
 \label{eq:boltz}
\end{align}
where $i=S,N$. Indeed, we have by our substitution transformed the partial differential equation in two variables into an equation containing a differential with respect to one variable only. Of course, the collision terms in their general form of \Equref{eq:Boltzmann_system} have to be transformed according to this change of variables. Note that this general form can be written down for both the scalar and the sterile neutrino, but it is only necessary for the former, as the sterile neutrino evolution equation is comparatively simple once $f_S$ is known. We will therefore concentrate on the case $i=S$ in the following.

Still, the main issue remains: the collision term is nonlinear in $f_S$, and it furthermore includes an integral over $\xi$, which renders this equation to be of a partial integro-differential type, which are very difficult to solve. While several strategies for certain types of integro-differential equations (such as the Volterra-type) are available in the literature, the equation under consideration turns out not to be of any such type, so that we have to find a dedicated workaround. We apply a trick to get at least an (arbitrarily good) approximation by discretising the equations and substituting the integral over $\xi'$ by a finite sum over $M$ discrete momentum values $\xi_i,\, i\in\{1,...,M\}$. This transforms the original non-linear integro-differential equation~\eqref{eq:boltz} into a system of coupled non-linear ordinary differential equations for the different modes $f_S^i\left(r\right)$:
\begin{align}
 \dd{r} f_S^i\left(r\right) = \tilde{\mathcal{D}}^i\left(T\right) f_S^i\left(r\right) 
 + \sum\limits_{j=1,...,M}^{}{\tilde{\mathcal{K}}^{ij}\left(r\right)} \left[f_S^i\left(r\right)f_S^j\left(r\right) - f_S^{i,\mathrm{eq}}\left(r\right)f_S^{j,\mathrm{eq}}\left(r\right)  \right], \quad \forall i \in\{1,\dots,M \}.
\end{align}
Note that the integral results in a ``non-local'' coupling, since it couples any mode $f_S^i$ not only to some ``neighbouring'' modes, but to \emph{all} modes instead.

Finally, we have to numerically solve the resulting system of differential equations. This step involves one more difficulty, since the most simple strategies (such as the Runge-Kutta method) fail for large parts of the parameter space due to the stiffness of the system. Another more technical issue is that packages such as \textsc{Mathematica} natively employ list-based algorithms, while matrix-based ones are much more appropriate for the equations under consideration. We have thus found it more convenient to use the built-in \textsc{Matlab} solver \textit{ode15s}~\cite{Matlab:ode15s:reference1,Matlab:ode15s:reference2},  which is particularly efficient when dealing with stiff problems. Nonetheless, one should try to tweak any solver algorithm to take advantage of our a-priori knowledge that a distribution function must never attain negative values.
This way, we have been able to solve the equations for the scalar distribution functions $f_S$ efficiently. The resulting numerical functions can then be inserted into Eq.~\eqref{eq:boltz} for $i=N$ or, alternatively, one can use the ``master formula'' as given in Eq.~(16) of Ref.~\cite{Merle:2015oja}, once the variables are transformed according to Eqs.~\eqref{eq:xi_and_r_definition}.

Obtaining the functions $f_N (r,\xi)$ was the goal we wanted to reach. These distribution functions form the basis to compute all properties of the DM species for a given point in the parameter space, see Ref.~\cite{Merle:2015oja} for details. Not only can we compute the DM abundance $\Omega_N h^2$, we can also use $f_N (r,\xi)$ to derive predictions for cosmic structure formation, which can then be matched to observations. We will explain what to do with $f_N$ in the next subsection. For instance, the particle number density of sterile neutrinos is computed as expected,
\begin{align}
 n_N\left(r\right)= \frac{g_N}{2\pi^2}\int\limits_{0}^{\infty}{\diffd \xi \DD{p}{\xi}p^2\left(\xi\right) f_N\left(\xi,r\right)} = \frac{g_N}{2\pi^2} \frac{g_S(T)}{g_S(T_0)}\left(\frac{m_0}{r}\right)^3\int\limits_{0}^{\infty}{\diffd \xi \; \xi^2 f_N\left(\xi,r\right)} \,,
\end{align}
where $g_N = 2$ counts the spin degrees of freedom of the sterile neutrino $N$. From the particle number density, the Dark Matter abundance follows as
\begin{align}
 \sub{\Omega}{DM}h^2 = \frac{s_0}{s\left(\sub{r}{prod}\right)}\cdot \frac{m_N n\left(\sub{r}{prod}\right)}{\sub{\rho}{crit}/h^2}\,,
 \label{eq:DMAbundance}
\end{align}
where $n\left(\sub{r}{prod}\right)$ and $s\left(\sub{r}{prod}\right)$ are the number and entropy desnsities at $r = \sub{r}{prod}$, respectively, $s_0 = 2891.2~{\rm cm}^{-3}$~\cite{Agashe:2014kda} is today's entropy density, and $\rho_{\rm crit}/h^2 = 1.054\cdot 10^{-2}~{\rm MeV}~{\rm cm}^{-3}$~\cite{Agashe:2014kda} is the critical density in units of the squared reduced Hubble constant $h$.

\subsection{\label{sec:Technicalities:Bounds}Relevant bounds}

By solving the set of Boltzmann equations for the scalar $S$ and for the sterile neutrino $N$, we obtain the momentum distribution function of the sterile neutrino, which contains all relevant information and must therefore be in accordance with a number of (partly model-dependent) observational or experimental constraints. In the following, we will present how this is tested and which assumptions go into these tests.

\paragraph{Mass of the sterile neutrino}
Since we neglected active-sterile mixing (as argued in Sec.~\ref{sec:QualitativeDiscussion} and in~\cite{Merle:2015vzu}), the momentum distribution function $f_N$ does not yet contain any information on $m_N$, but only on the number density of steriles present in the Universe. We can, however, fix the mass of the sterile neutrinos by demanding that they make up all (or a certain part) of the cosmic DM energy density, cf.~\equref{eq:DMAbundance}. This will later on result into the allowed regions marked in our plots.

\paragraph{Tremaine-Gunn bound}
The mass of the sterile neutrinos is also restricted by an absolute lower limit, the so-called Tremaine-Gunn (TG) bound~\cite{Tremaine:1979we}, which arises from fermions having a maximal phase space density. When applied to astrophysical objects such as the central regions of galaxies, the resulting mass turns out to be $m_N \gtrsim 0.5$~keV~\cite{Boyarsky:2008xj}. This bound will result into a relatively large excluded region in our plots, and it will in practice separate the regions of the scalars freezing out and freezing in.

\paragraph{Overclosure}
In principle, even for the lowest possible mass of $m_N = 0.5$~keV, a too large sterile neutrino number density will overclose the Universe (i.e., lead to $\Omega_{\rm tot} > 1$). Since this is not in agreement with observations, a certain patch of the parameter space is excluded by this bound. However, the overclosure bound turns out to be irrelevant in practice, since it is inferior compared to the TG bound.

\paragraph{Structure formation}
A simple estimator for the predictions of structure formation is the so-called free-streaming (FS) horizon $\sub{\lambda}{fs}$~\cite{Boyarsky:2008xj}, which yields an estimate of the average length scale (usually given at redshift $z=0$) that a DM particle would have travelled from the time of production $\sub{t}{prod}$ until today ($t_0$) -- had it not been gravitationally trapped. It can be computed via
\begin{align}
 \sub{\lambda}{fs} = \int\limits_{\sub{t}{prod}}^{t_0}{\diffd t \; \frac{\av{v\left(t\right)}}{a\left(t\right)}} \;.
 \label{eq:Def:LambdaFS}
\end{align}
A free-streaming horizon below something like $\unit{0.01}{Mpc}$ indicates a scenario that -- in terms of structure formation -- cannot be distinguished from cold Dark Matter (CDM), although the border is arbitrary to some extend (but reasonably well motivated to be about one order of magnitude below the border to HDM). A value of $\unit{0.1}{Mpc}$, the typical scale of dwarf galaxies, conventionally indicates the cross-over from models that suppress power on small scales (compared to CDM) in a way that is still in accordance with observational data from HDM models that suppress too much power. As already mentioned, in the literature this regime is mostly referred to by the very unfortunate term \emph{warm} DM, which seems to indicate a thermal spectrum. However, as we will see in the following, realistic distribution functions are in many cases (highly) \emph{non-thermal}, which may be very badly reflected by such an over-simplistic quantity like the FS horizon.

It is essential to bear in mind that $\sub{\lambda}{fs}$ is -- at best -- to be understood as an order-of-magnitude estimator. In fact, depending on the case under consideration, it might have to be modified by an $\orderof{1}$ factor to give more accurate classifications of models. Moreover, by its definition in \equref{eq:Def:LambdaFS}, it does not take into account the full spectral form of the DM distribution and will therefore never accurately treat non-thermal distributions. We have nevertheless exemplified this analysis (and its failure) for one selected mass of the scalar for the sake of a comparison, cf.\ App.~\ref{app:C:FSvsHalfMode}.

A more robust analysis uses the linear matter power spectrum $P\left(k\right)$, which reflects the full spectral information. For every point in the parameter space, we therefore compute the linear power spectrum using the \code{CLASS} code~\cite{Blas:2011rf,Lesgourgues:2011rh} for non-cold DM species, and then normalise it to the linear power spectrum of a perfectly cold distribution. This ratio, $P\left(k\right)/\sub{P}{CDM}\left(k\right)$ is also known as squared transfer function $\mathcal{T}^2\left(k\right)$, which gives information on how strongly the power (i.e., the number of halos formed) is suppressed compared to a perfectly cold distribution at the scale $k$. Another advantage is that this quantity can then be compared to limits on the squared transfer function obtained from Lyman-$\alpha$ data, displayed by a limiting squared transfer function $\sub{\mathcal{T}}{lim}^2\left(k\right)$.

Limiting functions $\sub{\mathcal{T}}{lim}^2\left(k\right)$ are usually obtained \emph{assuming} a thermal DM distribution (in that case truly warm DM). Accordingly, it is for principle reasons difficult to compare a given model to the observational limits. Ideally, one would re-analyse the Lyman-$\alpha$ data using the exact spectral shape of the DM distribution and the respective mass of the sterile neutrino. However, this is impossible given the virtually infinite number of possible distribution functions. Nonetheless, if $\mathcal{T}^2\left(k\right) \geq \sub{\mathcal{T}}{lim}^2\left(k\right)$ for all $k$, we can safely categorise the model as in agreement with the Lyman-$\alpha$ bound applied. If, on the other hand, $\mathcal{T}^2\left(k\right) < \sub{\mathcal{T}}{lim}^2\left(k\right)$ for all $k$, the model clearly has to be discarded.

\begin{figure}[t]
\begin{tabular}{lr}\hspace{-1cm}
 \includegraphics[width=8.3cm]{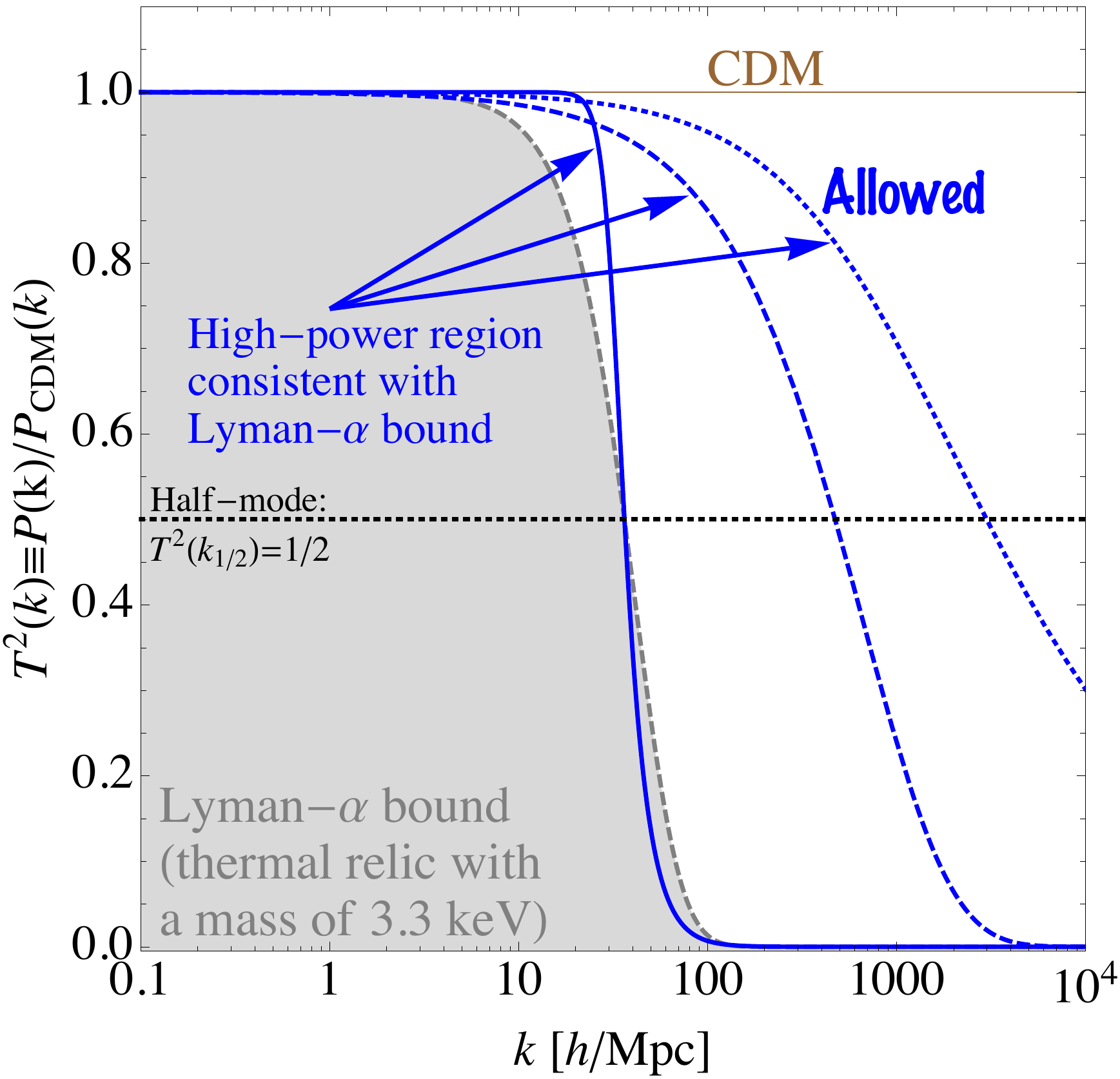} & \includegraphics[width=8.3cm]{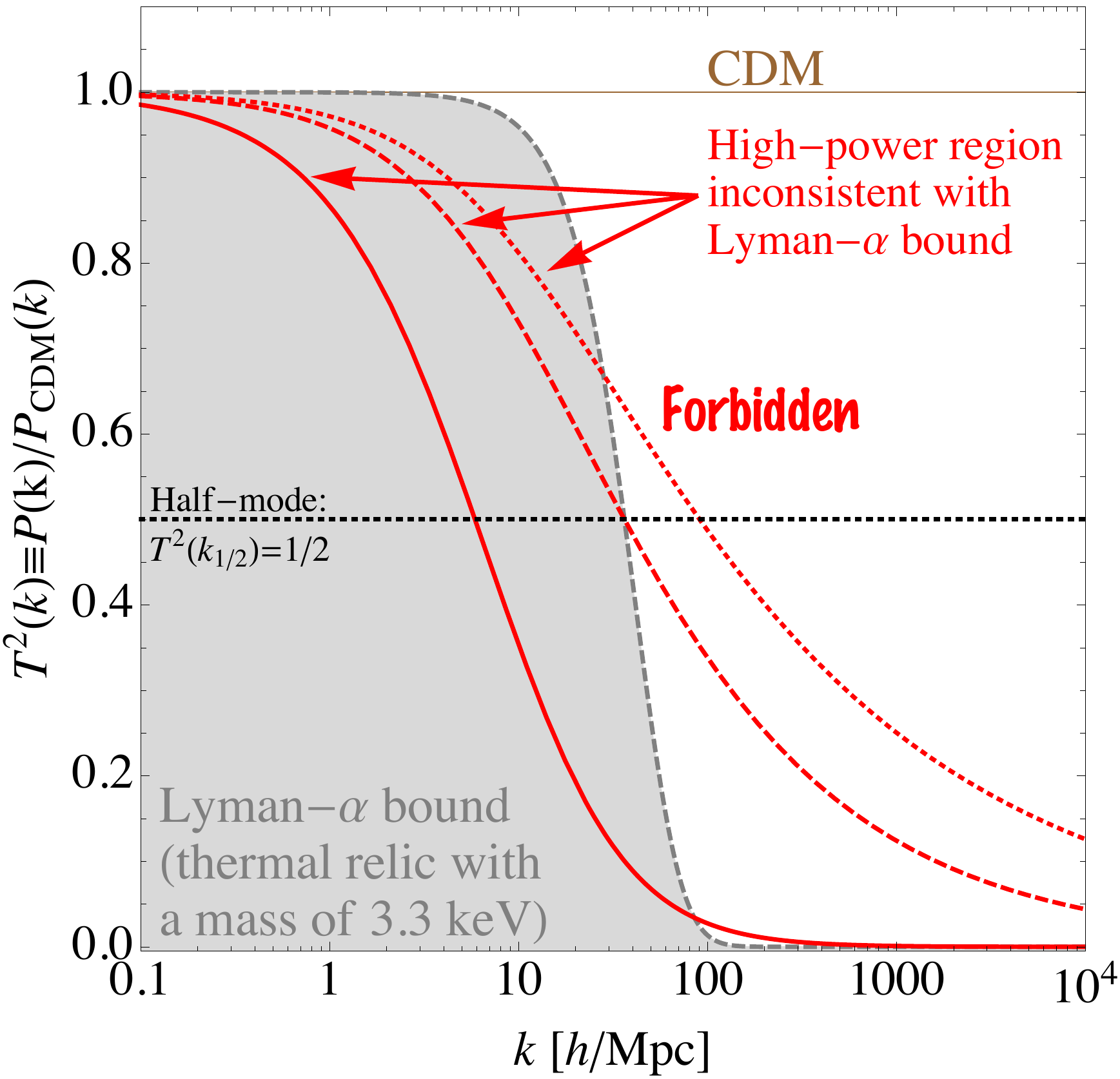}
\end{tabular}
\caption{\label{fig:Ly-alpha_illustration}Illustration of how we classify certain squared transfer functions as consistent (\emph{left}) or inconsistent (\emph{right}) with a certain bound.}
\end{figure}

There is a problem with this procedure, though. In some cases, $\mathcal{T}^2\left(k\right) \geq \sub{\mathcal{T}^2}{lim}\left(k\right)$ might only hold true for a certain interval of wave numbers $k$, but not for the  entire range. For example, $\mathcal{T}^2\left(k\right) \geq \sub{\mathcal{T}^2}{lim}\left(k\right)$ could be true for all $k$ smaller than a particular value $\tilde k$, but not anymore for $k > \tilde k$. This complication originates from the very fact that the limiting transfer functions are derived from thermal spectra, while realistic spectra can be highly non-thermal, which results in a difference in the exact evolution of the squared transfer function around its cutoff (e.g., the slope may be different).

In order to still use the limiting squared transfer functions, which are very good estimators accounting for observational relevance, we have developed the following procedure:
\begin{enumerate}

\item First, we compute the half-mode $k_{1/2}$, i.e., the wave number at which the squared transfer function has dropped to $1/2$:
\begin{align}
 k_{1/2}\ :\Longleftrightarrow\ \mathcal{T}^2\left(k_{1/2}\right) \stackrel{!}{=} 1/2 \;.
 \label{eq:Def:Half-mode}
\end{align}

\item Subsequently, we check whether $\mathcal{T}^2\left(k\right) \geq \sub{\mathcal{T}^2}{lim}\left(k\right)$ is fulfilled all $k \leq k_{1/2}$. If the condition is met within this range of small wave numbers (i.e., larger length scales), we consider the model as being consistent with a given bound. This is clearly an approximate method, since we intrinsically disregard some relevant information (namely the power spectrum below the half-mode). Furthermore, the value of $1/2$ is, ultimately, pure choice and one could equally well justify other values. However, it is clear that the border has to be somewhere between the two extreme cases of a certain squared transfer functions being completely untouched by the Lyman-$\alpha$ bounds (restrictive view) or by at least in small parts being allowed by the bound (conservation view), which is why a value of $1/2$ appears to be a fair compromise. We have in addition shown that our results are, in fact, quite robust with respect to this arbitrary choice: due to the power spectra considered still having a slope somewhat similar to that of a thermal relic, even a choice of $0.05$ instead of $0.5$ would not alter our results very significantly, cf.\ App.~\ref{app:D:HalfmodeThreshold}. We are currently working on more advanced analyses that will result into even more reliable procedures.

\end{enumerate}
This procedure is in fact rather simple, as can be seen from the cartoon-like illustration in Fig.~\ref{fig:Ly-alpha_illustration}. For the Lyman-$\alpha$ bounds, we use the squared transfer functions derived from thermal spectra with thermal masses of $m_{\rm lim} = \unit{2.0}{keV}$ ($m_{\rm lim} = \unit{3.3}{keV}$) in a conservative (more restrictive) scenario. Note that only the restrictive bound is displayed in Fig.~\ref{fig:Ly-alpha_illustration}, as the purpose of this figure is only to illustrate the principles behind the procedure, while later on we will display both bounds for completeness. The values quoted are motivated in~\cite{Viel:2005qj}, where the authors also provide an analytical fit formula for the transfer function of a given thermally distributed species of mass $m_{\rm lim}$, which is assumed to make up all the DM.

\paragraph{Bounds from the observed amount of dark radiation}
The combination of the momentum distribution function of the sterile neutrino and its mass is also constrained by the observation of the cosmic microwave background (CMB) and by the light element abundances related to big bang nucleosynthesis (BBN). The corresponding observables allow to constrain the effective number $\sub{N}{eff}$ of neutrinos and its deviation from the SM value of $3.046$~\cite{Mangano:2005cc}. The effective number of neutrinos is a measure of the radiation present in anything else than photons at a given epoch. It is therefore a measure of the Universe's expansion rate, which in turn leaves its imprint on both the CMB and the abundances of the elements produced during BBN. A similar analysis was shown in \cite[fig.~9]{Merle:2015oja}, however, with a small error in the numerical computation of the contribution to $\Delta\sub{N}{eff}$ from the non-cold sterile neutrinos. While this error led to an overestimation of their impact, these bounds would in any case only be relevant in the region that is already excluded by the requirement of DM not being classified as ``hot''. Of course the precise numbers have to be computed numerically, but it can easily be understood that the bounds from structure formation and from $\sub{N}{eff}$ must be closely related, since they both critically depend on how long DM stays ultra-relativistic.
  
Note that, apart from possibly impacting the region in the parameter space where the DM is produced extremely late, the dark radiation bound could in fact also be relevant for very small DM masses. However, as we have seen, any value of $m_N$ below $0.5$~keV is already excluded by the TG bound.

\paragraph{Model-dependent bounds on the most minimal particle physics setting}
So far, we have fixed the mass of the sterile neutrino by matching its number density to the observed DM energy density. In the most minimal setting, we could simultaneously demand that the mass of the sterile neutrino is generated by a VEV of the singlet scalar $S$, i.e., $m_N=y\av{S}$. This would allow us to use bounds from perturbative unitarity and bounds derived on the scalar mixing angle from its contribution to the $W$-mass. In principle, there are also direct collider bounds~\cite{Robens:2015gla}, but they are off our plots and do in practice not constrain the relevant region of the parameter space.

Following~\cite{Robens:2015gla}, we can bound the VEV of the scalar by:
\begin{align}
 \av{S} \geq \sqrt{\frac{3}{16 \pi}} m_S.
 \label{eq:PerturbativeUnitarity}
\end{align}
Using $\av{S} = \frac{m_N}{y}$ enables us to conclude an upper bound on the Yukawa coupling:
\begin{align}
 y \leq \frac{m_N}{m_S} \sqrt{\frac{16\pi}{3}} \,.
 \label{eq:PerturbativeUnitarity2}
\end{align}

A singlet scalar mixing with the Higgs via its VEV can also yield a radiative correction to the $W$-boson mass. Combining Eqs.~(8), (9), and~(11) from Ref.~\cite{Robens:2015gla}, we can deduce an upper limit on the Higgs portal coupling:
\begin{align}
 \lambda \leq \lambda^{\mathrm{max}} = y \sin^{\mathrm{max}}\left(2\alpha\right) \frac{|m_S^2 -m_h^2|}{2\sub{v}{EW} m_N} \,,
 \label{eq:W-mass-bound}
\end{align}
where $\sub{v}{EW}$ is the VEV of the SM-Higgs and $\alpha$ is related to the mixing of the scalars.\\

This completes our list of bounds which can possibly be relevant. As we will see, apart from the Lyman-$\alpha$ bound, the most relevant constraint arises from the TG bound. In the most minimal setting, collider-related constraints can become relevant, too, however these are strongly model-dependent and in fact very easy to circumvent. Finally, the overclosure and dark radiation bounds play no role in practice.


\section{\label{sec:Results}Results}

In this section, we present our main results for sterile neutrino Dark Matter. However, before discussing the sterile neutrinos themselves, it is very useful to understand the evolution of the singlet scalar in the early Universe, which will make many of our results much easier to grasp.

\begin{figure}[t]
\begin{tabular}{lcr}\hspace{-1cm}
 \includegraphics[width=5.5cm]{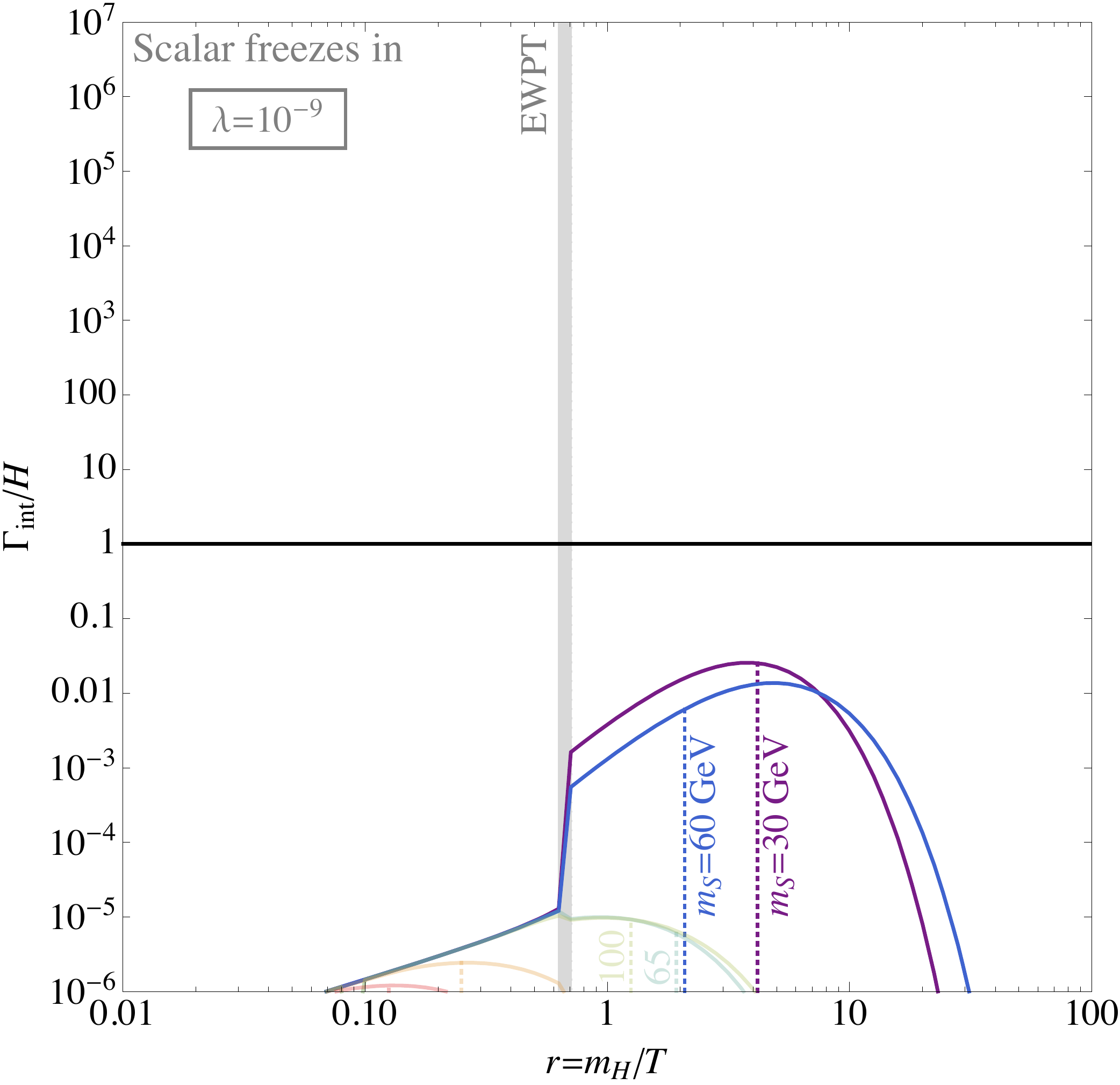} & \includegraphics[width=5.5cm]{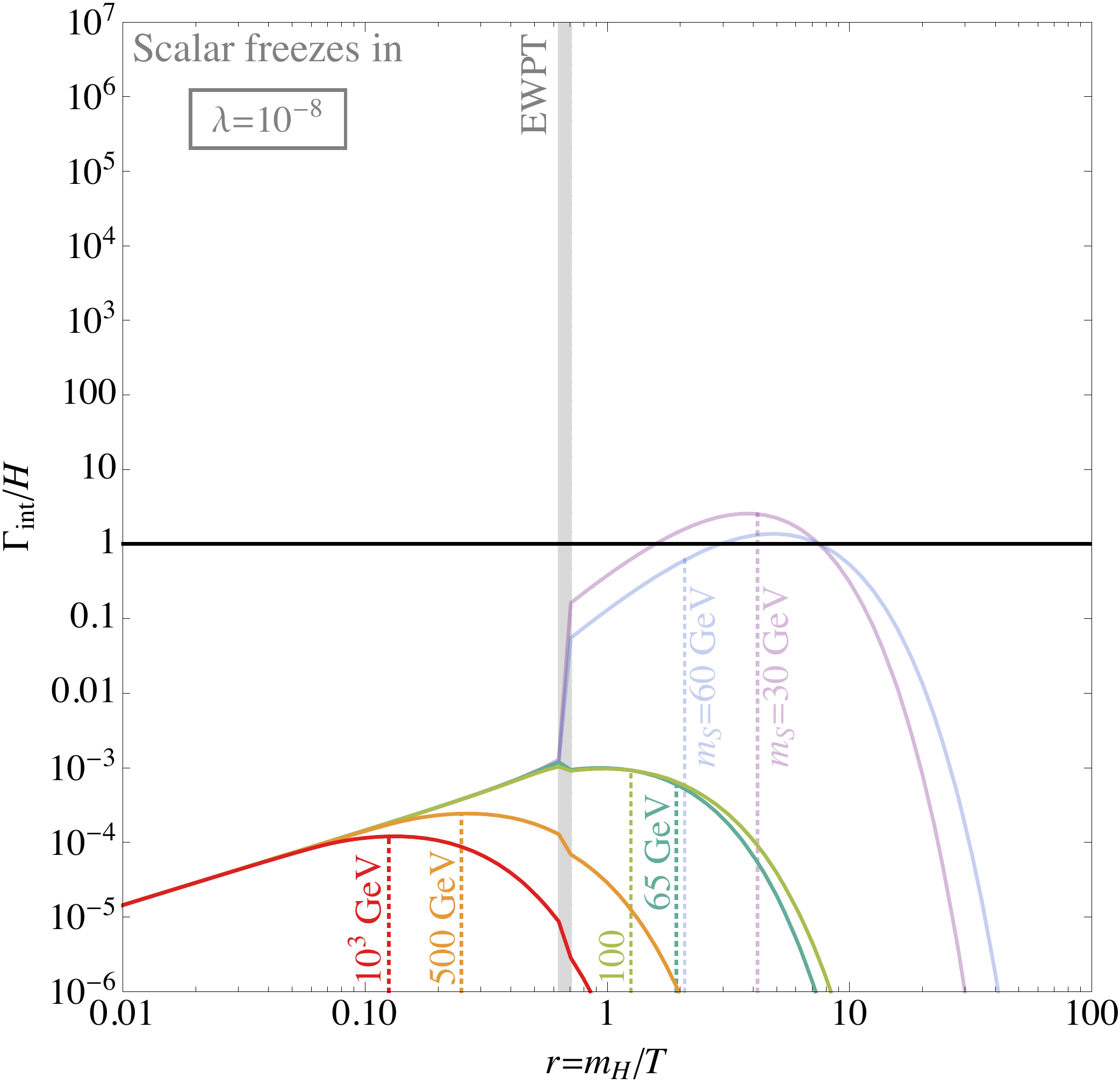} & \includegraphics[width=5.5cm]{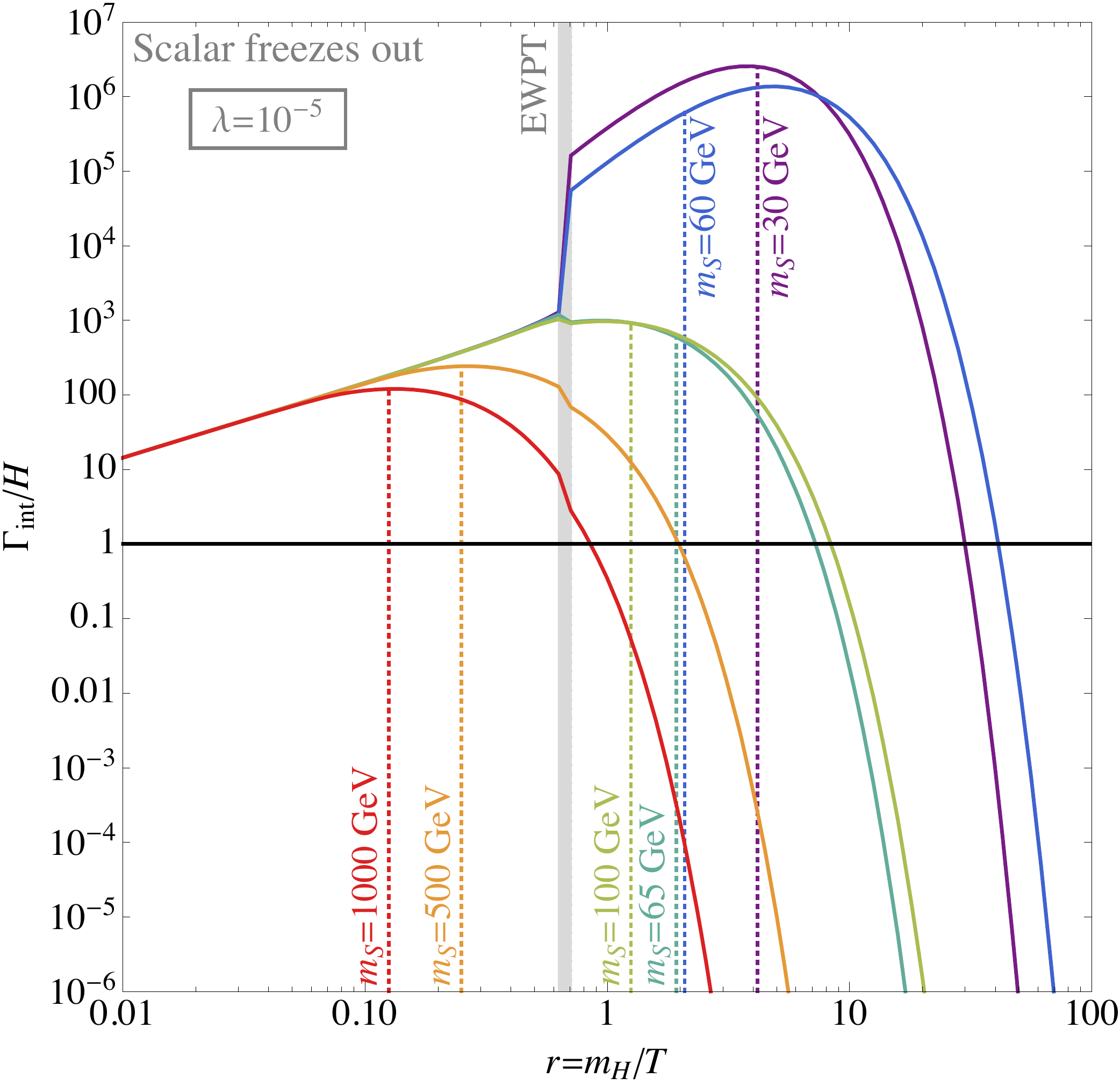}
\end{tabular}
\caption{\label{fig:interaction_rates}Interaction rates $\Gamma_{\rm int}$ compared to the Universe's expansion rate $H$, as functions of the time parameter $r=m_h/T$. We illustrate two freeze-in cases -- depending on the scalar mass we show either $\lambda=10^{-9}$ (\emph{left}) or $\lambda=10^{-8}$ (\emph{center}) -- as well as one freeze-out case, with $\lambda=10^{-5}$ (\emph{right}). Notably, in the latter, all scalars undergo a \emph{cold} freeze-out a temperatures within $[m_S/20, m_S/4]$, which explicitly proves that it is a good approximation to consider the scalar to be at rest at the point of freeze-out~\cite{Merle:2014xpa}, contrary to the claim made in~\cite{Bezrukov:2014qda}.}
\end{figure}

So let us, just for a moment, assume that the scalar is stable and understand what happens to it. As we had already mentioned, the scalar can either freeze in or freeze out in the early Universe, depending on its interaction strength $\lambda$. We have illustrated all the different cases in Fig.~\ref{fig:interaction_rates}, where we display the ratio between interaction rates $\Gamma_{\rm int}$ of the singlet scalar and the Universe's expansion rate $H$, as functions of the time parameter $r=m_h/T$. The interaction rates $\Gamma_{\rm int}$ are computed using all diagrams available in the different regimes, cf.\ Sec.~\ref{sec:QualitativeDiscussion}, which depending on the processes may receive contributions from different scattering or decay diagrams. As can be concluded from Tab.~\ref{tab:RegimesFeynman}, all diagrams are proportional to $\lambda$, which is why the evolution of the ratio $\Gamma_{\rm int}/H$ with $r$ looks identical for every plot in Fig.~\ref{fig:interaction_rates}, up to a rescaling. However, it is an important information whether $\Gamma_{\rm int}/H < 1$, in which case the scalar freezes in, or whether $\Gamma_{\rm int}/H>1$ and falls off only later on, in which case the scalar first thermalises and then freezes out.

\paragraph{Freeze-in} Let us start with the freeze-in case. This regime does depend somewhat sensitively on the scalar mass, which is why we have in Fig.~\ref{fig:interaction_rates} depicted the two cases of $\lambda=10^{-9}$ (\emph{left panel}: for $m_S = 30,~60$~GeV) and $\lambda=10^{-8}$ (\emph{central panel}: for $m_S = 65,~100,~500,~1000$~GeV). In all cases, we can clearly see that the scalars are always far from equilibrium.\footnote{In the central plot in Fig.~\ref{fig:interaction_rates}, one can also see the curves for $m_S = 30,~60$~GeV in light colours, which cases apparently can thermalise after a while. This would then lead to an ``overabundance'', at least for this part of the parameter space, which will later also be visible in our main plots -- the sterile neutrinos would need to be so light in order to compensate for the massive number density produced that they are already excluded by several bounds.} We can furthermore see the change in the interaction rates due to the new diagrams available after the EWPT. In particular for very small scalar masses, $m_S = 30$ or $60$~GeV, we can see that the interaction rate jumps up by some two orders of magnitude. This is because for such small masses the Higgs decay $h \to S S$ is available, which is the dominant contribution as noted in Ref.~\cite{Adulpravitchai:2014xna}. For larger scalar masses, in turn, there is hardly any difference visible (if at all, the interaction rates seem to decrease a little); however, the main reason is that larger scalar masses are thermally suppressed at this late stage, as most clearly visible for the largest scalar masses of $m_S = 500$ or $1000$~GeV. But even for medium-scale masses of $m_S = 65$ or $100$~GeV, a small drop in the rate is visible. This is in parts due to kinematics, since the $W$-, $Z$-, and Higgs bosons as well as the top quark are already thermally suppressed at this stage, cf.\ diagrams in Tab.~\ref{tab:RegimesFeynman}, but it happens also because of the different structures of the diagrams involved. In any case, the freeze-in of the scalar ceases at a temperature that is roughly equal to its mass, and indeed all interaction rates start to drop shortly after (or even before) $T = m_S$ is reached. A larger coupling $\lambda$ translates into a larger number density, as expected.

\paragraph{Freeze-out} For the larger coupling, $\lambda=10^{-5}$, scalars of all masses equilibrate. As visible in the \emph{right panel} of Fig.~\ref{fig:interaction_rates}, all interaction rates are larger than the expansion rate already for very early times. As long as the scalars are equilibrated, the actual interaction rate does not matter much. However, a boost in the rate -- as due to the EWPT for $m_S = 30$ and $60$~GeV -- can ``delay'' the freeze-out. This is particularly visible for a $m_S = 60$~GeV, which freezes out at $r \approx 41$ (or at $T\approx m_S/20$), and thus even later than the scalar with $m_S = 30$~GeV, for which freeze-out happens at $r\approx 30$ (or $T\approx m_S/7$). Thus, as we will later see to be correct, we can expect a much larger number density of scalars (and hence of sterile neutrinos) for $m_S = 30$~GeV than for $60$~GeV. Similarly, the case of $m_S = 65$~GeV should yield a larger abundance than the one for $m_S = 100$~GeV, which also turns out to be correct.

\paragraph{Sterile neutrinos} Let us now come to sterile neutrinos as DM. In order to include as much information as possible in the figures, without however making them too busy, we illustrate the allowed ranges and all relevant constraints as ``spaghetti plots'' displayed in Figs.~\ref{fig:verysmall_masses}, \ref{fig:small_masses},~\ref{fig:large_masses}. In these plots, we depict the regions in the $\lambda$--$y$ parameter space where the correct DM abundance is met (where only part of the abundance is produced) by the dark coloured solid lines (by the lightly coloured bands), for different values of the sterile neutrino mass: $m_N = 2, 7.1, 20, 50, 100$~keV. To obtain these regions, we have for each combination of the parameters $(\lambda, y, m_S)$ computed the resulting distribution function along the lines described in Sec.~\ref{sec:Technicalities}, and then integrated the result to obtain the DM abundance according to Eq.~\eqref{eq:DMAbundance}.

In addition, we display the TG phase space bound, as explained in Sec.~\ref{sec:Technicalities:Bounds}. The overclosure bound (gray area) marks the part of the parameter space where we would obtain overclosure of the Universe even for the minimum sterile neutrino mass of $0.5$~keV.

We also display the model-dependent bounds, which only hold in the most minimal setting, cf.\ Sec.~\ref{sec:Technicalities:Bounds}. While they may not even exist in more involved settings, they do serve the purpose of representing the \emph{maximal} effect collider-related bounds could possibly have. To indicate some numbers, the effect of the unitarity bound, Eq.~\eqref{eq:PerturbativeUnitarity2}, for a scalar mass 100~GeV is that the Yukawa coupling $y$ has to be smaller than about $8.2\cdot 10^{-8}$ [$2.9\cdot 10^{-7}$, $8.2\cdot 10^{-7}$] for $m_N = 2$~keV [$7.1$~keV, $20$~keV], in very good agreement with Figs.~\ref{fig:verysmall_masses}, \ref{fig:small_masses},~\ref{fig:large_masses}. For a scalar mass of 1000~GeV, these bounds get stronger by a factor of roughly~10.  The $W$-boson mass correction, Eq.~\eqref{eq:W-mass-bound}, is somewhat more subtle, since the scalar mixing angle $\alpha$ depends on the Higgs portal $\lambda$ which in turn influences the DM abundance and is thus related non-trivially to the Yukawa coupling $y$ and $m_N$. As discussed, these bounds do only exist in the most minimal model, which is why we have marked them by thin dashed lines (see areas left of the TG bound and down in the lower right corners). For some cases, they are even off the plot.

The main bound however arises from structure formation, where we take into account the two Lyman-$\alpha$ bounds derived in Ref.~\cite{Markovic:2013iza} and perform the half-mode analysis described in Sec.~\ref{sec:Technicalities:Bounds} to determine whether a given distribution function is consistent with the data, or not. We have for each scalar mass $m_S$ numerically computed the linear power spectrum for each pair $(\lambda, y)$. Reproducing the correct abundance results in a condition on the sterile neutrino mass $m_N$, so that every point with the correct abundance can be characterised by a point $(m_S, \lambda, y, m_N)$ in the parameter space. Depending on which Lyman-$\alpha$ bound we used, we have in our plots marked the following regions:
\begin{itemize}

\item \textcolor{red}{\bf forbidden}: If the upper half of the squared transfer function is forbidden by both the conservative and restrictive Lyman-$\alpha$ bounds, the respective point is \textcolor{red}{\bf red}.

\item \textcolor{purple}{\bf constrained}: If the upper half of the squared transfer function is only forbidden by the restrictive Lyman-$\alpha$ bound but allowed by the conservative one, the respective point is \textcolor{purple}{\bf purple}.

\item \textcolor{blue}{\bf allowed}: If the upper half of the squared transfer function is allowed by both the conservative and restrictive Lyman-$\alpha$ bounds, the respective point is \textcolor{blue}{\bf blue}.

\end{itemize}
This colour code is slightly reminiscent of the historically grown terms ``hot'', ``warm'', and ``cold'' DM, however, we would like to stress once more that the distributions obtained are \emph{non-thermal}, and one thus \emph{cannot} associate any temperatures with them; see App.~\ref{app:C:FSvsHalfMode} for an explicit counterexample showing that the classification drawn from thermal spectra is bound to fail for non-thermal distributions. Nevertheless it is correct to say that, roughly, the red regime is associated with rather high momenta compared to the DM mass, while the blue one tends to feature smaller momenta. By the light colours, we have indicated regions with a sizable but insufficient abundance for a given mass. Technically, these scenarios correspond to a slightly larger mass which yields the same effect as replacing a fraction of, say 10\% of the Dark Matter by a perfectly cold Dark Matter component. We have checked that this simplified procedure yields correct results up to the level of accuracy limiting analyses using Boltzmann equations anyway.

A final comment on the unavoidable but tiny contribution by Dodelson-Widrow production at temperatures of a few 100~MeV: as had been shown in Ref.~\cite{Merle:2015vzu}, this modification does not only hardly contribute to the abundance for sterile neutrino masses of 4~keV and higher, also the modification of the actual spectrum is completely negligible. Thus, in our plots, there would be no effect visible even if we did take into account the largest modification allowed by data. This may only be different for a sterile neutrino mass of 2~keV, but even then the error by not considering the modification is of about 10\%, which is in practice still negligible. For even smaller masses, the correction could be sizable, but in such cases a mixed scenario would be excluded by structure formation anyway.

\subsection{\label{sec:Results_VeryLight}Very light scalars: $\boldsymbol{m_S < m_h/2}$}

\begin{figure}[t]
\begin{tabular}{lr}\hspace{-1cm}
 \includegraphics[width=8.3cm]{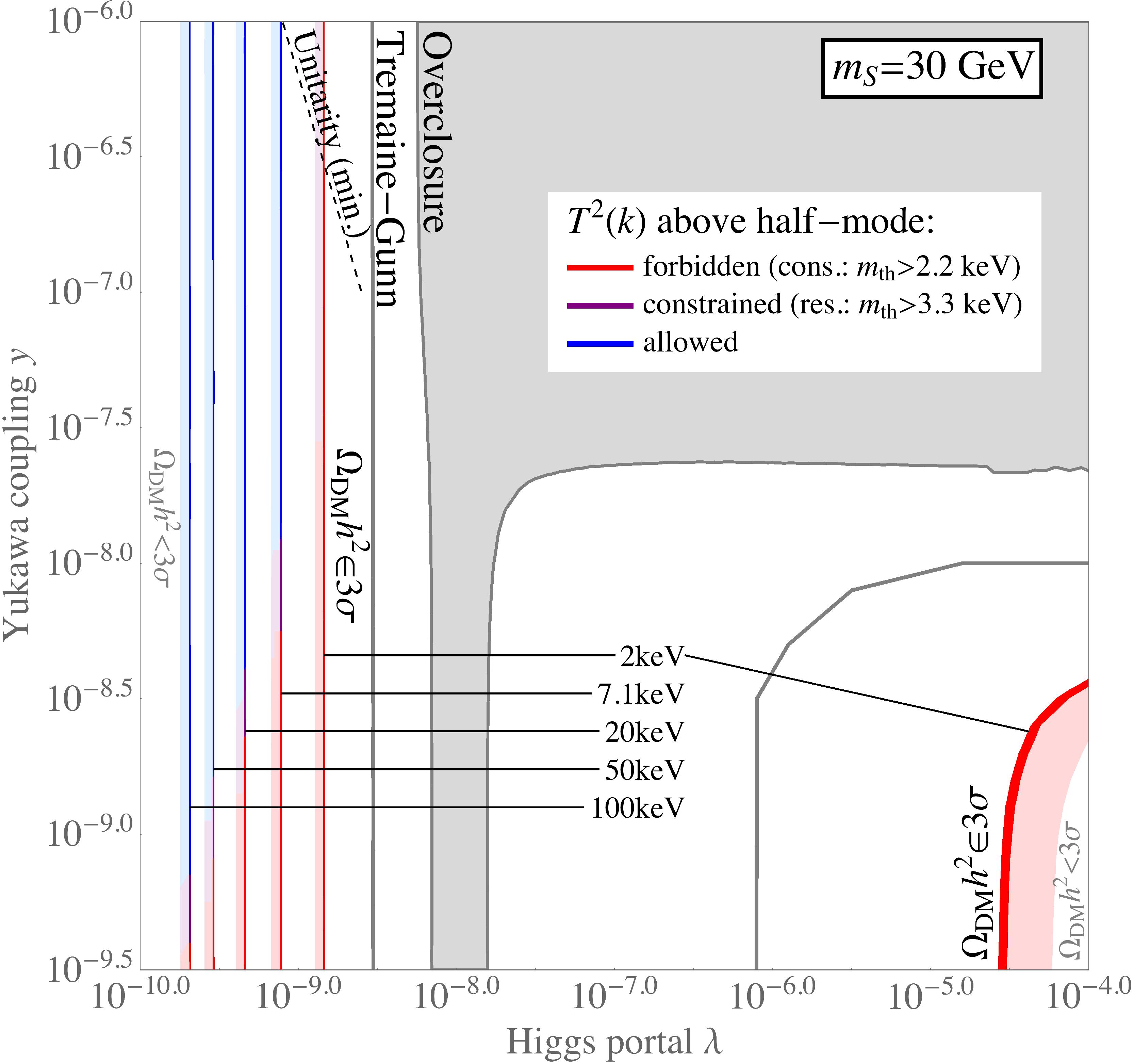} & \includegraphics[width=8.3cm]{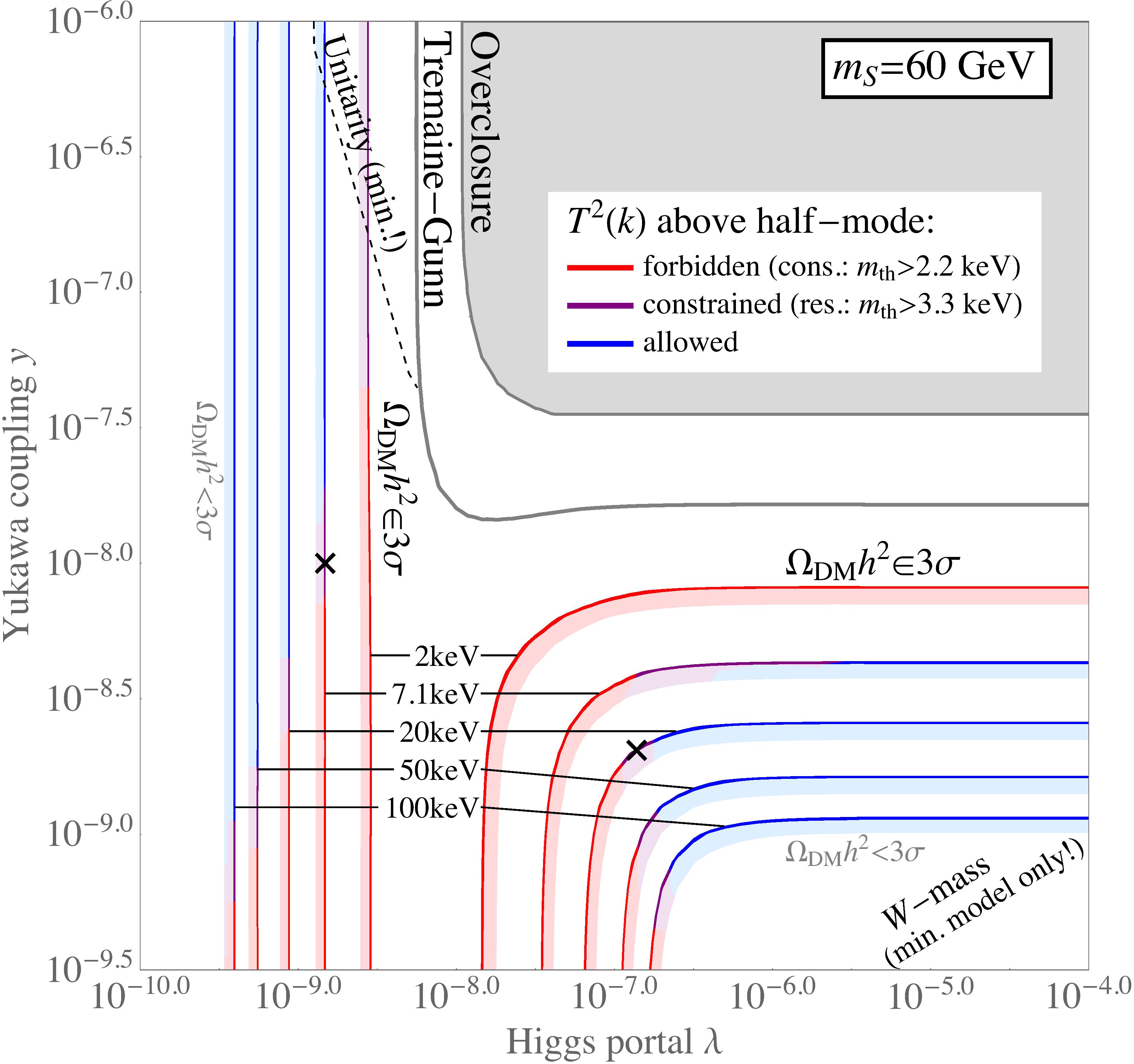}
\end{tabular}
\caption{\label{fig:verysmall_masses}Abundance and constraints for small scalar masses, $m_S < m_h/2$. The crosses mark the points displayed explicitly in Fig.~\ref{fig:FIMP_WIMP_60}.}
\end{figure}

Let us begin with the case of very light scalar masses, i.e., $m_S < m_h/2$, displayed in Fig.~\ref{fig:verysmall_masses}. As discussed in Sec.~\ref{sec:QualitativeDiscussion}, this leaves us with an interval of roughly $5~{\rm GeV} \lesssim m_S \lesssim 62~{\rm GeV}$. Here, the lower limit arises from the temperature where the relevant interactions are frozen out and/or not efficient anymore, while the upper limit is obtained from $m_h \simeq 125~{\rm GeV}$. Here, depending on the coupling $\lambda$, we may enter different regimes. First of all, for very small couplings $\lambda$, we will be in the FIMP region, i.e., in the left part of the plots. Freeze-in generically is most efficient at temperatures $T$ around the mass $m_S$ of the FIMP, i.e., in regime~III (see Fig.~\ref{fig:RegimeExampleCases} and Tab.~\ref{tab:RegimesFeynman}). Of course the production of scalars has started in regime~I, at high $T$, but in any case -- even with all diagrams of regime~III at work -- the interaction remains feeble. Once the scalar is produced, it decays with a certain lifetime proportional to $y^{-2}$ into sterile neutrinos. Thus, for $y$ large, the decays happen very close to the freeze-in of $S$ and thus the resulting DM particles will be more strongly redshifted and ``cooler''. For too small $y$, in turn, the scalars remain present in the Universe and decay only very late, thus injecting too much energy into the sterile neutrinos and rendering them HDM -- and thus excluded. Notably, as can be seen in the plots, the actual mass of the scalar within the allowed interval does not make much of a difference. The decay rate is proportional to $y^{2} m_S$, such that a larger scalar mass $m_S$ should translate into more red-shift, both because the correct abundance is obtained earlier and and on top of that the decay happens faster. However, for a fixed $y$, the unboosted decay rate also becomes smaller, such that the lifetime effectively increases, leaving less time for the steriles to redshift. To a first approximation, these two effects compensate each other, but of course there will be corrections due to scalars of different masses being boosted more or less (and thus having bigger or smaller lifetmes in the cosmic frame). Apart from these minor effects, the FIMP regions for $m_S=\unit{30}{GeV}$ and $m_S=\unit{60}{GeV}$ appear rather similar, up to a shift in the Higgs portal $\lambda$.

On the other hand, for larger values of $\lambda$, the scalar can enter thermal equilibrium and then freeze out, see the lower right regions of the plots. Depending on the exact combination of $(\lambda, y)$, the decay into sterile neutrinos happens while the scalar is in equilibrium (horizontal part of the isoabundance-lines -- which is independent of $\lambda$ from a certain value on), after freeze-out (vertical part -- which is similar to freeze-in in the sense that the scalars cool down before their decay and thus ``forget'' about their cosmological history for a long enough lifetime or small enough coupling $y$), or in between (kink-region -- where a strong dependence on both $\lambda$ and $y$ is present). For this part of the plot, the question of whether there is at all any allowed region depends strongly on $m_S$. This observation can be understood by realising that the scalar freezes out earlier and thus with a much larger abundance for $m_S = 30$~GeV, which is due to the dependence of the interaction rates on the scalar masses. Hence, given many more scalars than for $m_S = 60$~GeV, a too large abundance can only be avoided for sufficiently small sterile neutrino masses -- which is why only one strip is visible in the lower right corner of the left plot, and this one is forbidden by structure formation due to the small DM mass favouring a ``hotter'' spectrum, i.e., with a stronger tendency for large momenta.

\begin{figure}[t!]
\begin{tabular}{cc}
 \hspace{-1cm}\includegraphics[width=7.9cm]{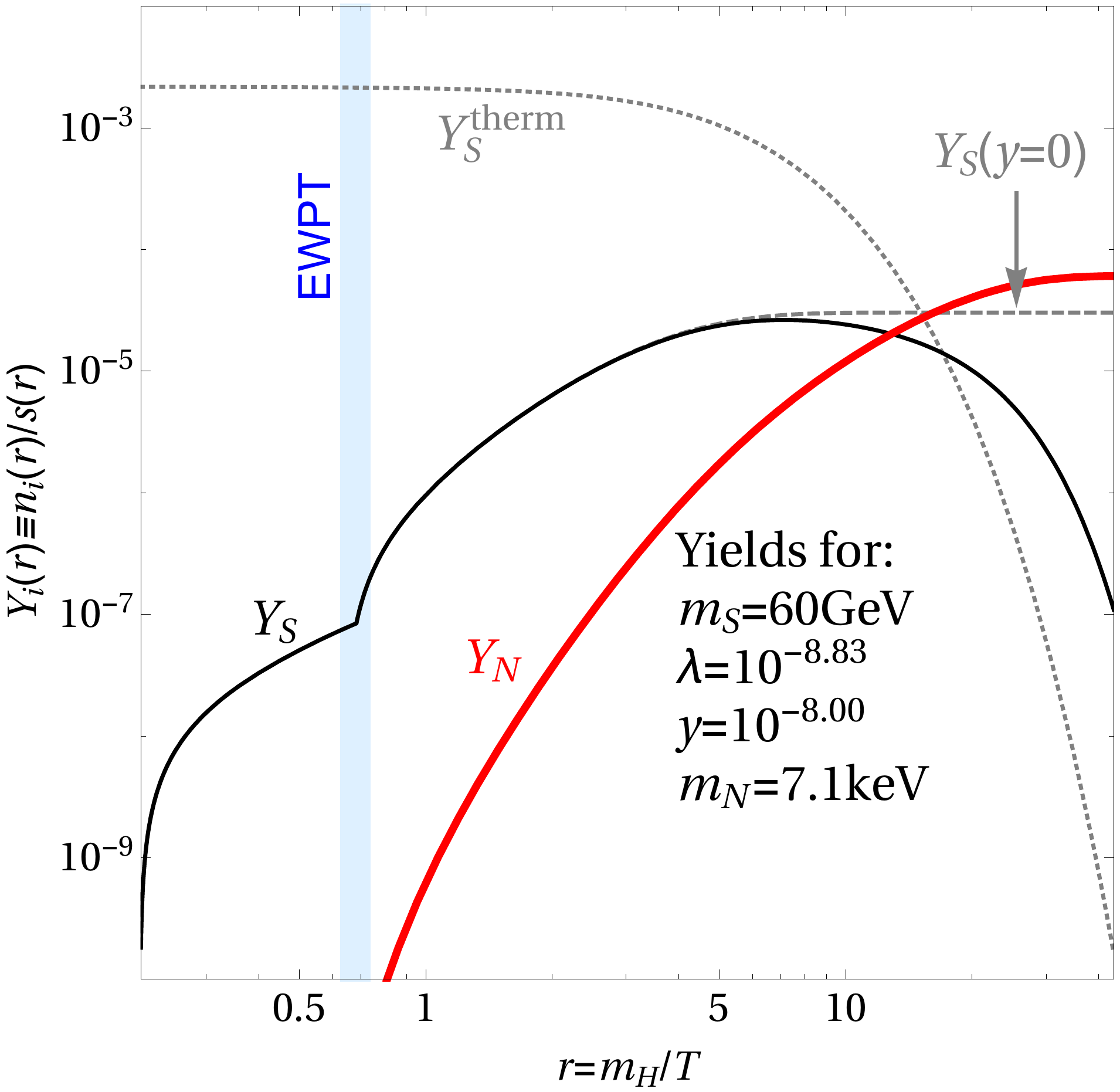} & \includegraphics[width=7.9cm]{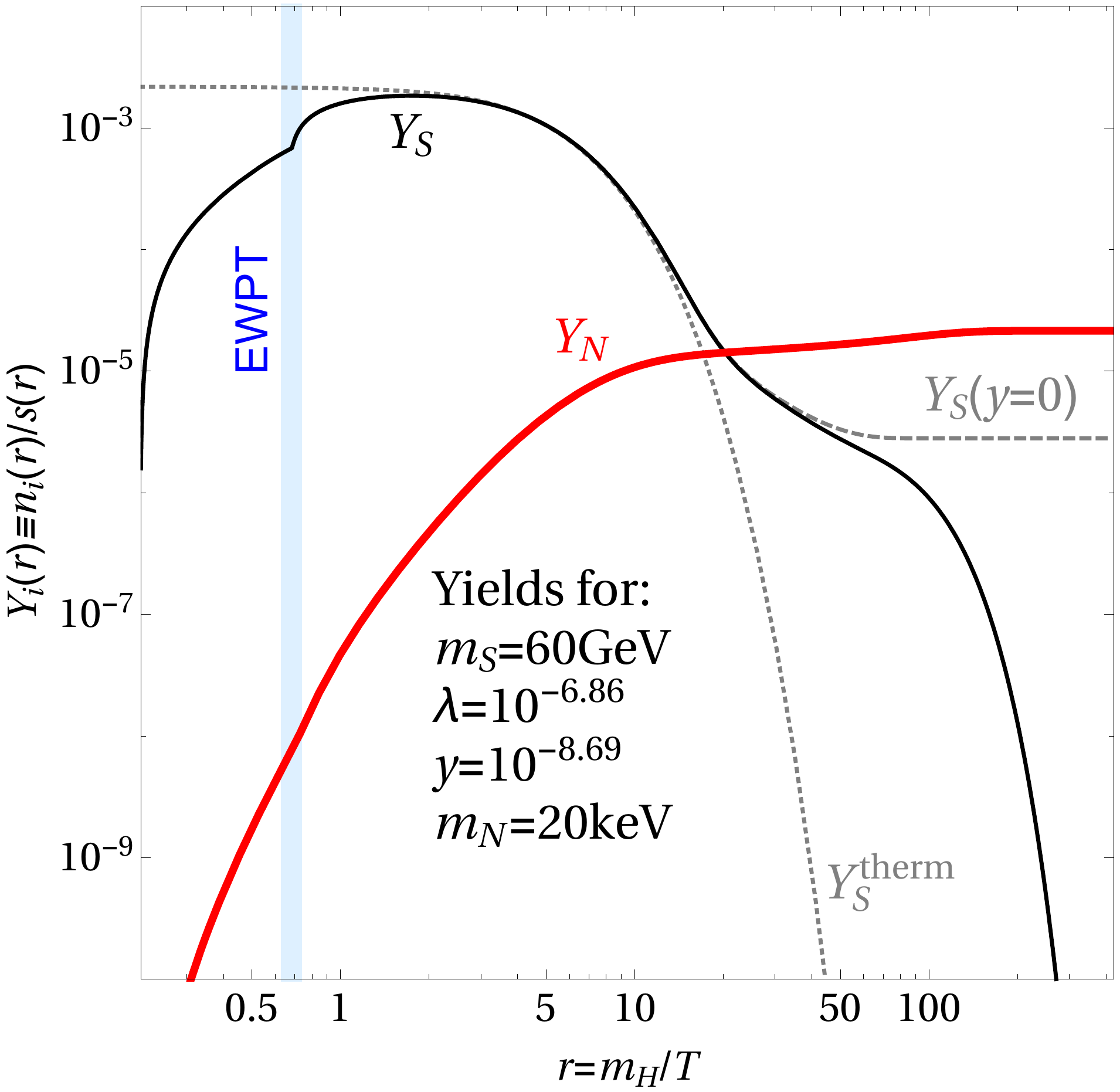}\\
 \hspace{-1cm}\includegraphics[width=8.3cm]{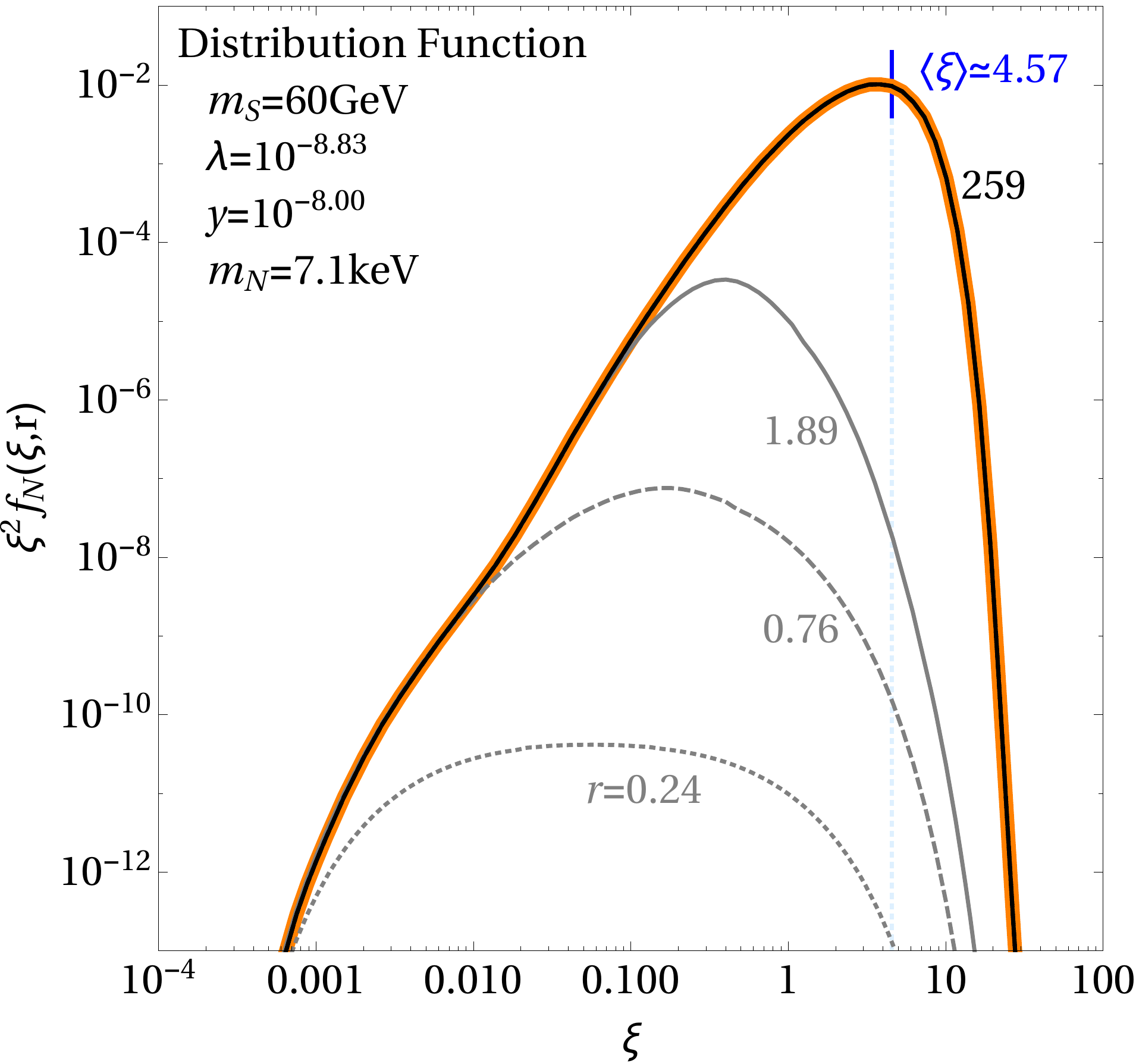} & \includegraphics[width=8.3cm]{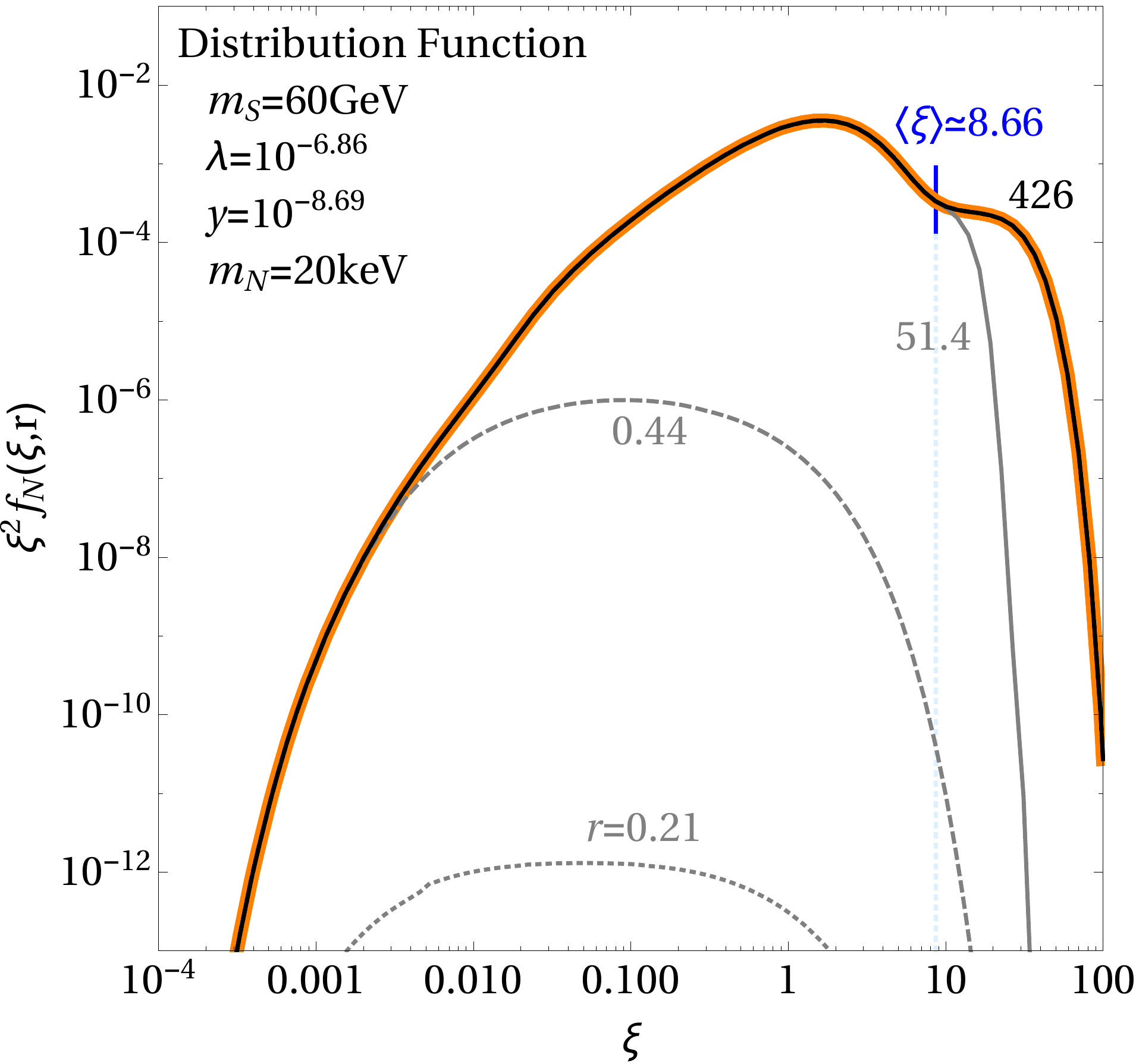}
\end{tabular}
\caption{\label{fig:FIMP_WIMP_60}Example evolutions of the yield (\emph{top row}) and sterile neutrino distributions (\emph{bottom row}) for a scalar with a mass of $60$~GeV undergoing freeze-in (\emph{left}) or freeze-out (\emph{right}) before decaying into sterile neutrinos, for the two points marked in the right Fig.~\ref{fig:verysmall_masses}.}
\end{figure}

Two representative example evolutions of the yield can be seen in the top row of Fig.~\ref{fig:FIMP_WIMP_60}, both for the scalar freezing in (\emph{left}) or out (\emph{right}), which correspond to the crosses marked in the right Fig.~\ref{fig:verysmall_masses}. If the scalar was stable (gray dashed line), it would compare trivially to the hypothetical equilibrium curve (gray dotted line): as a FIMP (\emph{left}), it would gradually be produced but never reach a thermal abundance before freezing in, while as WIMP (\emph{right}) it would quickly thermalise before freezing out. Once the decay into sterile neutrinos is switched on, the true abundance of the scalar (black solid line) again increases initially but ultimately decreases to zero. Note that, in both plots, one can see the increase in the interaction rate due to the EWPT, cf.\ Fig.~\ref{fig:interaction_rates}. The sterile neutrino abundance (red solid line), in turn, increases until the point where ultimately all scalars available have decayed.

Let us make a quick comparison to the results obtained in Ref.~\cite{Adulpravitchai:2014xna}. Looking at the top two plots in their Fig.~2, which feature the same scalar masses as our top two plots in Fig.~\ref{fig:FIMP_WIMP_60}, our yields look qualitatively very similar except for a missing ``bump'' in the yield of the scalar for temperatures around $10$~GeV, which appears in Ref.~\cite{Adulpravitchai:2014xna} but not in our result. While we cannot explain this feature, we do however have good agreement with the overall abundance obtained in~\cite{Adulpravitchai:2014xna},\footnote{To see this, one has to adjust the pair $(\lambda,y)$ to the case from~\cite{Adulpravitchai:2014xna}, though, taking into account the different normalisation of $\lambda$ in their Eq.~(5) compared to our Eq.~\eqref{eq:ModelLagrangian}.} up to some 20\% difference, which can be attributed to applying different numerical procedures and also due to Ref.~\cite{Adulpravitchai:2014xna} using some approximations such as rate equations assuming a (suppressed) thermal shape.

The evolution of the resulting sterile neutrino distributions for the same two points is displayed in the bottom row of Fig.~\ref{fig:FIMP_WIMP_60}, with all spectra being functions of $\xi = \left( \frac{g_{s}(T_0)}{g_{s}(T)} \right)^{1/3}\;\frac{p}{T}$, cf.\ the second Eq.~\eqref{eq:xi_and_r_definition}. We can see a certain deformation in both curves, although the one on the left (for the scalar being a FIMP), it is probably a small bump rather than an actual double peak. In both cases, these shapes come from two different production phases. In the FIMP case (\emph{left}), the two phases correspond to the production before and after the EWPT, whereas the visible bump for larger $\xi$ for the WIMP case (\emph{right}) rather comes from production while the scalar is still in equilibrium and after its freeze-out, respectively.\footnote{Note that, in the spectrum on the right, a tiny bump stemming from the changes during the EWPT is nevertheless visible at around $\xi \sim 0.03$, thereby implying a third momentum scale. However it is so tiny that it does not have a real influence on the spectrum.}

\begin{figure}[t]
\begin{tabular}{lr}\hspace{-1cm}
 \includegraphics[width=8.3cm]{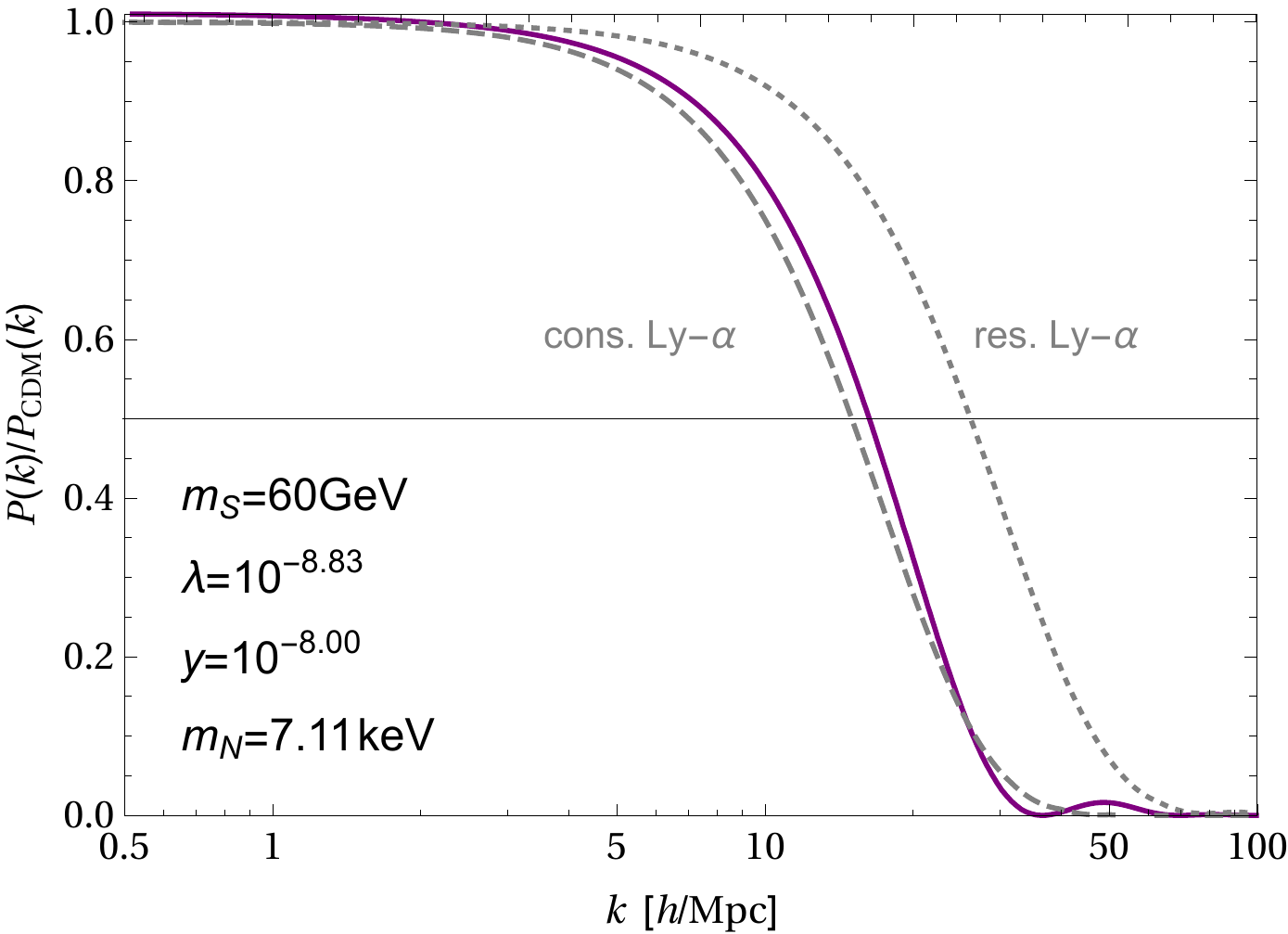} & \includegraphics[width=8.3cm]{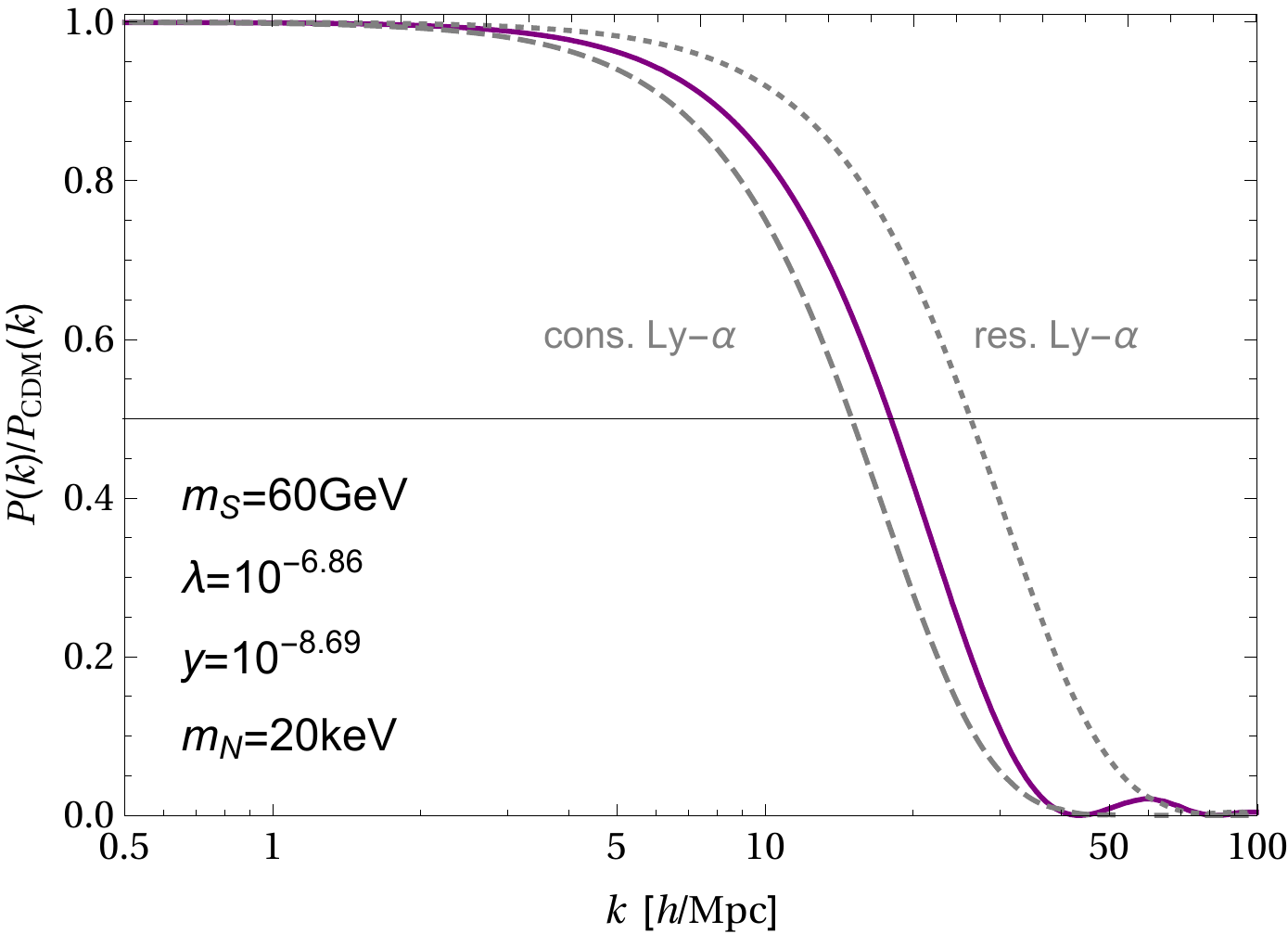}
\end{tabular}
\caption{\label{fig:TF_60}Example squared transfer functions for a scalar with a mass of $60$~GeV undergoing freeze-in (\emph{left}) or freeze-out (\emph{right}) before decaying into sterile neutrinos, for the two points marked in the right Fig.~\ref{fig:verysmall_masses}.}
\end{figure}

Finally, in Fig.~\ref{fig:TF_60}, we display the squared transfer functions for the two example points. Glancing at the right Fig.~\ref{fig:verysmall_masses}, we can see that both points lie inside ``purple'' regions, i.e., they are constrained but not excluded by the Lyman-$\alpha$ bounds. According to the procedure detailed in Sec.~\ref{sec:Technicalities:Bounds}, this would imply that the parts of the squared transfer functions with $k \leq k_{1/2}$ should both be consistent with the conservative Lyman-$\alpha$ bound, but inconsistent with the restrictive one. This is just what we can see in Fig.~\ref{fig:TF_60}. It is also visible that, in particular for the FIMP case depicted on the left, the slope of the functions is different from that of the bounds, for which the distributions were thermal by construction. Furthermore, note that the sterile neutrino mass on the left plot, $7.1$~keV, is significantly different from the mass of $2$~keV corresponding to the conservative bound, which originates from the non-thermal shape of the scalar-decay produced DM spectrum. Obviously, according to the left Fig.~\ref{fig:TF_60}, one would classify the case displayed as ``warmer'' than the restrictive bound corresponding to a mass of $3.3$~keV, even though the true DM mass is even larger than that. This is one more reflection of conclusions which do hold for thermal spectra being invalidated once the shape of the momentum distribution function is more complicated. Thus, any ``translation'' of the structure formation properties to a hypothetical spectrum of thermal shape, as attempted for DW-produced sterile neutrinos on several occasions in the literature (e.g.\ Refs.~\cite{Colombi:1995ze,Viel:2013apy,Markovic:2013iza,Abada:2014zra,Popa:2015eta,Bozek:2015bdo,Baur:2015jsy}), must be treated with extreme care and should not be regarded as a precision statement.

We can also see that, although both of the points that we had explicitly illustrated above fall into the same category in what concerns structure formation, the FIMP case tends to be slightly ``warmer'', i.e., closer to being excluded than the WIMP example. This may come as a small surprise, given that the average momentum of the WIMP-produced sterile neutrino is nearly two-times larger than the FIMP-produced one.\footnote{Note that, while one has to be slightly careful when converting the quantity $\xi$ to a physical momentum, cf.\ Sec.~\ref{sec:Technicalities:BoltzmannEquation-solution}, it can still be used to compare the relative average momenta of two given spectra.} However, when dividing the average momenta by the sterile neutrino masses, hence looking at the average velocities, the factor of nearly two remains present -- just that in this case the FIMP-produced particle yields the larger value.

Looking back at Fig.~\ref{fig:verysmall_masses}, to get a more global picture, it is visible in the freeze-in (\emph{left}) regions that for larger Yukawa couplings $y$ the DM particles get ``colder'' (i.e., shifted to smaller momenta). This is due to the parent particles decaying earlier, thus leaving more time for the DM particles to redshift. Of course, this effect is more pronounced when the DM particles are heavier, see upper left corners of the plots. Interestingly, the unitarity bound would cut off part of that otherwise allowed parameter space (but only if $m_N = y \langle S \rangle$). For the freeze-out (\emph{right}) regions, instead, the case of small Yukawa coupling corresponds to a very late decay, always far after the freeze-out. This leaves the DM particles no time to cool down, so that they will generically be threatened by structure formation. The second limit of large $\lambda$ keeps the scalars in equilibrium for a very long time, so that practically all sterile neutrinos are produced while the scalar is still equilibrated. That leads to a ``colder'' spectrum, because the DM particles have more time to redshift. The turnover, corresponding to the double peak structure observed in Ref.~\cite{Merle:2015oja} for the first time, can be somewhere in between. Depending on the (non-trivial) interplay between the different parameters and in particular on the relative strength of the two peaks, it can be ruled in or out by structure formation.

\subsection{\label{sec:Results_Light}Light scalars: $\boldsymbol{m_h/2 < m_S < m_h}$}

\begin{figure}[t]
\begin{tabular}{lr}\hspace{-1cm}
 \includegraphics[width=8.3cm]{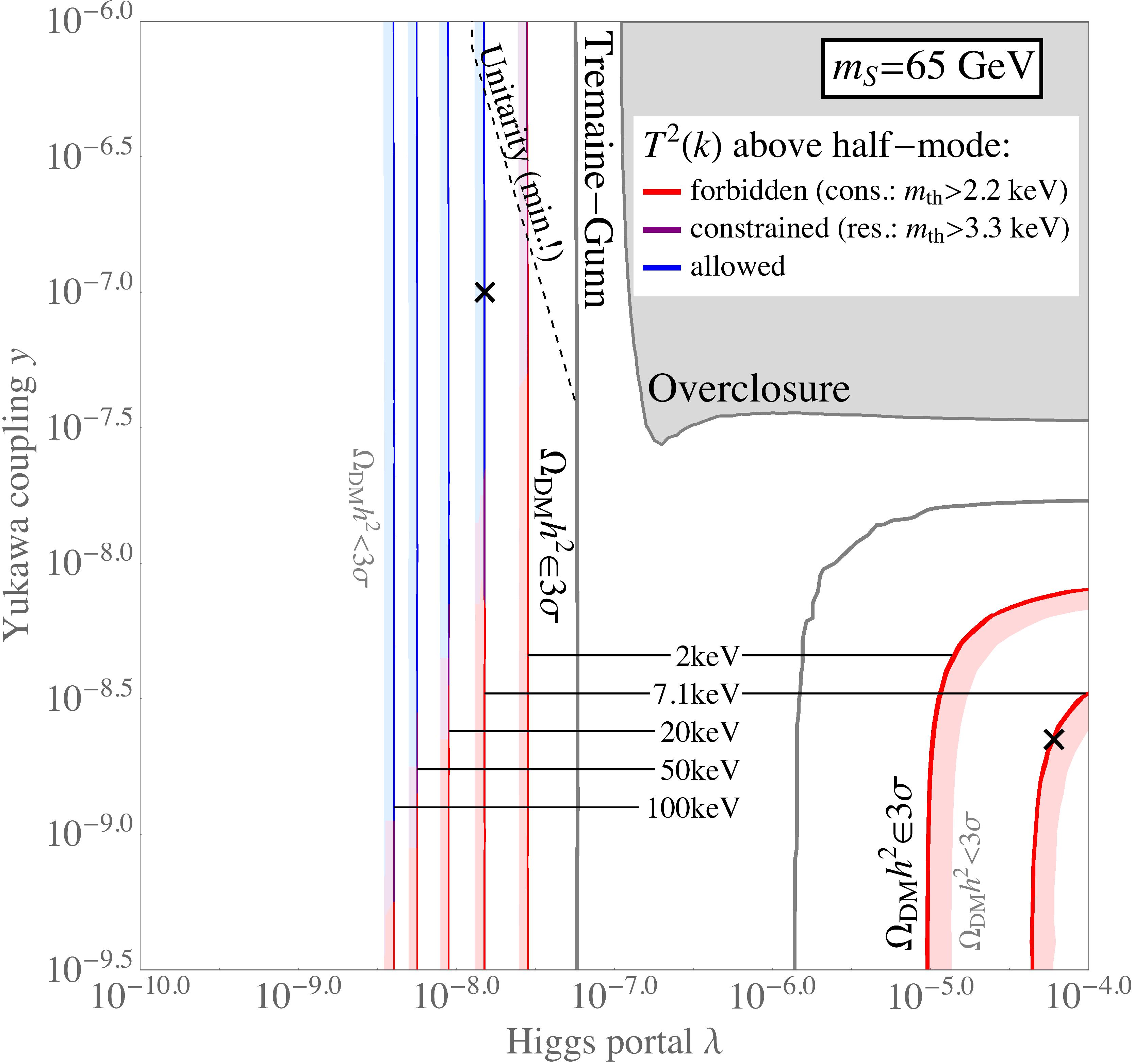} & \includegraphics[width=8.3cm]{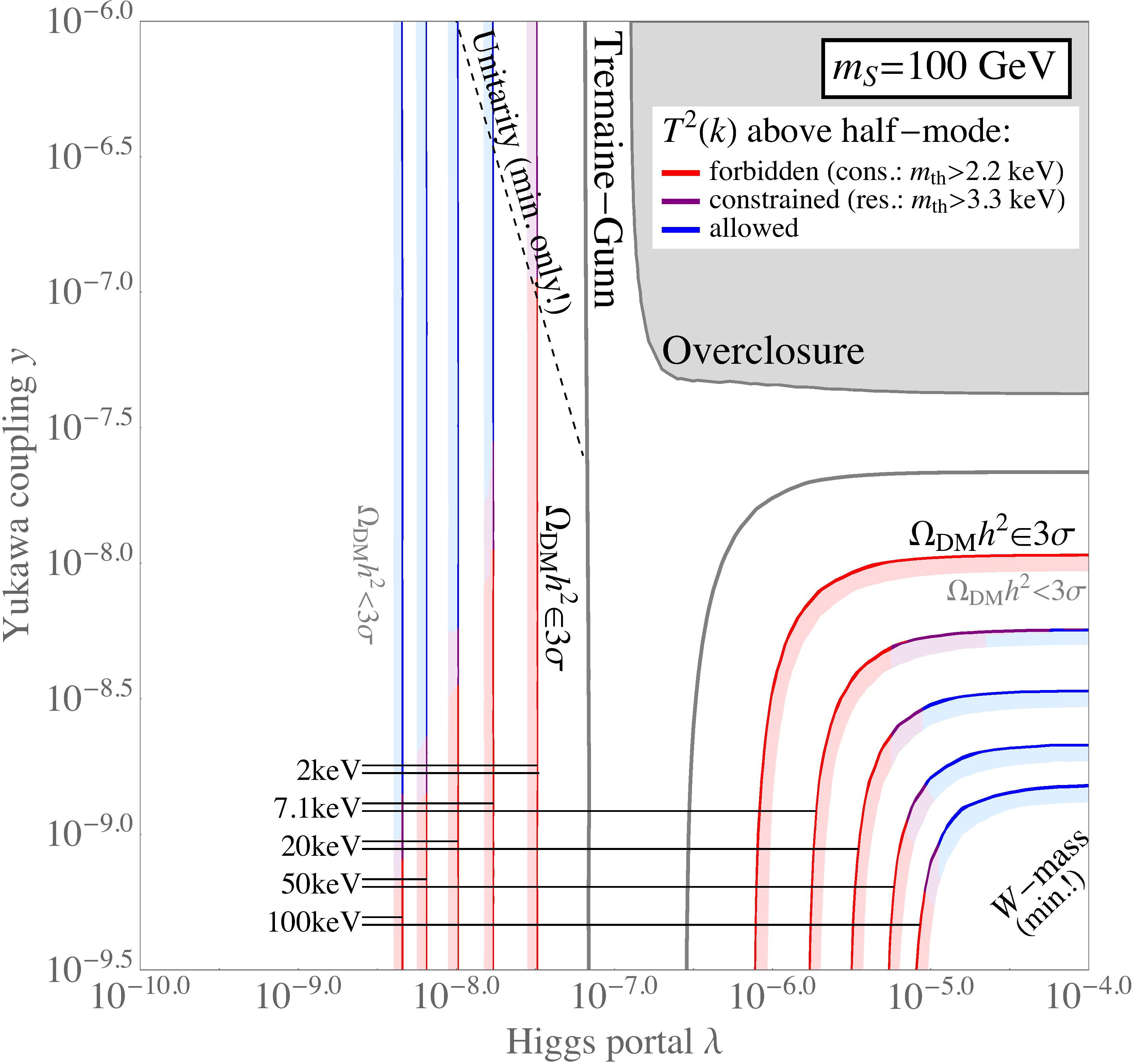}
\end{tabular}
\caption{\label{fig:small_masses}Abundance and constraints for intermediate scalar masses, $m_h/2 < m_S < m_h$. The crosses mark the points displayed explicitly in Fig.~\ref{fig:FIMP_WIMP_65}.}
\end{figure}

The next case to be discussed, cf.\ Fig.~\ref{fig:small_masses}, is that of ``light'' scalars, in the sense that their mass is still less than the mass $m_h$ of the Higgs boson, but larger than half the Higgs mass. This corresponds to regime~II in Tab.~\ref{tab:RegimesFeynman} for low temperatures $T$, while the production nevertheless starts at high $T$ and thus in regime~I, cf.\ Fig.~\ref{fig:RegimeExampleCases}. Here, the main characteristic is that, at the EWPT, all of a sudden lots of new channels open up. For the freeze-in case, i.e.\ small $\lambda$, this means that it is easier to produce scalars at a temperature just above their mass, which will result in an increase in their abundance and thus also in that of sterile neutrinos -- however this increase is much less dramatic than for the very light scalars, and hence hardly visible in the plots, due to the Higgs decay into two scalars not anymore being accessible in this mass range. For the freeze-out case, i.e.\ large $\lambda$, this means that the scalars will be kept in equilibrium for a longer time in regime~II, resulting into a large overall number of sterile neutrinos produced while the scalar is still equilibrated.

\begin{figure}[t!]
\begin{tabular}{cc}
 \hspace{-1cm}\includegraphics[width=7.9cm]{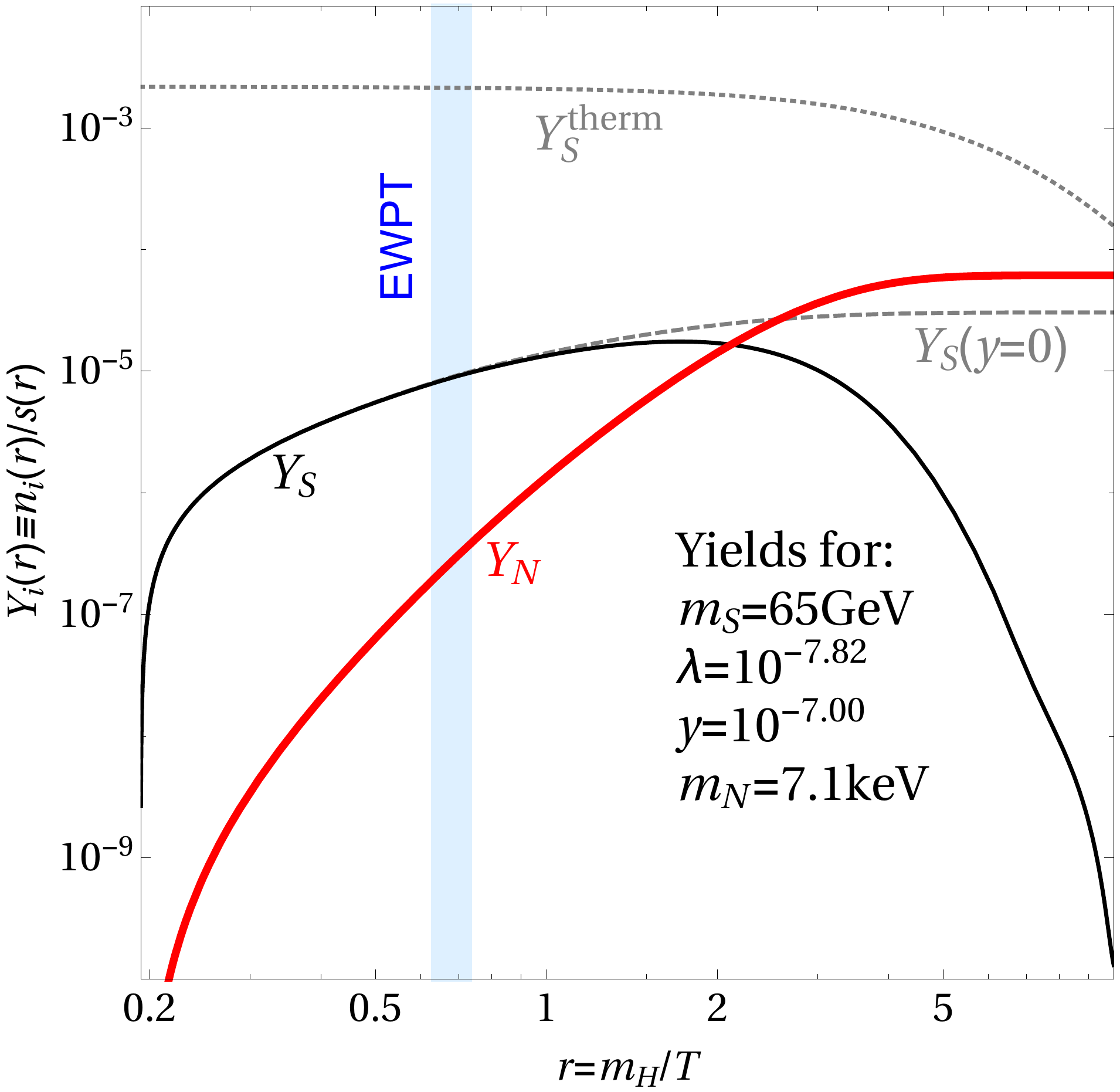} & \includegraphics[width=7.9cm]{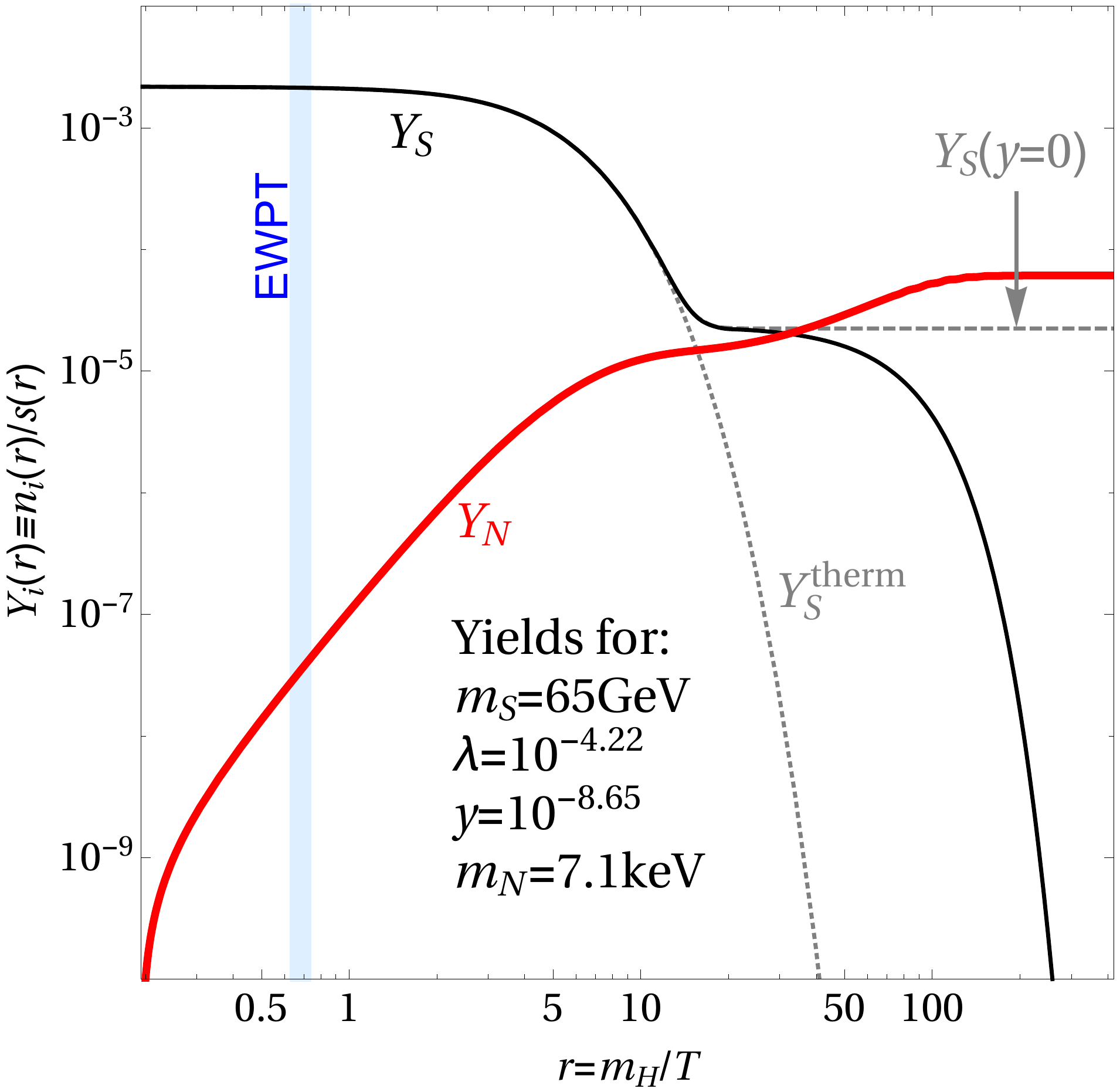}\\
 \hspace{-1cm}\includegraphics[width=8.3cm]{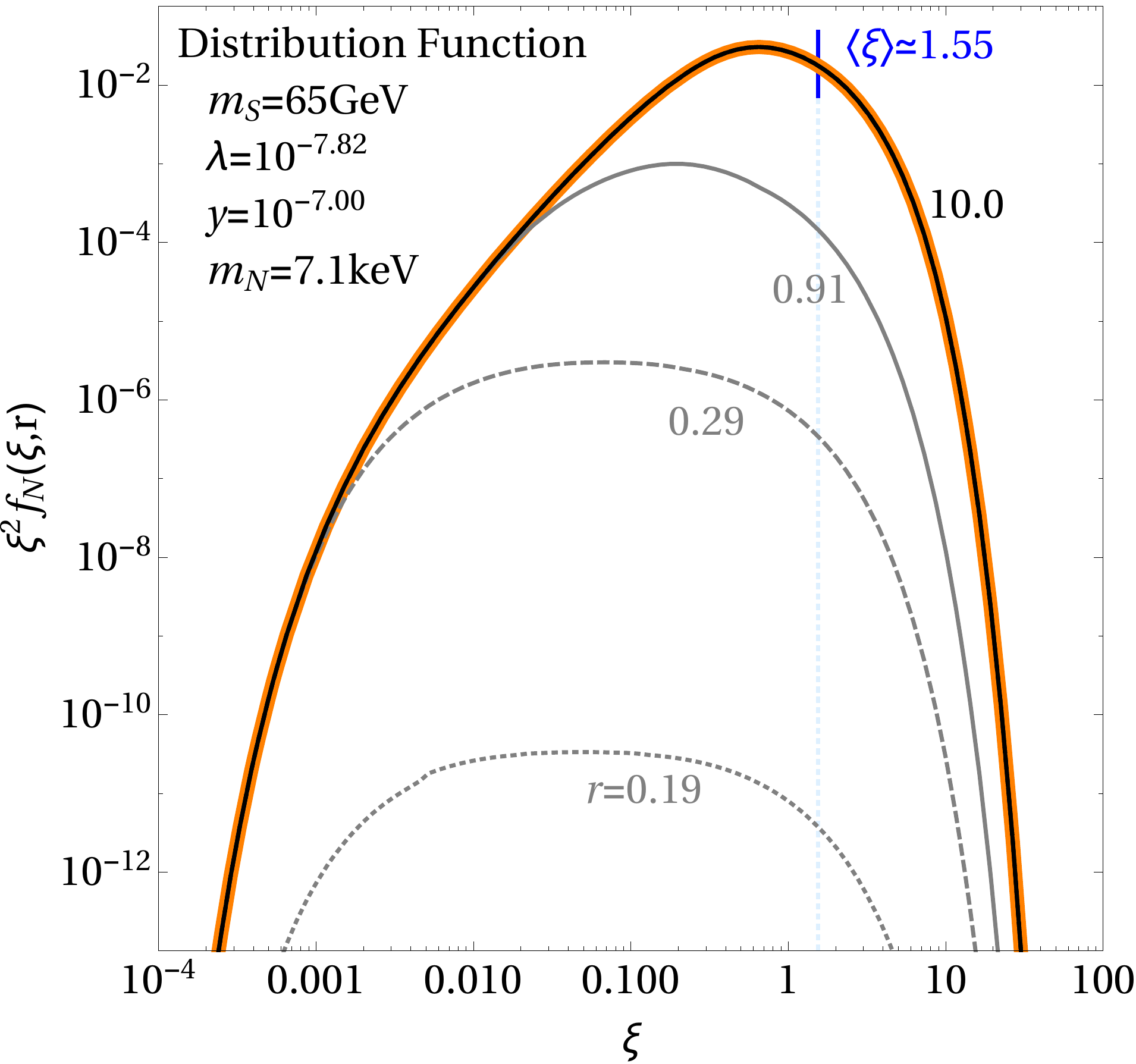} & \includegraphics[width=8.3cm]{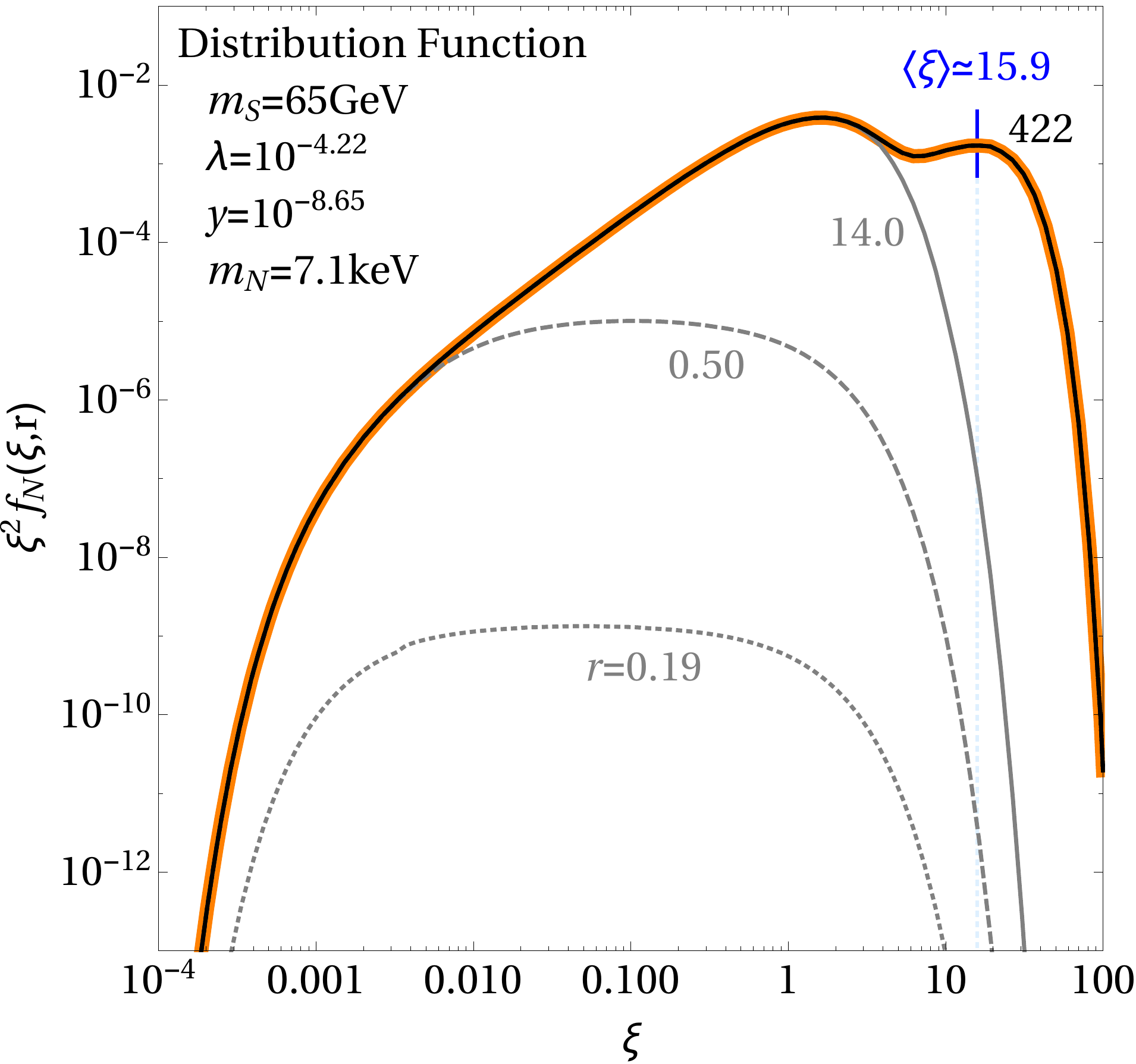}
\end{tabular}
\caption{\label{fig:FIMP_WIMP_65}Example evolutions of the yield (\emph{top row}) and sterile neutrino distributions (\emph{bottom row}) for a scalar with a mass of $65$~GeV undergoing freeze-in (\emph{left}) or freeze-out (\emph{right}) before decaying into sterile neutrinos, for the two points marked in the left Fig.~\ref{fig:small_masses}.}
\end{figure}

The spaghetti plots for two different scalar masses, $m_S = 65$ and $100$~GeV, are depicted in Fig.~\ref{fig:small_masses}. At first sight, these plots appear qualitatively similar to those depicted in Fig.~\ref{fig:verysmall_masses}, except for an overall shift towards larger values of $\lambda$. However, an interesting observation is that, for the FIMP case, the spectrum seems to become ``warmer'' when going from $m_S = 65$ to $100$~GeV, while the trend was opposite when going from $m_S = 30$ to $60$~GeV. The reason is that, for the case at hand, Higgs decay does not play a role in the production of the scalar. Instead, many scalars are already produced rather early (see upper left panel of Fig.~\ref{fig:FIMP_WIMP_65}), such that they can indeed decay earlier (and are less boosted), so that in this case a smaller scalar mass results into smaller initial sterile neutrino velocities and this effect dominates the production. 

As already anticipated, the lighter scalar ($m_S = 65$~GeV) freezes out \emph{earlier}, namely at $T\sim 17$~GeV, than the heavier one ($m_S = 100$~GeV), which does so at $T\sim 15$~GeV, cf.\ Fig.~\ref{fig:interaction_rates}. This may look surprising at first, since usually heavier particles tend to freeze out earlier due to the thermal suppression for non-relativistic particles. But the thermal suppression is similar for both cases, $e^{-65/100} \sim 0.5$, so that the actual interactions are decisive. Already at $T\sim 50$~GeV, all heavy bosons and the top are thermally suppressed, so that they cannot easily be produced anymore. However, a scalar with a mass of $100$~GeV can even at rest annihilate into both $W^+ W^-$ and $Z^0 Z^0$, while a lighter scalar cannot. Thus, the interaction rate of the heavier scalar is considerably larger, thereby keeping it in equilibrium significantly longer. It follows that the number density of the lighter scalar, and thus of the sterile neutrinos originating from its decay, is higher and hence must be compensated by a smaller sterile neutrino mass -- which is the origin of only small masses being allowed in the freeze-out region of the left Fig.~\ref{fig:small_masses}.

\begin{figure}[t]
\begin{tabular}{lr}\hspace{-1cm}
 \includegraphics[width=8.3cm]{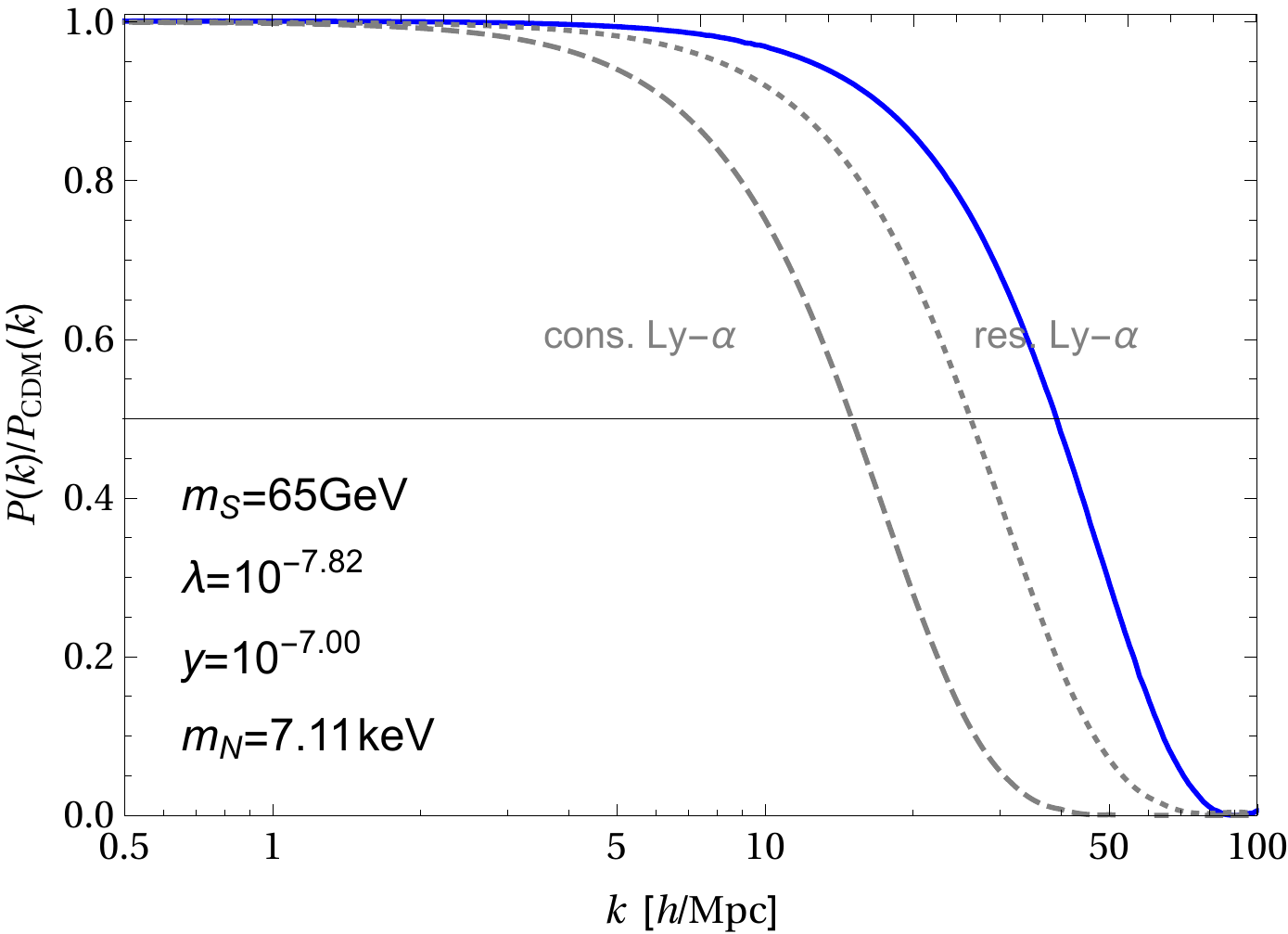} & \includegraphics[width=8.3cm]{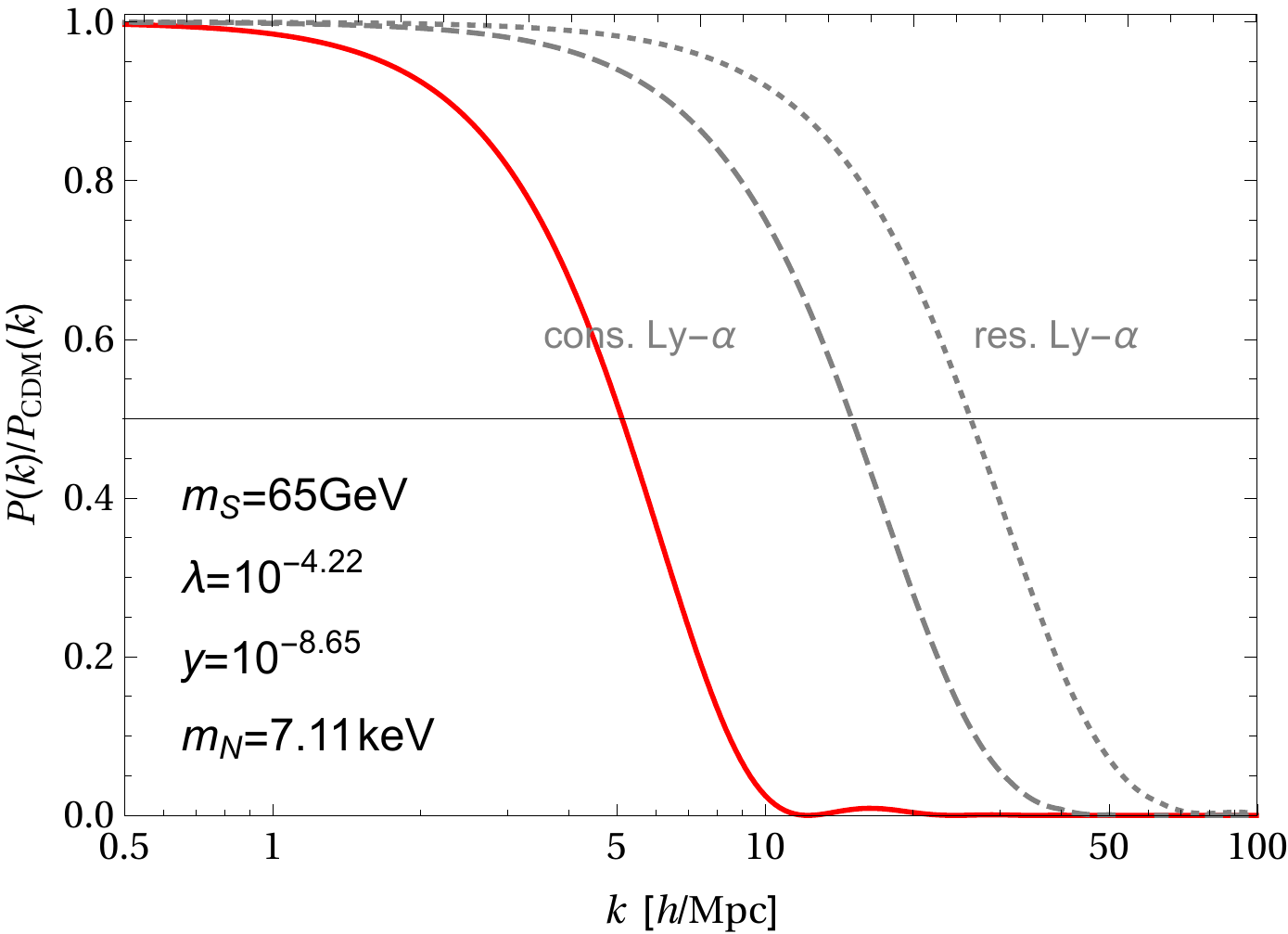}
\end{tabular}
\caption{\label{fig:TF_65}Example squared transfer functions for a scalar with a mass of $65$~GeV undergoing freeze-in (\emph{left}) or freeze-out (\emph{right}) before decaying into sterile neutrinos, for the two points marked in the left Fig.~\ref{fig:small_masses}.}
\end{figure}

Again, two example points are depicted in Fig.~\ref{fig:FIMP_WIMP_65}, with the order of the plots as in Fig.~\ref{fig:FIMP_WIMP_60}. It is clearly visible that, in the yields of the scalars, the effect of the EWPT is much less dramatic than for the very light scalars. In the resulting sterile neutrino spectra, the effect of the EWPT is thus much less pronounced, although little bumps can be spotted on the left sides of both spectra in the bottom row of the figure. Qualitatively, the course of the DM production is similar to case of very light scalars: again, if the scalar freezes in, it gradually decays into sterile neutrinos; if it freezes out, then it may decay either in or out of equilibrium, which a gradual transition between the two regimes. Note that, for the masses displayed, the comparison to Ref.~\cite{Adulpravitchai:2014xna} is in fact much more difficult. While the same scalar masses are used in that reference, we strongly suspect that paper to have a small error in the degrees of freedom of the Higgs field above the EWPT. While the Higgs field does possess four physical components in the unbroken phase, in Ref.~\cite{Adulpravitchai:2014xna}, only one seems to have been used. In other words, at high $T$, while the gauge boson contributions are switched off, the contributions of the would-be Goldstone bosons are \emph{not} switched on, as one would need to do. We will confirm this claim later in Sec.~\ref{sec:Results_large}, where it results in a factor of four between our results and those obtained in Ref.~\cite{Adulpravitchai:2014xna}. However, this factor is only close to four for very heavy scalar masses (even when taking into account the difference in the normalisation of $\lambda$, which also happens to be a factor of four), where the production almost entirely happens in regime~I. But for the comparatively light scalars treated in this subsection, the production is spread out over regimes~I and~II, cf.\ Fig.~\ref{fig:RegimeExampleCases}, such that a discrepancy by less than a factor of four but no perfect agreement can be expected. This is confirmed by our numerical results, making us confident that our treatment is the correct one for high temperatures.

The squared transfer functions (i.e., the ratios of the resulting linear power spectra to the CDM case) are depicted in Fig.~\ref{fig:TF_65}. As to be anticipated from the left Fig.~\ref{fig:small_masses}, the FIMP point is located in a blue region, and it should thus be perfectly allowed by all bounds. The WIMP point, in contrast, is in a region that is entirely red -- and it should thus correspond to DM that is by far too hot. Indeed, the squared transfer functions displayed in Fig.~\ref{fig:TF_65} confirm this conclusion. The whole curve of the FIMP case is in the allowed region, whereas the whole curve of the WIMP case is forbidden by far, even by the conservative bound. For such clear cases, an elaborate analysis would not even be necessary. Again we observe that, although both points feature a sterile neutrino mass of $7.1$~keV, the predictions for cosmic structure formation are different by far. This clearly illustrates that the DM mass itself is not at all a good way to discriminate between two different DM spectra, even if viewed at one and the same temperature.

\subsection{\label{sec:Results_large}Heavy scalars: $\boldsymbol{m_h < m_S}$}

\begin{figure}[t]
\begin{tabular}{lr}\hspace{-1cm}
 \includegraphics[width=8.3cm]{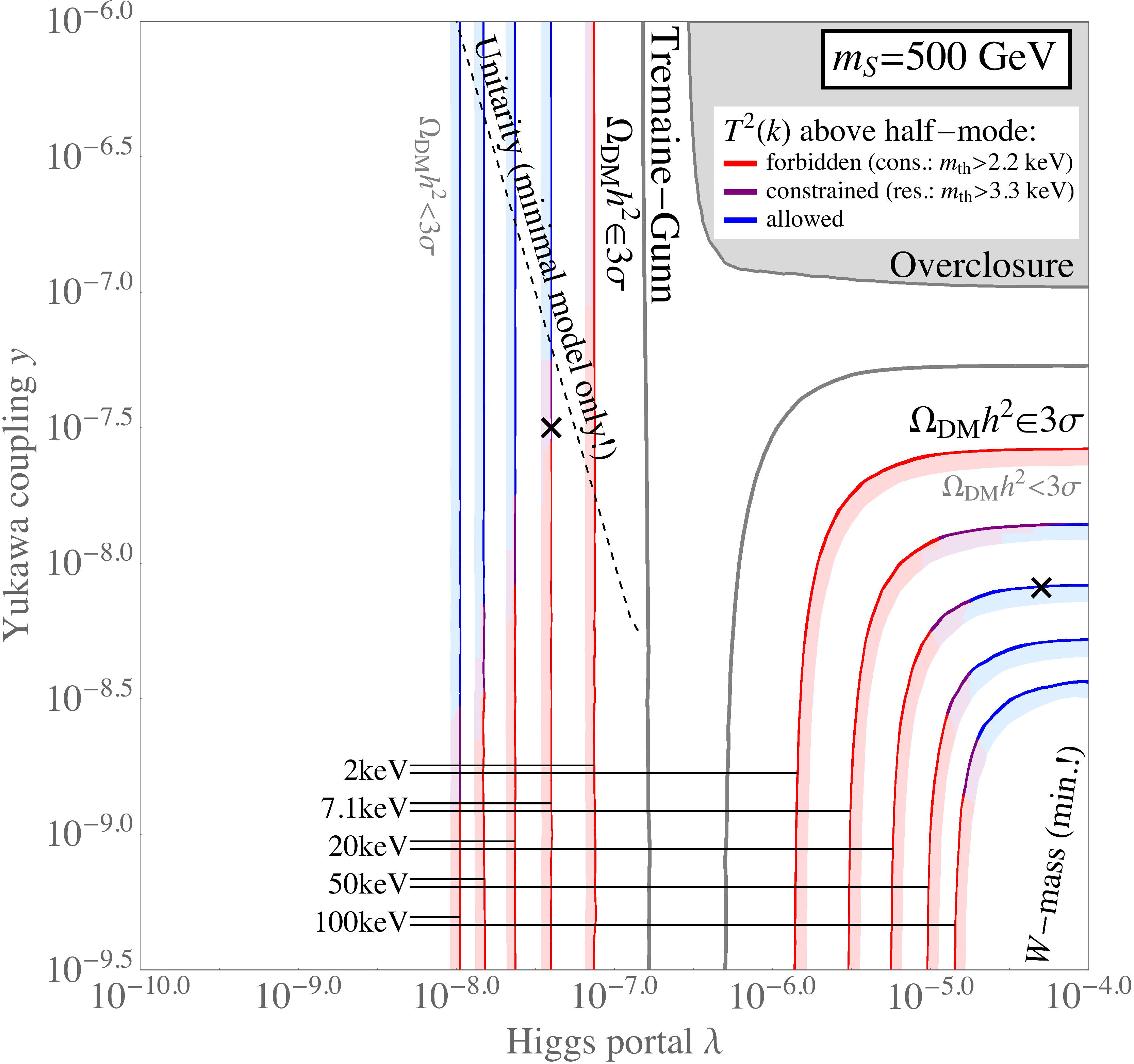} & \includegraphics[width=8.3cm]{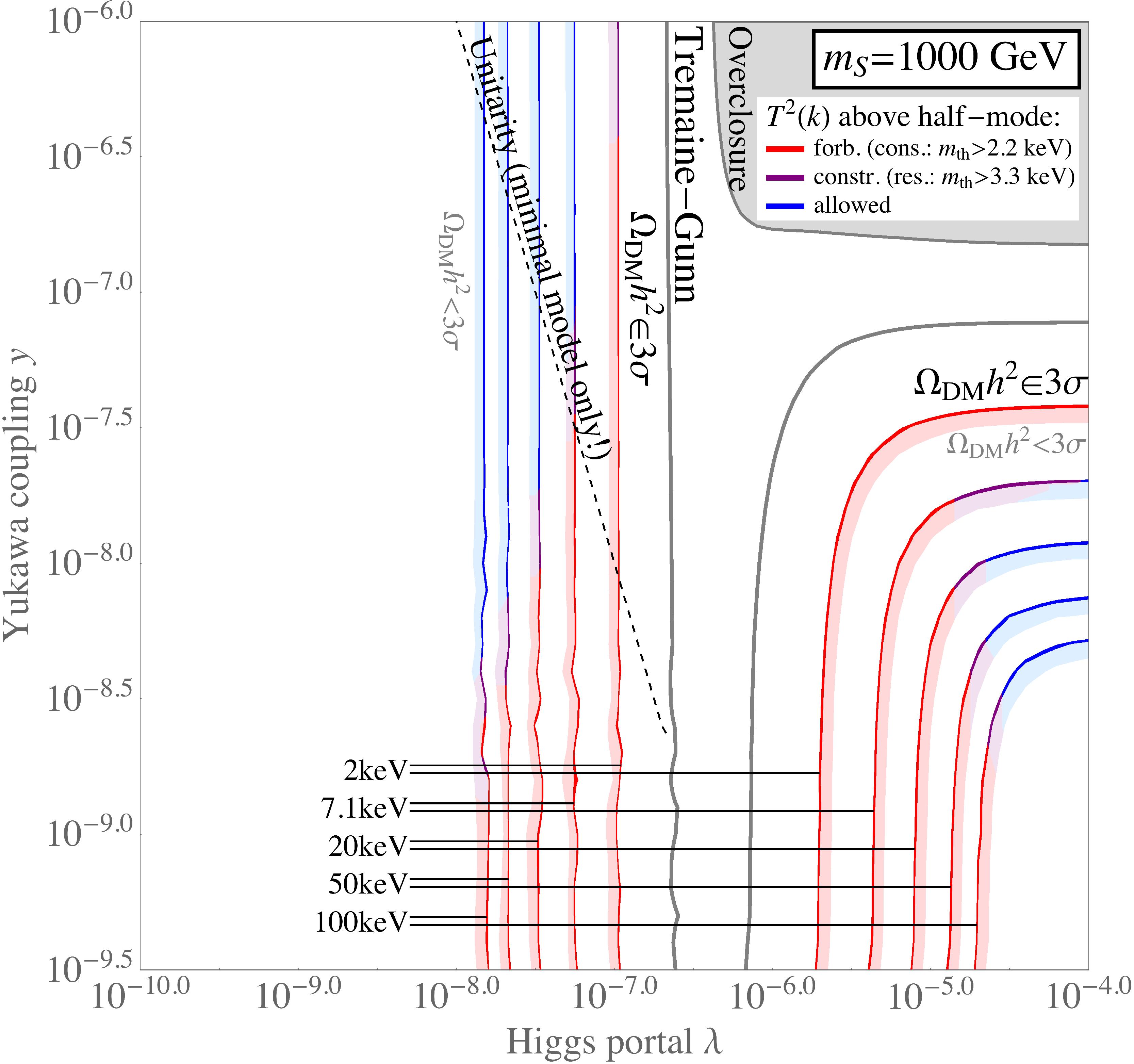}
\end{tabular}
\caption{\label{fig:large_masses}Abundance and constraints for large scalar masses, $m_h < m_S$. The crosses mark the points displayed explicitly in Fig.~\ref{fig:FIMP_WIMP_500}.}
\end{figure}

Finally, the scalars may also be heavier than the Higgs, see Fig.~\ref{fig:large_masses}, which is the case that had already been discussed in our earlier reference~\cite{Merle:2015oja}. In this case, given that both freeze-in and freeze-out are correlated to the mass of the scalar, most of the production of scalars happens above the EWPT -- we basically remain in regime~I in Fig.~\ref{fig:RegimeExampleCases} during the whole production. Only for somewhat light scalar masses, such that $m_S/20 \lesssim T_{\rm EWPT}$, a small amount may be produced after the EWPT. However, for most of the production, only the simple 4-scalar interaction in the top row of Tab.~\ref{tab:RegimesFeynman} plays a role. In particular, one can greatly simplify the equations involved for $(m_h/m_S)^2 \ll 1$, see Ref.~\cite{Merle:2015oja} for details. As we can see from the central and right panels of Fig.~\ref{fig:interaction_rates}, it is in fact a good approximation to assume that the production of sterile neutrinos from very heavy scalars happens entirely before the EWPT -- thereby providing a clear justification of the assumptions made previously in~\cite{Merle:2015oja}.

Understanding the results for this case is somewhat less subtle than for the previous two. Comparing the two plots depicted in Fig.~\ref{fig:large_masses}, one can see that there are no overly drastic changes. The reason is that, as argued in Ref.~\cite{Merle:2015oja}, there is a trade-off between a heavier scalar and an earlier decay: a larger scalar mass means that the production of scalars ceases at higher temperatures. In the freeze-in case, this indicates a lower overall production, which is what the vertical lines in that regime slightly shift to the right when going from $m_S = 500$~GeV to $1000$~GeV. However, given that no new production channels open up while staying in regime~I, the change is rather mild. In the other limiting case, where the scalar freezes out, a larger scalar mass means a larger abundance for the case where the scalar decays mainly after freeze-out (i.e., the vertical lines on the bottom right of the plots). Thus, these vertical lines should slightly shift to the left to compensate for that, which they indeed do, as visible in the plots. For the horizontal parts of the lines, where the scalar decays while in equilibrium, however, freezing out earlier means less sterile neutrinos unless this is compensated by a larger decay rate: thus, these lines should shift to slightly larger values of the Yukawa coupling $y$ when going from $m_S = 500$~GeV to $1000$~GeV, which is just what is visible in the plot.

\begin{figure}[t!]
\begin{tabular}{cc}
 \hspace{-1cm}\includegraphics[width=7.9cm]{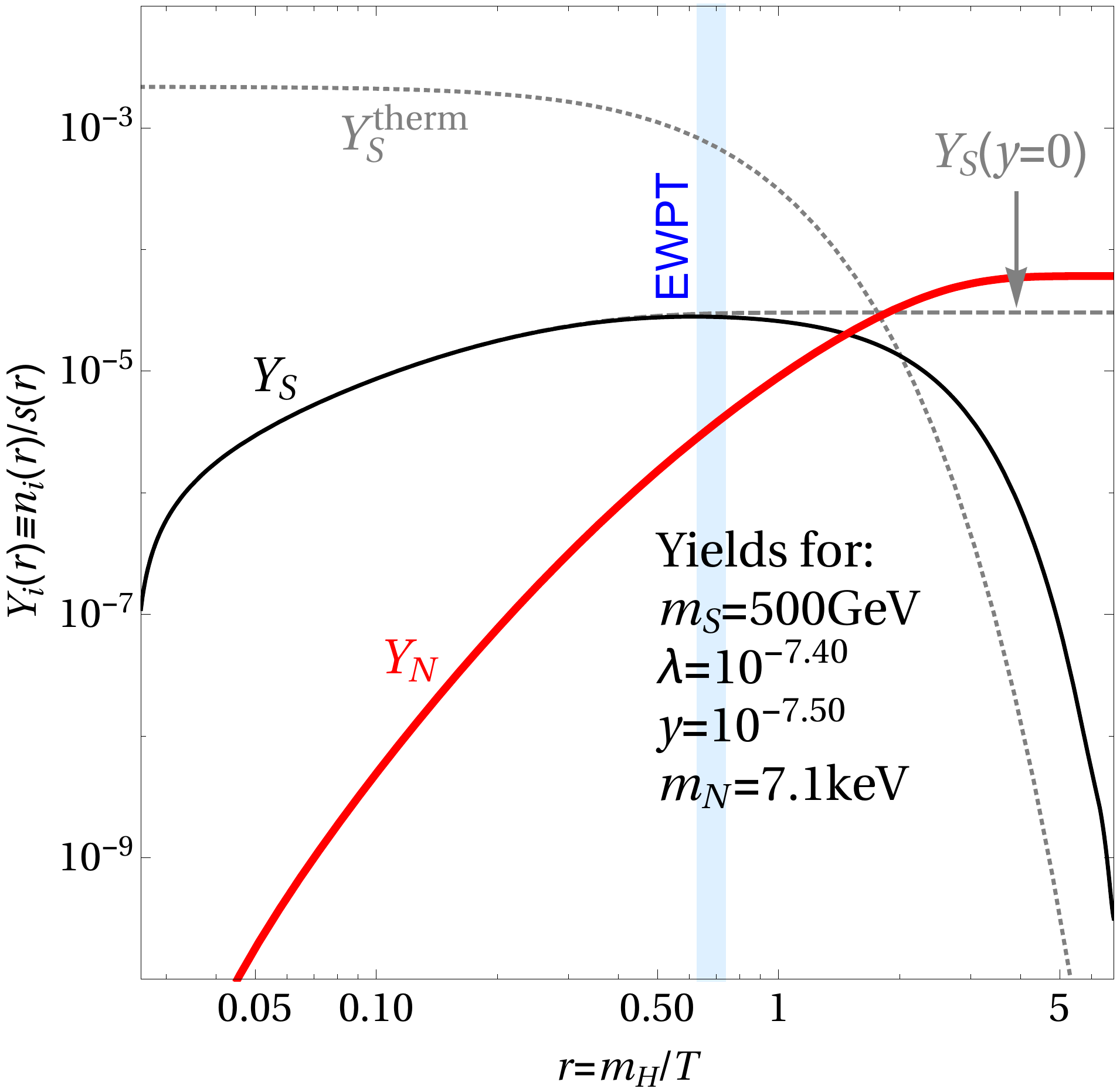} & \includegraphics[width=7.9cm]{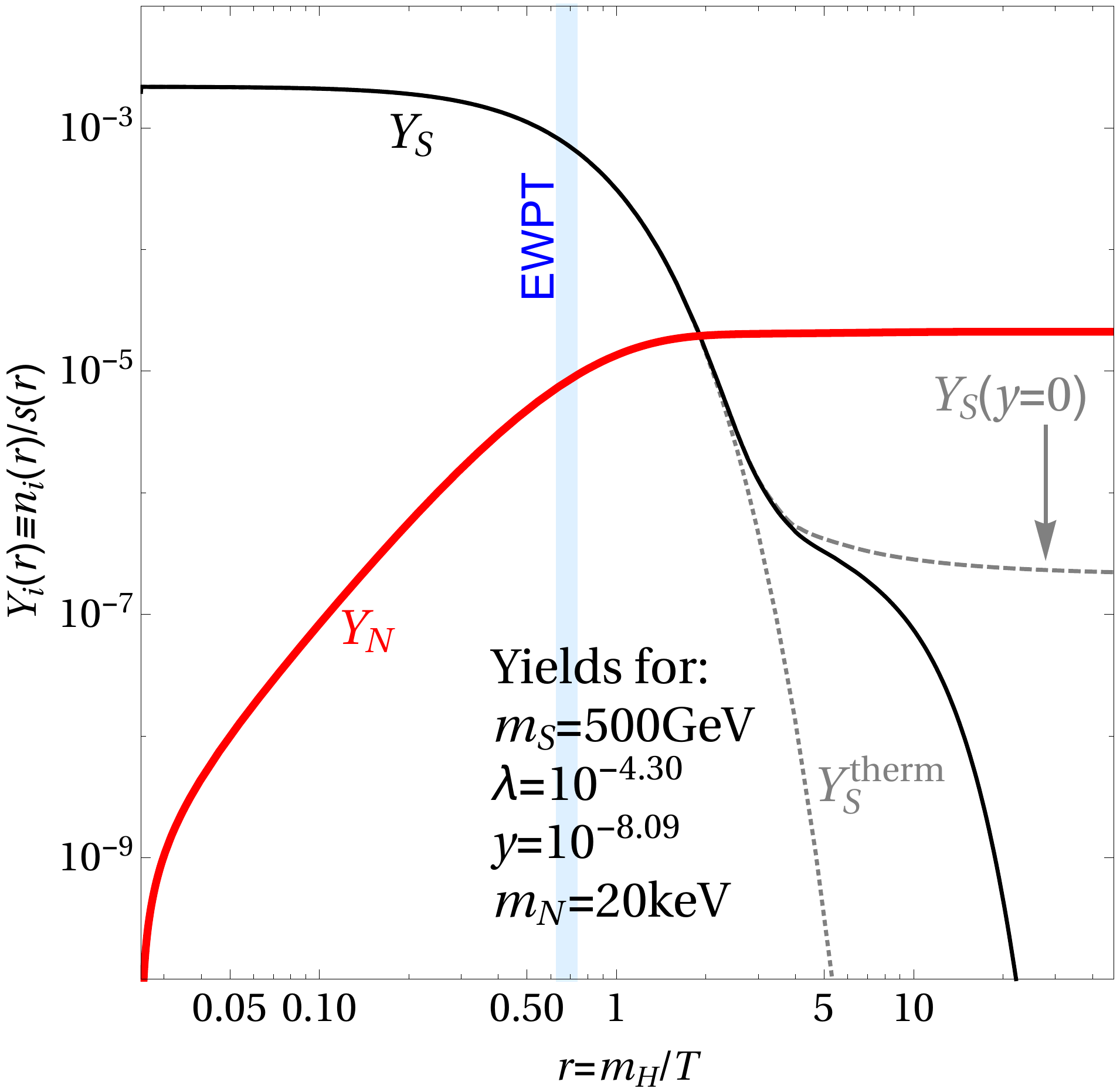}\\
 \hspace{-1cm}\includegraphics[width=8.3cm]{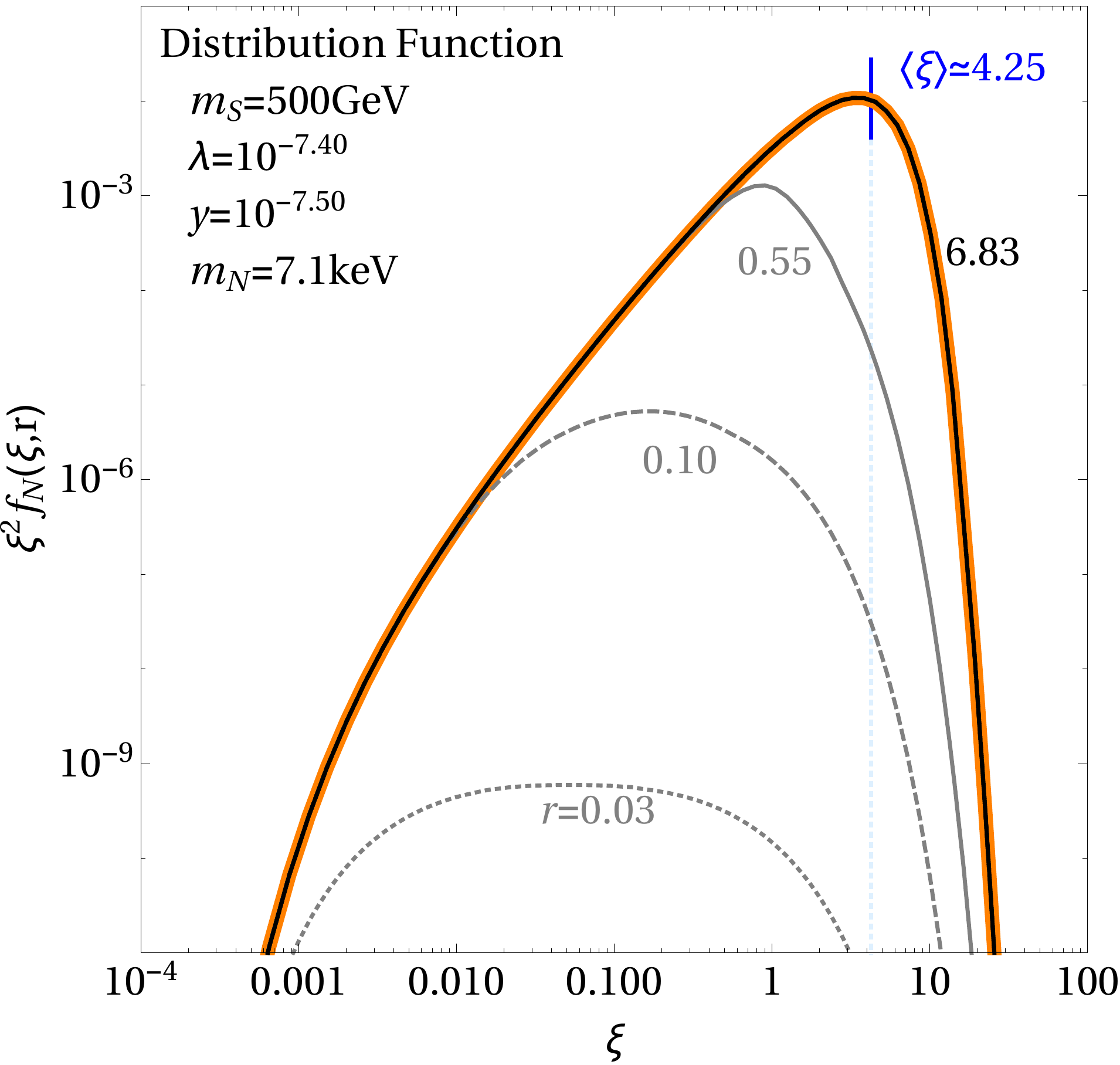} & \includegraphics[width=8.3cm]{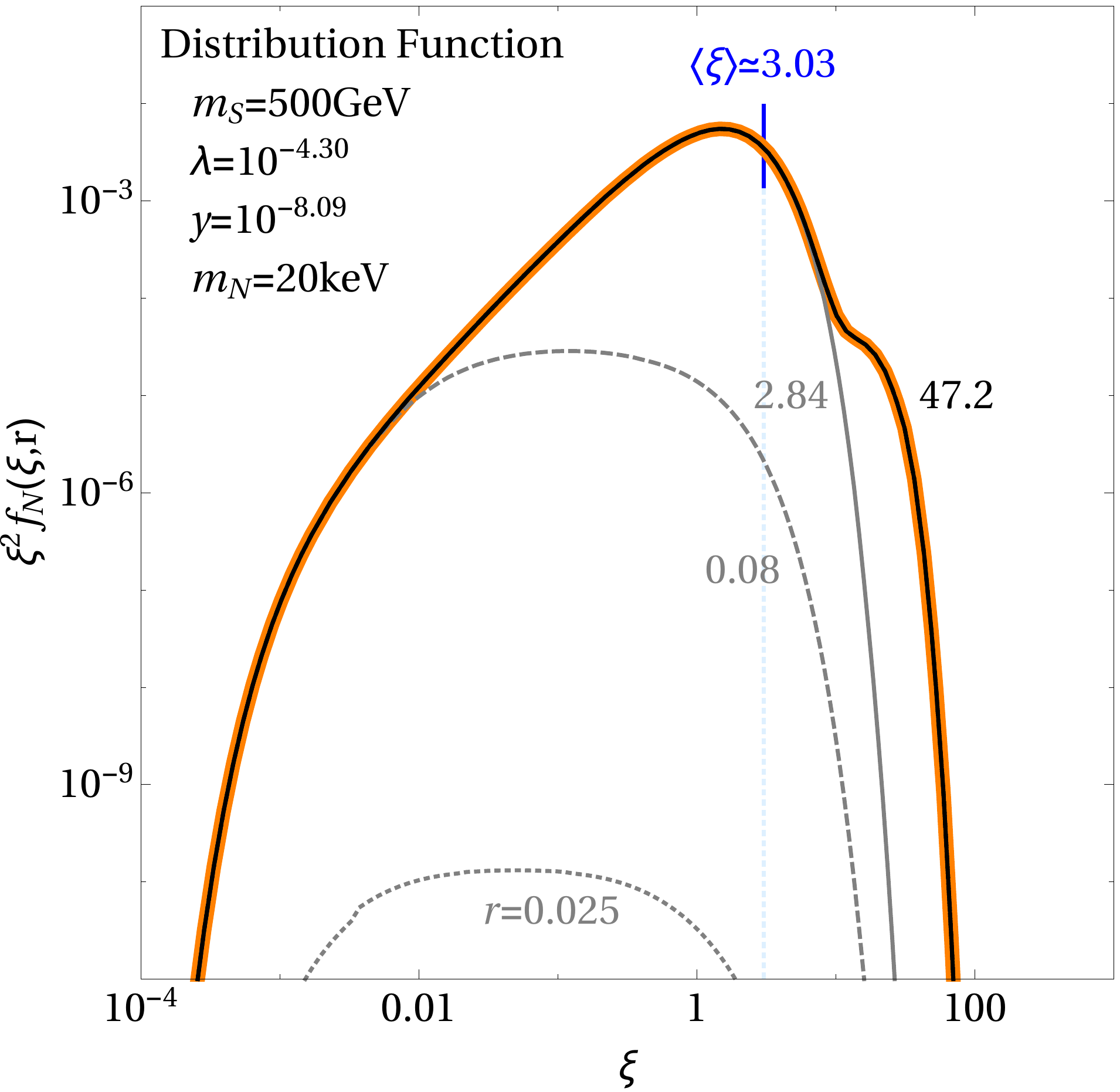}
\end{tabular}
\caption{\label{fig:FIMP_WIMP_500}Example evolutions of the yield (\emph{top row}) and sterile neutrino distributions (\emph{bottom row}) for a scalar with a mass of $500$~GeV undergoing freeze-in (\emph{left}) or freeze-out (\emph{right}) before decaying into sterile neutrinos, for the two points marked in the left Fig.~\ref{fig:large_masses}.}
\end{figure}

Two representative example evolutions of the yield can be seen in the upper row of Fig.~\ref{fig:FIMP_WIMP_500}, both for the scalar freezing in (\emph{left}) or out (\emph{right}). Just as before, in the FIMP case, the scalar is gradually produced but never reaches a thermal abundance before freezing in (while constantly decaying), whereas for the WIMP-case it quickly thermalises before freezing out. Also here, the decay can happen earlier or later, depending on the value of $y$. As already hinted in Sec.~\ref{sec:Results_Light}, we expect a factor four difference compared to the results obtained in Ref.~\cite{Adulpravitchai:2014xna}, as DM production happens nearly completely in regime~I. This approximation is best for very large scalar masses, however, the most extreme case discussed in~\cite{Adulpravitchai:2014xna} involves a scalar with $m_S = 500$~GeV. Taking the same couplings as in the reference and taking into account the different normalisation of $\lambda$, the discrepancy obtained numerically is already larger than a factor of three, and should converge to four for even larger scalar masses (which we however cannot explicitly check, given the absence of this case in~\cite{Adulpravitchai:2014xna}). Yet, given that no distribution functions had been computed in Ref.~\cite{Adulpravitchai:2014xna}, their results can be rectified by simply including the missing factor.

\begin{figure}[t]
\begin{tabular}{lr}\hspace{-1cm}
 \includegraphics[width=8.3cm]{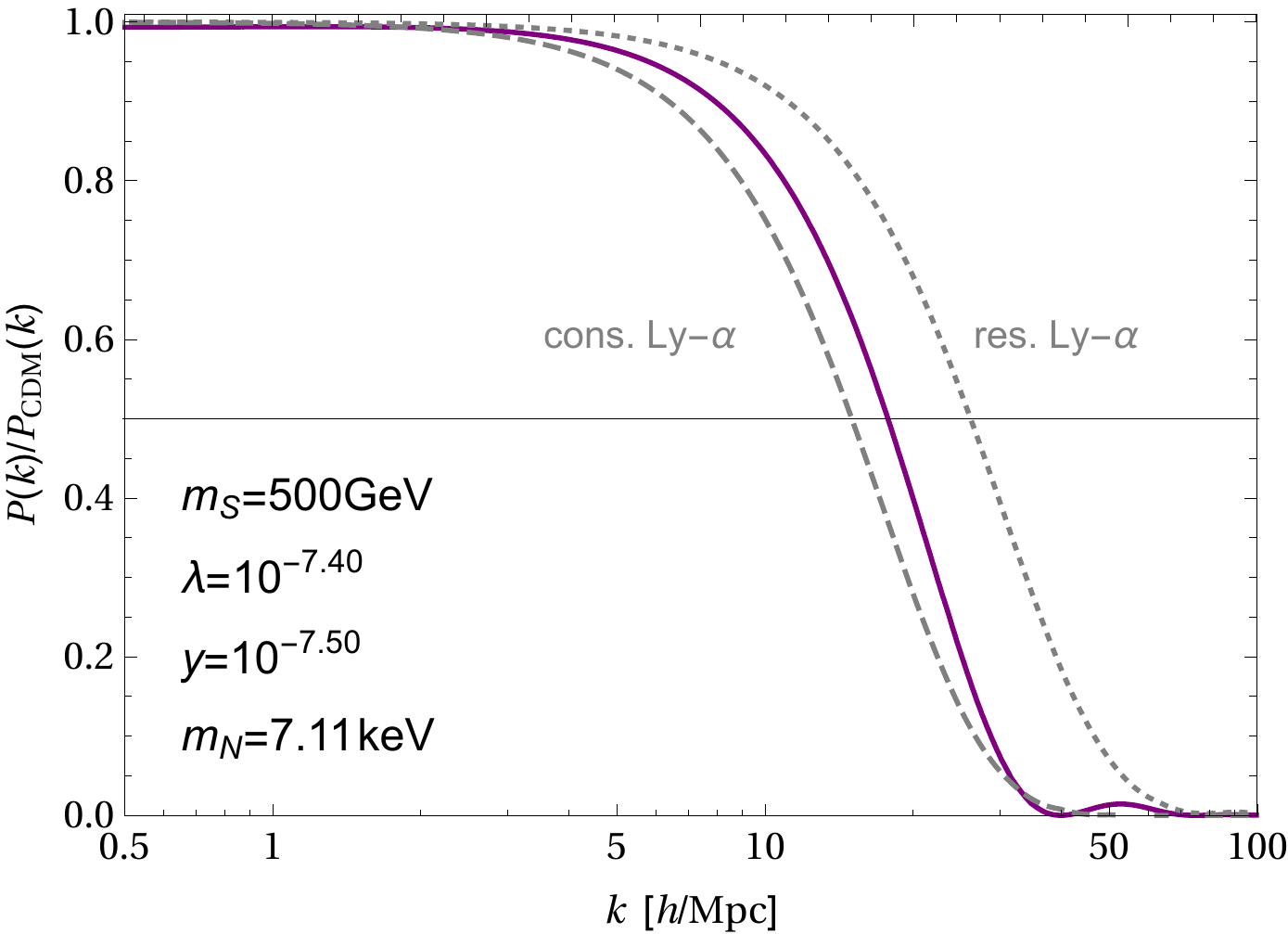} & \includegraphics[width=8.3cm]{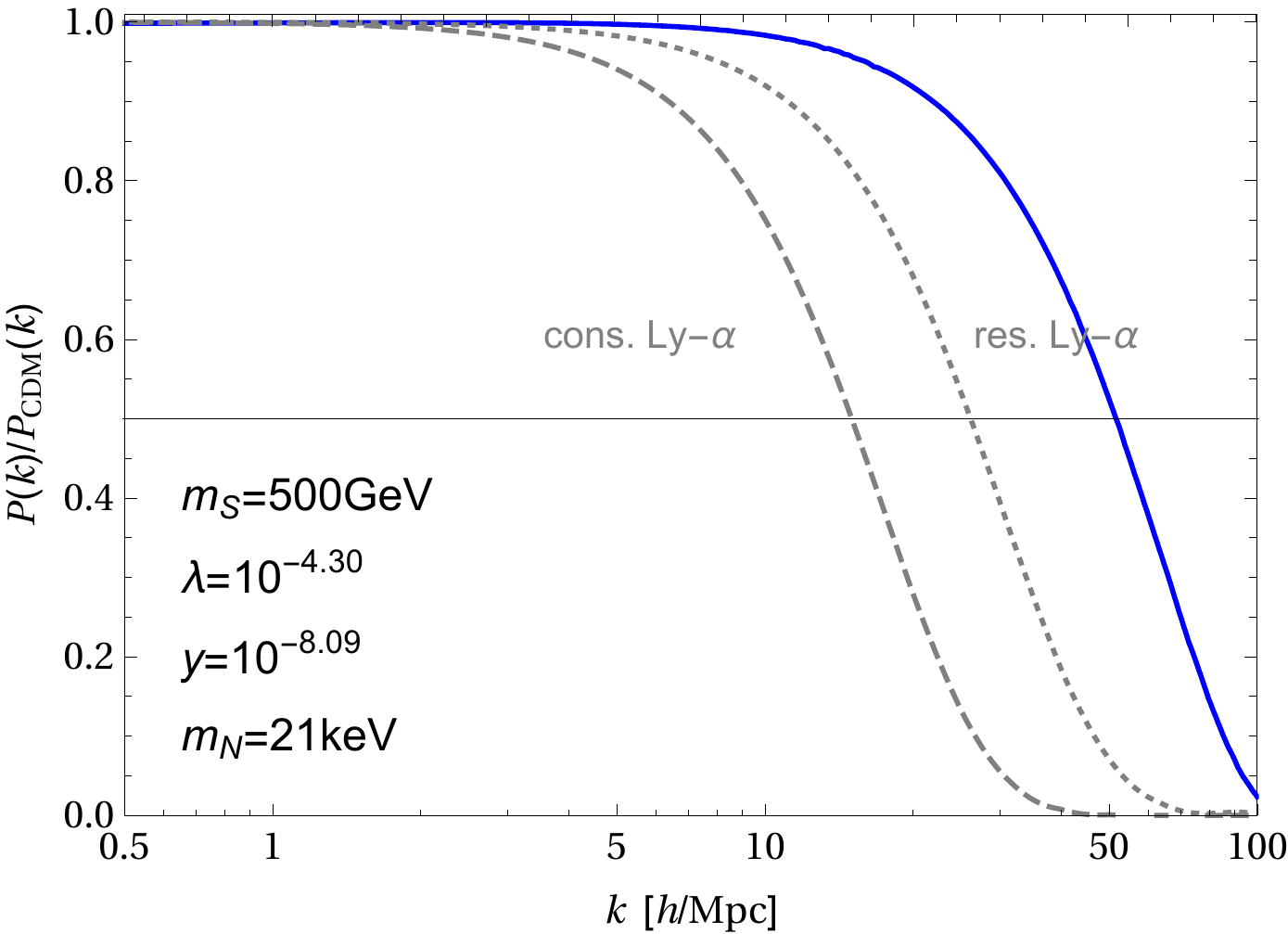}
\end{tabular}
\caption{\label{fig:TF_500}Example squared transfer functions for a scalar with a mass of $500$~GeV undergoing freeze-in (\emph{left}) or freeze-out (\emph{right}) before decaying into sterile neutrinos, for the two points marked in the left Fig.~\ref{fig:large_masses}.}
\end{figure}

The corresponding evolutions of the distribution functions are depicted on the bottom row of Fig.~\ref{fig:FIMP_WIMP_500}, with similar characteristics as for smaller masses. One big difference, though, is that the DM production is finished much earlier, which comes from heavier scalars decaying faster. This allows more time to redshift, but it is compensated to some extend by a larger initial velocity.

In what concerns structure formation, however, these shifts do not change very much. In the freeze-out case, while the sterile neutrinos are produced by a heavier scalar and are thus faster, they are also produced earlier (since the decay rate of the scalar is proportional to its mass), so that hardly any change in the colouring can be seen for the freeze-in case when comparing the two plots in Fig.~\ref{fig:large_masses}. For the case of the scalar freezing out,  only the in-equilibrium decays are phenomenologically relevant, as the late decays in any case produce too ``hot'' Dark Matter. However, for scalars decaying while in equilibrium, a larger mass translates into larger Yukawa couplings and thus into earlier decays so that, again, the resulting sterile neutrinos have a larger momentum at production but they have also more time to redshift, such that the prediction for structure formation remains basically unaffected -- hence the similarity between both plots. Note that, for large scalar masses, none of the ``kink'' regions on the bottom right of the plot is allowed, which would correspond to a double peak structure in the sterile neutrino's momentum distribution function (due to nearly equal contributions from the decays before and after freeze-out of the scalar) -- such distributions do exist but they are already peaking into the region of the restrictive Lyman-$\alpha$ bounds. This result is consistent with the findings from Ref.~\cite{Merle:2015oja}, which analysed just that case, and it is also backed up by the example squared transfer functions, cf.\ Fig.~\ref{fig:TF_500}.


\section{\label{sec:CandO}Conclusions \& Outlook}

In this work, we have completed the investigation of sterile neutrino Dark Matter production from scalar decays. We have shown how to perform the full computation using Boltzmann equations on the level of momentum distribution functions. While this would in principle amount to numerically solving a coupled system of partial integro-differential equations of a rather uncommon (and possibly unknown) type, we have argued why it is allowed to decouple the equations, how to transform the partial into an ordinary integro-differential equation, and which steps to take to finally solve it using an algorithm based on a discretisation in one variable. All that comprised pioneering work, which had previously not been available without drastic assumptions or simplifications. While our findings are of course somewhat specific to the examples studied, our methods carry over to more general settings such as decays of non-scalar particles or multiple generations of sterile neutrinos: any reader inclined to investigate such scenarios can adopt our methods and will be able to apply the same strategies to crack the equations involved.

The importance of investigating sterile neutrino Dark Matter on the level of momentum distribution functions lies in these quantities carrying \emph{all} the information about the Dark Matter model at hand. Not only can they be used to extract binary information such as the Dark Matter number or energy densities, but having the distribution functions is vital to derive predictions for and apply constraints from cosmic structure formation. Many previous works tried to go around this by solving oversimplified rate equations and using a simplistic estimate of the free-streaming horizon to conclude about whether or not a setting is consistent. However, we have shown that this procedure can fail once non-thermal distribution functions are involved. In these cases, it is just not a very reliable strategy. We have instead proposed to compute the linear power spectrum (from which we derive the squared transfer function), for which task dedicated tools are available, and to use the half-mode analysis developed in this work to obtain a fairly good estimate of the compatibility of a certain distribution function with data~\cite{MenciSchneiderViel}.

Having investigated the production of sterile neutrino Dark Matter from scalar decay in full generality, the final question is where to go from here. While we have shown how to obtain a reliable constraint from Lyman-$\alpha$ data using the half-mode analysis presented in Sec.~\ref{sec:Technicalities:Bounds}, in principle, one could refine the analysis presented, as we intend to do, and/or study further halo properties of settings like the one presented, as done e.g.\ for mixed Dark Matter settings~\cite{Maccio:2012uh,Anderhalden:2012jc}. The distribution functions obtained from scalar decay are conceptually not very different, however, as we have seen, they are highly non-thermal in both shape and number of momentum scales, so that even a description by a superposition of two thermally-shaped distributions may in fact not be sufficient. Various constraints can be applied, with maybe the most generic apart from the Lyman-$\alpha$ forest being dwarf satellite galaxy counts~\cite{Schneider:2014rda}.\footnote{Note that this has nothing to do with the missing satellite problem. On the contrary, for non-cold Dark Matter, it could possibly happen that \emph{not enough} satellites are produced, instead of producing too many. This yields quite a strong bound: at least the number of observed satellites has to be met, since satellites cannot be produced from larger structures.}

The main limitation of our method is that, currently, we are using only the linear power spectrum~\cite{Ma:1995ey}, while more considerable differences of decay-produced Dark Matter to thermal spectra may be visible in the non-linear regime. Ideally, one would perform and $N$-body simulation for each distribution presented here, however, this is clearly not doable given that these computations are numerically very expensive. Thus, a better strategy is to first obtain a pre-selection of cases which could be potentially interesting for a full $N$-body simulation, while discarding those which are clearly ruled out, or in practice no different from cold Dark Matter. One way to do such an analysis is to estimate the effect of the non-linearities, by the extended Press-Schechter approach~\cite{Press:1973iz,Sheth:1999mn}, modified by incorporating a collapse model that describes how the free-streaming Dark Matter particles get bound to form structures~\cite{Schneider:2013ria,Schneider:2014rda}. This is the next step to be done, which we will leave for future work~\cite{MenciSchneiderViel}. Given that present-day data is already sufficient to yield vastly different constraints on different sterile neutrino Dark Matter production mechanisms~\cite{Merle:2014xpa} it can be expected that -- with more detailed computations -- we may be able to clearly distinguish various production mechanisms by observational data.


\section*{Acknowledgements}
We are grateful to M.~Volpp for very useful discussions concerning the numerical solution of integro-differential equations, and to N.~Menci, A.~Schneider, and M.~Viel for sharing their insights on cosmic structure formation. We would furthermore like to thank A.~Adulpravitchai and M.~A.~Schmidt for giving us detailed information on their earlier work. We also received valuable comments on our manuscript from M.~A.~Schmidt and A.~Schneider. AM acknowledges partial support by the Micron Technology Foundation, Inc. AM furthermore acknowledges partial support by the European Union through the FP7 Marie Curie Actions ITN INVISIBLES (PITN-GA-2011-289442) and by the Horizon 2020 research and innovation programme under the Marie Sklodowska-Curie grant agreements No.~690575 (InvisiblesPlus RISE) and No.~674896 (Elusives ITN). MT acknowledges support by Studienstiftung des deutschen Volkes as well as support by the IMPRS-EPP.
\begin{appendices}
\section{\label{app:A:DetailsComputation}Details on the Boltzmann equation}
\renewcommand{\theequation}{A-\arabic{equation}}
\setcounter{equation}{0}  

The purpose of this appendix is to discuss some technical details related to the Boltzmann equations used in the text. Apart from reporting the explicit forms of the collision terms, we will also show how to perform a convenient transformation of variables, which enables us to deal with considerably simpler equations.

\subsection{\label{app:coll_terms}Collision Terms}

Let us first present the explicit versions of the collision terms. Note that we will only very briefly describe the basic form of the Boltzmann equation, as information on this part can be found in many textbooks, such as Refs.~\cite{Bernstein,Kolb:1990vq}. Constructing explicit forms for the collision terms is in fact somewhat trivial, although one of course has to be careful to include all relevant factors. Example derivations can be found in the appendix of Ref.~\cite{Merle:2015oja}, however, we will for illustration nevertheless present one explicit derivation, while we only list the results for all other cases.

As explained in the main text, we have two decisive structures, namely ``$2 \rightarrow 2$'' (scattering) and ``$1\rightarrow 2$'' (decay) processes. The question is how to extract their forms from a general collision term. The most general form possible for a collision term describing the reaction $\psi + a + b + ... \leftrightarrow \alpha + \beta + ...$ is~\cite{Kolb:1990vq}:
\begin{eqnarray}
 \mathcal{C}[f_\psi] &=& \frac{1}{2 E_p} \int {\rm d} P_a {\rm d} P_b ... {\rm d} P_\alpha {\rm d} P_\beta ... \times (2\pi)^4 \delta^{(4)}(\hat p + \hat p_a +\hat p_b +... -\hat p_\alpha - \hat p_\beta - ...) \times |\mathcal{M}|^2 \nonumber\\
&& \times \left[f_\alpha f_\beta ... \left(1\pm f_a\right) \left(1\pm f_b\right) ... \left(1 \pm f_\psi\right) - f_a f_b ... f_\psi \left(1 \pm f_\alpha\right) \left(1\pm f_\beta\right) ... \right] ,
 \label{eq:collision-general}
\end{eqnarray}
with $\hat p$ being the 4-momentum and $E_p$ the energy of the particle $\psi$ under consideration. The symbol $|\mathcal{M}|^2$ denotes the (initial and final) spin-averaged matrix element, which also contains symmetry factors for identical final and initial states,\footnote{This is in accordance with the conventions from Ref.~\cite{Kolb:1990vq}. The reason behind symmetry factors also appearing for initial state is that, to arrive at rate equations, we have to integrate over \emph{all} phase space elements, for both initial and final state particles. Thus, to avoid double counting, the symmetry factors have to be included here. This is different from, e.g., the situation at a particle collider, where the initial state is prepared in a certain way, but never integrated over.} as well as multiplicity factors for reactions involving multiple particles per process. Otherwise, the standard definitions apply: the phase space element is ${\rm d} P_X = g_X \frac{{\rm d}^3 p_X}{2 E_X (2\pi)^3}$, for a particle $X$ with $g_X$ internal degrees of freedom, momentum $p_X$, and energy $E_X = \sqrt{m_X^2 + p_X^2}$ with mass $m_X$.

To give one concrete example, we will now explicitly derive the collision term describing the decay of a SM-like Higgs into two singlet scalars $S$, $\mathcal{C}^S_{h \leftrightarrow SS}[f_S](p,T)$. The following quantities label the 4- and 3-momenta, as well as the absolute value of the latter:
\begin{itemize}
\item $\hat p$, $\mathbf{p}$, $p\ (\equiv |\mathbf{p}|)$: for either one of the scalars,
\item $\hat p'$, $\mathbf{p'}$, $p'\ (\equiv |\mathbf{p'}|)$: for the remaining scalar,
\item $\hat q$, $\mathbf{q}$, $q\ (\equiv |\mathbf{q}|)$: for the Higgs boson.
\end{itemize}
We furthermore abbreviate $E^i_{p_j}\equiv \sqrt{m_i^2+p_j^2}$. Starting from Eq.~\eqref{eq:collision-general}, we obtain:
\begin{eqnarray}
&& \mathcal{C}^S_{h\leftrightarrow SS}[f_S](p,T) = \frac{1}{2 E^S_p} \int \frac{\mathrm{d}^3p'}{(2 \pi)^3 2 E^S_{p'}} \frac{\mathrm{d}^3 q}{(2 \pi)^3 2 E^h_{q}} ( 2 \pi )^4 \delta^{(4)} \left( \hat p+ \hat p'- \hat q\right) \; \nonumber\\
&&\times |\mathcal{M}|^2 \left[ f^\text{eq}_h(q,T) - f_S(p,T) f_S(p',T)  \right],
\end{eqnarray}
where $|\mathcal{M}|^2 = 16\lambda^2v^2$, cf.\ Eq.~\eqref{eq:matrix_element_example}. The 3-dimensional $\delta$-function eliminates $q$:
\begin{eqnarray}
&&\mathcal{C}^S_{h\leftrightarrow SS} [f_S](p,T)= \frac{1}{2 E^S_p} \int \frac{\mathrm{d}^3 p'}{(2 \pi)^2 4 E^S_{p'} E^h_{|\mathbf{p}+\mathbf{p'}|}} \delta (E^S_p+E^S_{p'} - E^h_{|\mathbf{p}+\mathbf{p'}|}) \;\nonumber\\
&&\times |\mathcal{M}|^2 \left[ f^\text{eq}_h(|\mathbf{p}+\mathbf{p'}|,T) - f_S(p,T) f_S(p',T)  \right].
\end{eqnarray}
Due to the remaining $\delta$-function, the term in parentheses can be cast into:
\begin{eqnarray}
&&\mathcal{C}^S_{h\leftrightarrow SS} [f_S](p,T) = \frac{1}{2 E^S_p} \int \frac{\mathrm{d}^3p'}{(2 \pi)^2 4 E^S_{p'} E^h_{|\mathbf{p}+\mathbf{p'}|}} \delta (E^S_p+E^S_{p'} - E^h_{|\mathbf{p}+\mathbf{p'}|}) \;\nonumber\\
&&\times |\mathcal{M}|^2 \left[ f^\text{eq}_S(p,T)f^\text{eq}_S(p',T) - f_S(p,T) f_S(p',T)  \right].
\end{eqnarray}
Using spherical coordinates, $\mathrm{d}^3p'=2\pi p'^2\mathrm{d}p'\mathrm{d}(\cos\alpha)$, where $\alpha$ is the angle between $\mathbf{p}$ and $\mathbf{p'}$, we can rewrite the $\delta$-function to obtain:
\begin{equation}
\delta(E^S_p+E^S_{p'} - E^h_{|\mathbf{p}+\mathbf{p'}|}) =\delta(\cos\alpha-\cos\alpha_0)\frac{E^h_{|\mathbf{p}+\mathbf{p'}|}|_{\alpha=\alpha_0}}{pp'}.
\end{equation}
Here, $\alpha_0$ is the value of $\alpha$ such that $f(\cos\alpha_0)=0$. Using these results, we find
\begin{eqnarray}
&& \mathcal{C}^S_{h\leftrightarrow SS} [f_S](p,T) = \frac{1}{2 E^S_p} \int \limits_0^\infty \frac{2 \pi p'^2\mathrm{d}p'}{(2 \pi)^2 4 E^S_{p'} pp'} \underbrace{\int \limits_{-1}^1 \mathrm{d}(\cos\alpha)\; \delta (\cos\alpha-\cos\alpha_0)}_{
\footnotesize
\begin{matrix}
=1\text{ if }\cos\alpha_0\in [-1,1]\text{, }\\
=0 \text{ otherwise.}\hfill \hfill \hfill
\end{matrix}
} \;\nonumber\\
&&\times |\mathcal{M}|^2 \left[ f^\text{eq}_S(p,T)f^\text{eq}_S(p',T) - f_S(p,T) f_S(p',T) \right].
\end{eqnarray}
The result of the integration $\int_{-1}^1 \mathrm{d}(\cos\alpha) \;\delta (\cos\alpha-\cos\alpha_0)$ can be encoded in the limits of the integration over $p'$, i.e., by restricting that integral to those values of $p'$ for which the $\cos\alpha$-integral yields one. So we need to solve
\begin{align}
\pm 1= \cos\alpha_0 =\frac{\left( E^S_p + E^S_{p'} \right)^2 - m_h^2 - p^2 -p'^2 }{2 p p'}
\end{align}
for $p'$. The solutions are
\begin{align}
p' = \frac{(m_h^2-2m_s^2) p \pm m_h \sqrt{(m_h^2 - 4 m_S^2) ( m_S^2 + p^2 )}}{2 m_S^2}\ \ \text{ for }\cos\alpha_0 = +1,\\
p' = \frac{-(m_h^2-2m_s^2) p \pm m_h \sqrt{(m_h^2 - 4 m_S^2) ( m_S^2 + p^2 )}}{2 m_S^2}\ \ \text{ for }\cos\alpha_0 = -1.
\end{align}
As can readily be seen, these amount not to four but to only two distinct values:
\begin{align}
p' _1= \frac{(m_h^2-2m_s^2) p + m_h \sqrt{(m_h^2 - 4 m_S^2) ( m_S^2 + p^2 )}}{2 m_S^2},\\
p' _2= \left|\frac{(m_h^2-2m_s^2) p - m_h \sqrt{(m_h^2 - 4 m_S^2) ( m_S^2 + p^2 )}}{2 m_S^2}\right|.
\end{align}
Employing arguments on continuity and limits of $\cos\alpha_0$ for $p'\rightarrow \infty$ and $p'\rightarrow 0$, these two have to be the boundaries of the $p'$-integral, so the final form of this collision term is:
\begin{align}
\mathcal{C}^S_{h\leftrightarrow SS} [f_S](p,T) = \frac{1}{16 \pi p E^S_p} \int \limits_{p'_2}^{p'_1} \frac{p' \mathrm{d}p'}{ E^S_{p'} }  \; |\mathcal{M}|^2 \left[ f^\text{eq}_S(p,T)f^\text{eq}_S(p',T) - f_S(p,T) f_S(p',T)  \right].
\end{align}

Having seen one concrete example, we will now list the full set of collision terms needed to reproduce our results:
\begin{itemize}

\item $2 \rightarrow 2$-scattering processes:\\

All scattering processes are of the form
\begin{eqnarray}
&&\mathcal{C}^S_{ii \leftrightarrow SS}[f_S] (p,T)= \label{eq:all-scattering}\\
&&= \frac{g_i^2}{16 \sqrt{m_S^2 + p^2} (2 \pi)^3} \int\limits_0^\infty \frac{p'^2 \mathrm{d}p' }{\sqrt{m_S^2 + p'^2}}\int_{-1}^{\cos\alpha_\text{max}}\mathrm{d}(\cos\alpha) \sqrt{ 1 - \frac{4 m_i^2}{\hat s(p,p',\cos\alpha,m_S)}} \;\nonumber\\
&&\times |\mathcal{M}_{SS \rightarrow ii}(p,p',\cos\alpha)|^2 \left(\vphantom{\frac{ f_S^{\rm eq}(p,T)}{ 1}} f_S^\mathrm{eq}(p,T) \; f_S^\mathrm{eq}(p',T) - f_S(p,T) \; f_S(p',T) \right),\nonumber
\end{eqnarray}
where $ii = \phi\phi$,\footnote{Here, $\phi$ denotes any component of the SM-Higgs doublet $\Phi$.} $hh, t\bar{t}, W^+W^-, ZZ$ and $f_S^{\rm eq}(p) = \exp \left( -\sqrt{p^2 + m_S^2}/T \right)$ is the (would-be) equilibrium distribution of the scalar $S$, which is of Boltzmann-shape by virtue of the principle of detailed balance. Furthermore, $\hat s$ is the square of the centre-of-mass energy, explicitly given by:
\begin{align}
\hat s(p,p',\cos\alpha,m_S) = 2 ( m_S^2 + \sqrt{ ( m_S^2 + p^2 ) ( m_S^2 + p'^2 ) } - p p' \cos\alpha ).
\end{align}
In addition, $p'$ is the momentum of the second scalar participating in the process and $\alpha$ is the angle between $\mathbf{p}$ and $\mathbf{p'}$, whose maximum value is given by $\cos\alpha_\text{max} = \text{min}\{\text{max}\{ \cos\alpha_\text{im}, -1 \}, 1 \}$, defined by $4 m_i^2 = \hat s(p,p',\cos\alpha_\text{im},m_i,m_S)$. This upper integration boundary excludes all values of $\cos\alpha$ for which the integral becomes imaginary due to the square root contained in Eq.~\eqref{eq:all-scattering}.

The spin-averaged matrix elements including a factor of 2 in all of the following because two scalars are annihilated or produced, respectively, in each process, are given by (assuming $CP$-invariance):\footnote{The $hh\leftrightarrow SS$ result is only to leading order in $\lambda$. All others are full tree-level results. These values also include appropriate factors to account for identical particles in the initial or final state.}
\begin{eqnarray}
 |\mathcal{M}_{SS\rightarrow \phi\phi}|^2 = |\mathcal{M}_{\phi\phi\rightarrow SS}|^2 &=& 32\lambda^2 , \label{eq:ii_to_pp}\\
 |\mathcal{M}_{SS\rightarrow hh}|^2 = |\mathcal{M}_{hh\rightarrow SS}|^2 &=& 32 \lambda^2 \left( \frac{s + 2 m_h^2}{s - m_h^2} \right)^2 , \label{eq:ii_to_hh}\\
 |\mathcal{M}_{SS\rightarrow t\bar{t}}|^2 = |\mathcal{M}_{t\bar{t}\rightarrow SS}|^2 &=& 8 \lambda^2 m_t^2 \frac{s - 4 m_t^2}{(s - m_h^2 )^2 + m_h^2 \Gamma_h^2} , \label{eq:ii_to_tt}\\
 |\mathcal{M}_{SS\rightarrow W^+W^-}|^2 = |\mathcal{M}_{W^+W^-\rightarrow SS}|^2 &=& \frac{16}{9} \lambda^2 \frac{ s^2 - 4 m_W^2 s + 12 m_W^4 }{ ( s - m_h^2 )^2 + m_h^2 \Gamma_h^2 } , \label{eq:ii_to_WW}\\
 |\mathcal{M}_{SS\rightarrow ZZ}|^2 = |\mathcal{M}_{ZZ\rightarrow SS}|^2 &=& \frac{8}{9} \lambda^2 \frac{ s^2 - 4 m_Z^2 s + 12 m_Z^4 }{ ( s - m_h^2 )^2 + m_h^2 \Gamma_h^2 } . \label{eq:ii_to_ZZ} 
\end{eqnarray}

\item $1 \rightarrow 2$-decay processes:\\

Several different decays have to be taken into account. Starting with the decay of a SM-Higgs into two singlet scalars, the corresponding collision term is given by
\begin{eqnarray}
&&\mathcal{C}^S_{h \leftrightarrow SS}[f_S]( p , T ) = \frac{|\mathcal{M}_{h \rightarrow SS}|^2}{16\pi \; p \; \sqrt{m_S^2 + p^2} }\times\nonumber\\
&&\times \int_{p'_\mathrm{min}}^{p'_\mathrm{max}} \frac{p' \mathrm{d}p' }{\sqrt{m_S^2 + p'^2}} \left(\vphantom{\frac{ f_S^{eq}(\xi,r)}{ 1}} f_S^\mathrm{eq}(p,T) \; f_S^\mathrm{eq}(p',T) - f_S(p,T) \; f_S(p',T) \right),
\end{eqnarray}
with boundaries $p'_\text{min} = \left| \frac{m_h \varsigma - ( m_h^2 - 2 m_S^2 ) p}{2 m_S^2} \right|$ and $p'_\text{max} = \frac{m_h \varsigma + ( m_h^2 - 2 m_S^2 ) p}{2 m_S^2}$, where $\varsigma \equiv \sqrt{(m_h^2 - 4 m_S^2) ( m_S^2 + p^2 )}$, as well as the matrix element including a factor 2 because of annihilation/production of two scalars:
\begin{align}
|\mathcal{M}_{h \rightarrow SS}|^2=|\mathcal{M}_{SS \rightarrow h}|^2 = 16\lambda^2v^2.
\label{eq:matrix_element_example}
\end{align}
Furthermore, there are two collision terms related to the decay of a singlet scalar $S$ into two sterile neutrinos $N$. The first is the one used in the Boltzmann equation for $S$,
\begin{align}
\mathcal{C}^S_{S \rightarrow NN}[f_S]( p , T ) = -\frac{m_S}{\sqrt{m_S^2 + p^2}} \Gamma_{S \rightarrow NN} f_S(p, T),
\end{align}
with the decay widths $\Gamma_{S \rightarrow NN} = y^2 m_S/(16 \pi)$.  The second version is the one used in the Boltzmann equation for $N$,
\begin{align}
\mathcal{C}^N_{S \rightarrow NN}[f_S](p , T) = \frac{m_S \Gamma_{S \rightarrow NN}}{p^2}\int\limits_{p'_\text{min,N}}^\infty \frac{\mathrm{d}p' \; p' \; f_S(p',T)}{\sqrt{m_S^2 + p'^2}},
 \label{eq:CT_N_App}
\end{align}
with $p'_\text{min,N} = \left| p - \frac{m_S^2}{4 p} \right|$. Note that the two collision terms $\mathcal{C}^S_{S \rightarrow NN}$ and $\mathcal{C}^N_{S \rightarrow NN}$ appear to be somewhat different, although one may very naively expect one to be just the negative of each other. However, we should keep in mind that we are working on the level of momentum distribution functions, which implies that the collision terms look rather different depending on whether or not the desired distribution function is integrated over.

\end{itemize}
In these equations it is understood that $p$, $p'$, and $T$ are substituted in favour of $\xi$, $\xi'$, and $r$, as specified in Eq.~\eqref{eq:xi_and_r_definition}.

\subsection{\label{app:transf_variables}Transformation of Variables}

As stated in the main text, we perform a transformation of variables in order to bring the Liouville operator $\hat{L} = \frac{\partial}{\partial t} - H p \frac{\partial}{\partial p}$ into a more convenient form. To see how this results into Eq.~\eqref{eq:liouville_final_form}, consider a general transformation into new variables $r$ and $\xi$:
\begin{align}
 \left.
 \begin{matrix}
 t\\
 p
 \end{matrix} \right\} \to \left\{
 \begin{matrix}
 r = r(t, p),\\
 \xi = \xi (t,p).
 \end{matrix} \right.
\end{align}
These new variables can be inserted into the Liouville operator $\hat{L}$:
\begin{align}
\hat{L} = \frac{\partial r}{\partial t} \frac{\partial }{\partial r} +  \frac{\partial \xi}{\partial t} \frac{\partial }{\partial \xi}- H p(r,\xi) \left(  \frac{\partial r}{\partial p} \frac{\partial}{\partial r}+ \frac{\partial \xi}{\partial p} \frac{\partial}{\partial \xi} \right).
\end{align}
To simplify $\hat{L}$, we need to get rid of one of the two differential operators. In other words, the new variables would be most useful if we could choose them in such a way that they transform the partial differential equation into an effective ordinary differential equation. The first step is to demand that $r$ does not depend on $p$, which eliminates one term:
\begin{align}
\hat{L}=\frac{\partial r}{\partial t} \frac{\partial }{\partial r} +  \left[ \frac{\partial \xi}{\partial t} - H p(r,\xi)  \frac{\partial \xi}{\partial p}\right] \frac{\partial}{\partial \xi}.
\end{align}
Next, we demand
\begin{align}
\frac{\partial \xi}{\partial t}  =  H p(r,\xi)  \frac{\partial \xi}{\partial p} .
\end{align}
This is a rather simple partial differential equation. Fixing the initial condition
\begin{align}
\xi(p,t_0)=\xi_0(p),
\end{align}
where $\xi_0$ is some arbitrary $C^1$-function, it has the simple solution
\begin{align}
 \xi(p,t)=\xi_0\left(\frac{a(t)}{a(t_0)}\; p\right).
 \label{eq:xi_dependence}
\end{align}
This implies that, if we fulfill the requirements that $r$ only depends on $t$ and the dependence of $\xi$ on $p$ and $t$ is given by Eq.~\eqref{eq:xi_dependence}, the Liouville operator in terms of the new coordinates will have the simple form
\begin{align}
 \hat{L} = \frac{\partial r}{\partial t}\frac{\partial}{\partial r}.
 \label{eq:Liouville-simpler}
\end{align}
We can make our life even easier by a smart choice for the functions $r(t)$ and $\xi_0$. Exploiting the one-to-one correspondence between temperature $T$ and time $t$, one possible choice is:
\begin{align}\nonumber
 r &= \frac{m_0}{T}\ \ \ {\rm and}\\
 \xi &= \frac{1}{T_0}\frac{a(t)}{a(t(T_0))} \; p = \left( \frac{g_s(T_0)}{g_s(T)} \right)^{1/3}\;\frac{p}{T},
 \label{eq:xi_and_r_definition_appendix}
\end{align}
for some reference mass $m_0$ and some reference temperature $T_0$, both of which we choose to equal the Higgs mass:
\begin{align}
m_0 = T_0 = m_h.
\end{align}
For the last equality in Eq.~\eqref{eq:xi_and_r_definition_appendix}, we have used the fact that the comoving entropy density $s$ is constant,
\begin{align}
 s(T)a(T)^3 = \frac{2\pi^2}{45} g_s(T) \; T^3 \; a^3(T) = \text{const.},
 \label{eq:constant-entropy}
\end{align}
which allows to relate the scale factor $a(T)$ to the effective number $g_s(T)$ of relativistic entropy degrees of freedom. Eq.~\eqref{eq:constant-entropy} can also be used to derive the time-temperature relation
\begin{align}
\frac{dT}{dt}=-HT\left( \frac{T g_s'(T)}{3 g_s(T)} + 1 \right)^{-1}.
\end{align}
Plugging this into the Liouville operator from Eq.~\eqref{eq:Liouville-simpler} yields its final form,
\begin{align}
\hat{L} = r H \left( \frac{T g_s'}{3 g_s} + 1 \right)^{-1}\frac{\partial}{\partial r}.
\end{align}
This completes the proof of Eq.~\eqref{eq:liouville_final_form}.
\section{\label{app:B:Higgs-FO}The freeze-out of the Higgs boson}
\renewcommand{\theequation}{B-\arabic{equation}}
\setcounter{equation}{0}  

We argued that we do not have to solve a system of Boltzmann equations for all species in the early Universe, but only for the scalar singlet $S$ and the sterile neutrino $N$, since we rely on the SM particles sourcing the production of $S$ being in thermal equilibrium. This assumption is for sure good for $m_S \gg m_H$, but we should assess its quality if we want to proceed to smaller $m_S$. The smaller $m_S$ the later (in cosmic time) the scalar will be produced in general, and hence the less reliable the assumption of particles like the Higgs or gauge bosons being in equilibrium could be. In fact, all species relevant for sourcing $S$ in the broken phase ($W^\pm$, $Z$, $h$, $t$) have similar masses and are therefore expected to decouple from the thermal plasma at a similar time.

Still, if we restrict ourselves to $m_S > \unit{30}{GeV}$, it is enough to know whether the source particles are still in equilibrium at the corresponding temperature. Even if they are in equilibrium much longer, all the particles mentioned will by then have disappeared due to their masses resulting in strong Boltzmann-suppressions. Since we have not found any detailed source to cross-check this assumption, we have assessed it ourselves with the following analysis.

Let us assume that the Higgs boson $h$ and the top quark $t$ decouple from the plasma before the gauge-bosons, since their mass is larger by a factor of $\orderof{1}$.
We therefore looked at the following system of coupled Boltzmann equations:
\begin{align}
 \hat{L} f_h  &= \CTabs{h}{h}{\mathrm{SM'}\overline{\mathrm{SM'}}}\left[f_h\right]
  + \CTabs{h}{hh}{\mathrm{SM'}\overline{\mathrm{SM'}}}\left[f_h\right]
  + \CTabs{h}{h}{t\bar{t}}\left[f_h,f_t\right] \, , \nonumber \\
  \hat{L} f_t &= \CTabs{t}{t}{\mathrm{SM'}\overline{\mathrm{SM'}}}\left[f_h\right]
  + \CTabs{t}{tt}{\mathrm{SM'}\overline{\mathrm{SM'}}}\left[f_h\right]
  + \CTabs{t}{h}{t\bar{t}}\left[f_h,f_t\right] \, ,
\end{align}
where $\mathrm{SM'}$ denotes all SM degrees of freedom except for the $t$ and the $h$. The matrix elements going into all the collision terms were computed at tree-level only. The Higgs decay width was taken from Ref.~\cite{Agashe:2014kda}.

Solving them on the level of rate equations, we find that in this system, both the Higgs and the top closely track their equilibrium abundance all the way down to temperatures of about $\unit{1}{GeV}$, where their abundances are exponentially suppressed already. When switching off inverse decays and taking into account only two-to-two-processes, the top would freeze out around $\unit{5}{GeV}$ while the Higgs would freeze out around $\unit{3}{GeV}$. Note that the massive gauge bosons can also be produced via inverse decays, which is another argument to assume that they will decouple only after the Higgs and the top.

\section{\label{app:C:FSvsHalfMode}Failure of the free-streaming horizon as measure for non-thermal Dark Matter}
\renewcommand{\theequation}{C-\arabic{equation}}
\setcounter{equation}{0}  

Here, we would like to discuss one important point which we had stressed on several occasions throughout the manuscript. Given that we are dealing with highly non-thermal DM spectra, we cannot expect paradigms developed for thermal relics to carry over to this much more general case. In particular, a non-thermal spectrum cannot be described very well by a \emph{single} number, such as a temperature, an average momentum, or a velocity. Given that, we have in fact no reason to believe that a quantity based on an average velocity, such as the free-streaming (FS) horizon $\lambda_{\rm fs}$ as defined in Eq.~\eqref{eq:Def:LambdaFS}, should give us any countable information on the DM spectrum. Yet, it is incorrectly used in many occasions in the literature.

\begin{figure}[ht]
\begin{tabular}{lr}\hspace{-1cm}
 \includegraphics[width=8.3cm]{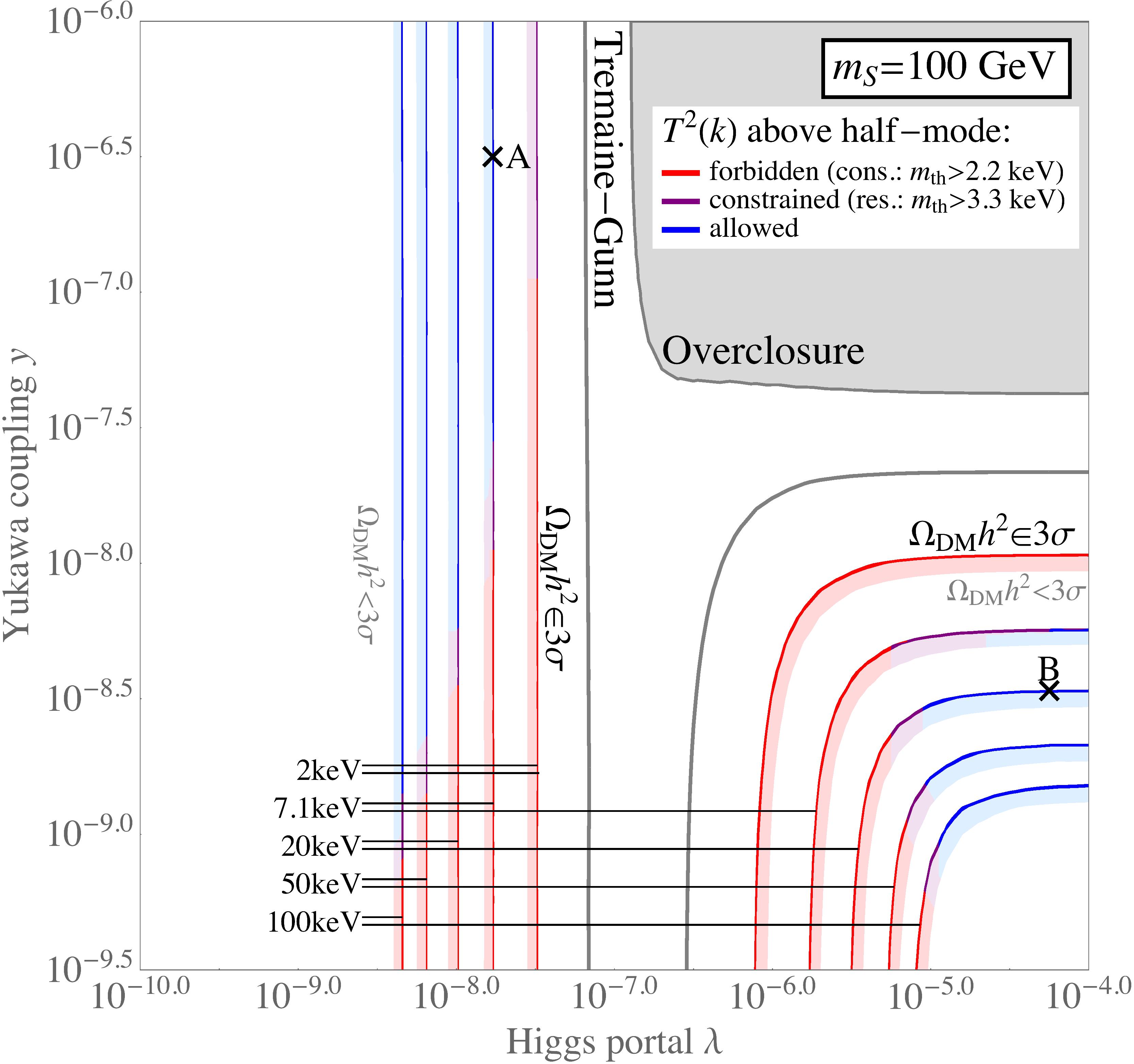} & \includegraphics[width=8.3cm]{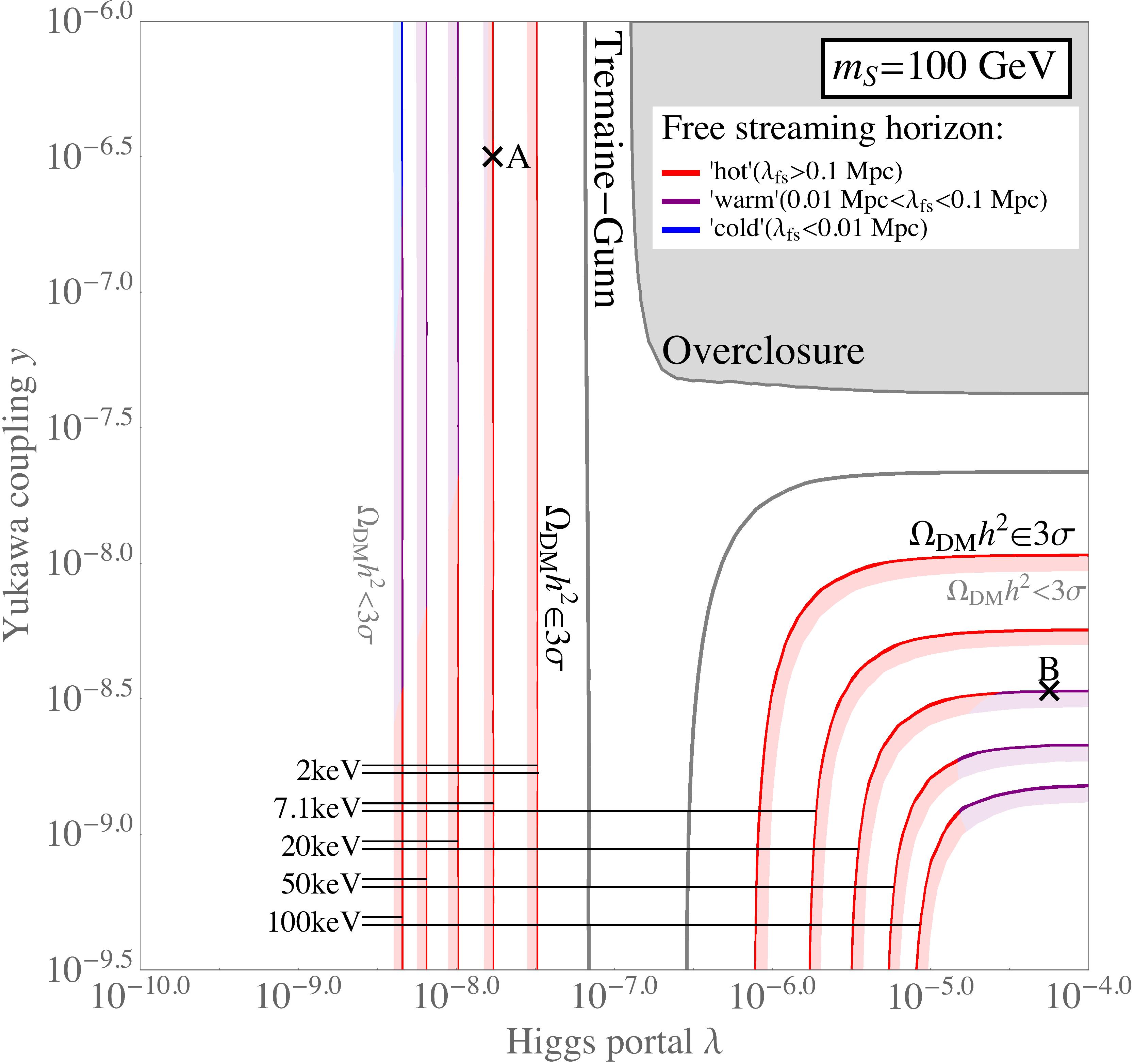}
\end{tabular}
\caption{\label{fig:FS-comparison}Comparison of half-mode analysis (\emph{left}; figure identical to the right Fig.~\ref{fig:small_masses}, except for the model-dependent bounds not being displayed here for simplicity) with the computation of the free-streaming horizon (\emph{right}).}
\end{figure}

In order to clearly illustrate how the FS horizon fails, we compare two versions of an example spaghetti plot, which are depicted in Fig.~\ref{fig:FS-comparison}. Here, on the left panel, we can see the plot for $m_S = 100$~GeV how we obtained it in Sec.~\ref{sec:Results_Light} (this figure is basically identical to the right Fig.~\ref{fig:small_masses}, apart from the missing collider-related bounds which we skipped here in order not to distract the reader). The identical patch of the parameter space is displayed on the right panel of Fig.~\ref{fig:FS-comparison}, however, this time with an analysis based on the FS horizon, classifying the different points as ``cold'', ``warm'', or ``hot'' -- even though, as already explained, these categories do not really suit any non-thermal spectrum. Comparing both plots, one can immedately see that the analysis based on the FS horizon\footnote{This we computed following the numerical computation of the FS horizon as reported in Ref.~\cite{Adulpravitchai:2014xna}, just with a small numerical error in $g_s$ which we have corrected in this version -- the resulting plot would be in between the ``numerical'' and ``analytic'' versions of $\lambda_{\rm fs}$ used in~\cite{Adulpravitchai:2014xna}.} is much more pessimistic than the result based on the half-mode analysis described in Sec.~\ref{sec:Technicalities:Bounds}. While there is no perfect correspondence of the colours red, purple, and blue between the two plots, it is nevertheless evident that some points which are not even constrained by current data in the left plot, seem to be excluded completely when looking at the right plot.

This is particularly true for the two points marked in the plots:
\begin{equation}
 \left\{
 \begin{matrix}
 {\rm A:}\hfill & & m_N = 7.1~{\rm keV},\hfill\hfill\hfill & & \lambda = 10^{-7.77},\hfill\hfill\hfill & & y = 10^{-6.50},\hfill\hfill\hfill\hfill\\
 {\rm B:}\hfill & & m_N = 20~{\rm keV},\hfill\hfill\hfill & & \lambda = 10^{-4.25},\hfill\hfill\hfill & & y = 10^{-8.47}.\hfill\hfill\hfill\hfill
 \end{matrix}
 \right.
 \label{eq:FS-points}
\end{equation}
Obviously, both these points are unconstrained (blue) in the half-mode analysis, while point~A would be classified as ``hot'' (red) in the analyses based on the FS horizon, and even point~B would still be labeled as ``warm'' (purple). This second classification is based on the results obtained for the FS horizon, which turn out to be:
\begin{equation}
 \left\{
 \begin{matrix}
 {\rm A:}\hfill\hfill\hfill & & \lambda_{\rm fs} = 0.117~{\rm MPc} \hfill\hfill\hfill & & \Rightarrow \text{``hot'',}\hfill\hfill\hfill\\
 {\rm B:}\hfill\hfill\hfill & & \lambda_{\rm fs} = 0.089~{\rm MPc} \hfill\hfill\hfill & & \Rightarrow\text{``warm'',}\hfill\hfill\hfill
 \end{matrix}
 \right.
 \label{eq:FS-classification}
\end{equation}
which perfectly matches the classification from the right panel of Fig.~\ref{fig:FS-comparison}.

\begin{figure}[ht]
 \centering
 \includegraphics[width=10cm]{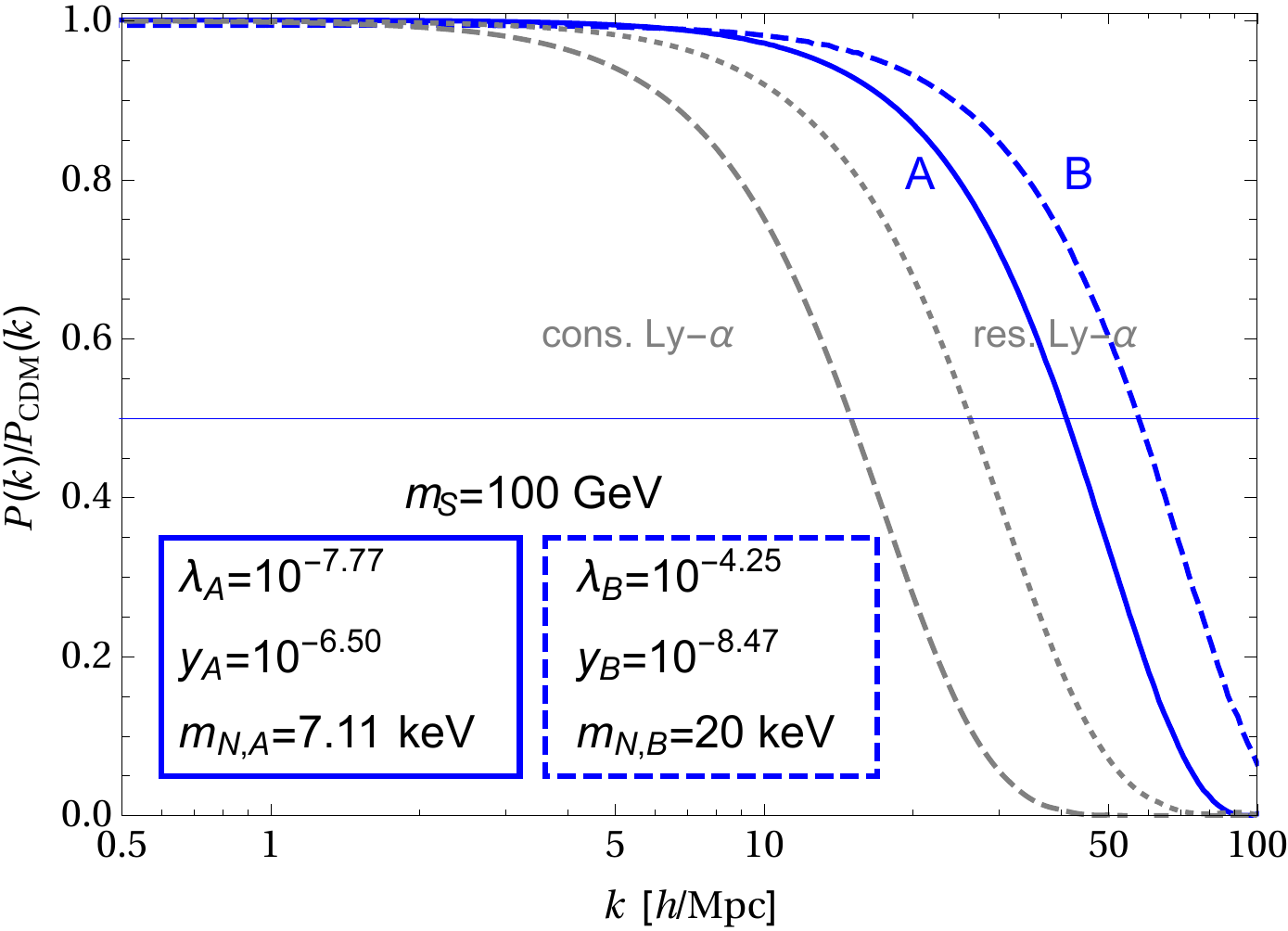}
 \caption{\label{fig:HM-resolution}Squared transfer functions for the points~A and~B marked in Fig.~\ref{fig:FS-comparison}. Even though point~A (point~B) would be completely discarded (strongly constrained) in an analysis based on the FS horizon, both points are in fact in full agreement with data.}
\end{figure}

But how can we be sure that this classification is insufficient? The simplest way is to confront the DM distribution functions with the actual data, which we can do by explicitly displaying the corresponding squared transfer functions in comparison to the Lyman-$\alpha$ data, as done in Fig.~\ref{fig:HM-resolution}. Looking at the curves for both points~A \&~B, we can immediately see that both of them are \emph{not at all} constrained by the data. Thus, both these points are, in fact, even indistinguishable from cold DM, from a structure formation point of view.

Given that this is truly obvious for the two example cases, it serves as a clear example for the FS horizon analysis leading to a conclusion that would be completely incorrect (namely discarding point~A all along, while viewing point~B as borderline case). Thus, as should now be obvious, \emph{the free-streaming horizon is not at all a suitable measure when applied to non-thermal DM distributions}.

\section{\label{app:D:HalfmodeThreshold}Robustness of the halfmode analysis}
\renewcommand{\theequation}{D-\arabic{equation}}
\setcounter{equation}{0}  

\begin{figure}[ht]
\begin{tabular}{lr}\hspace{-1cm}
 \includegraphics[width=8.3cm]{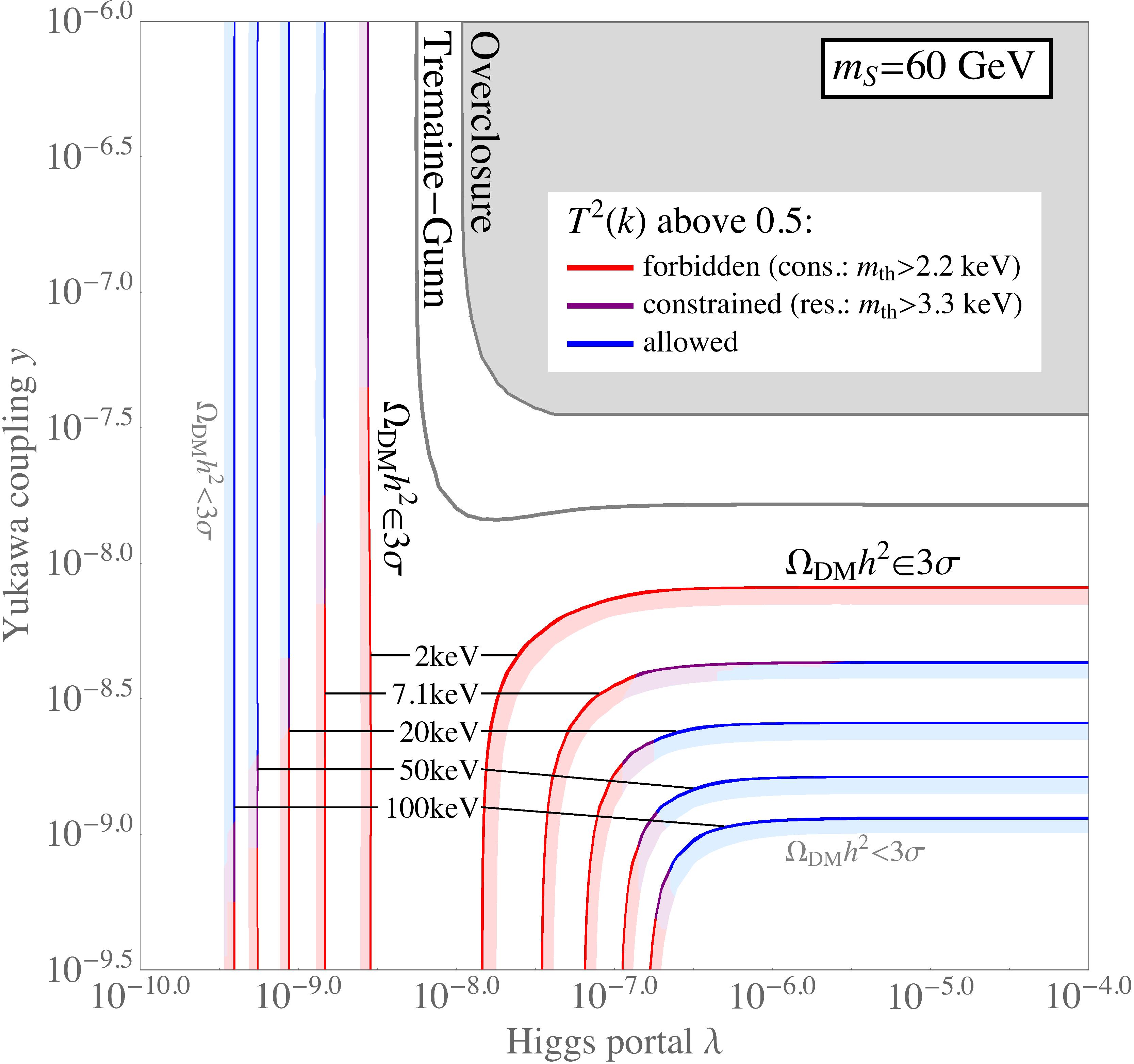} & \includegraphics[width=8.3cm]{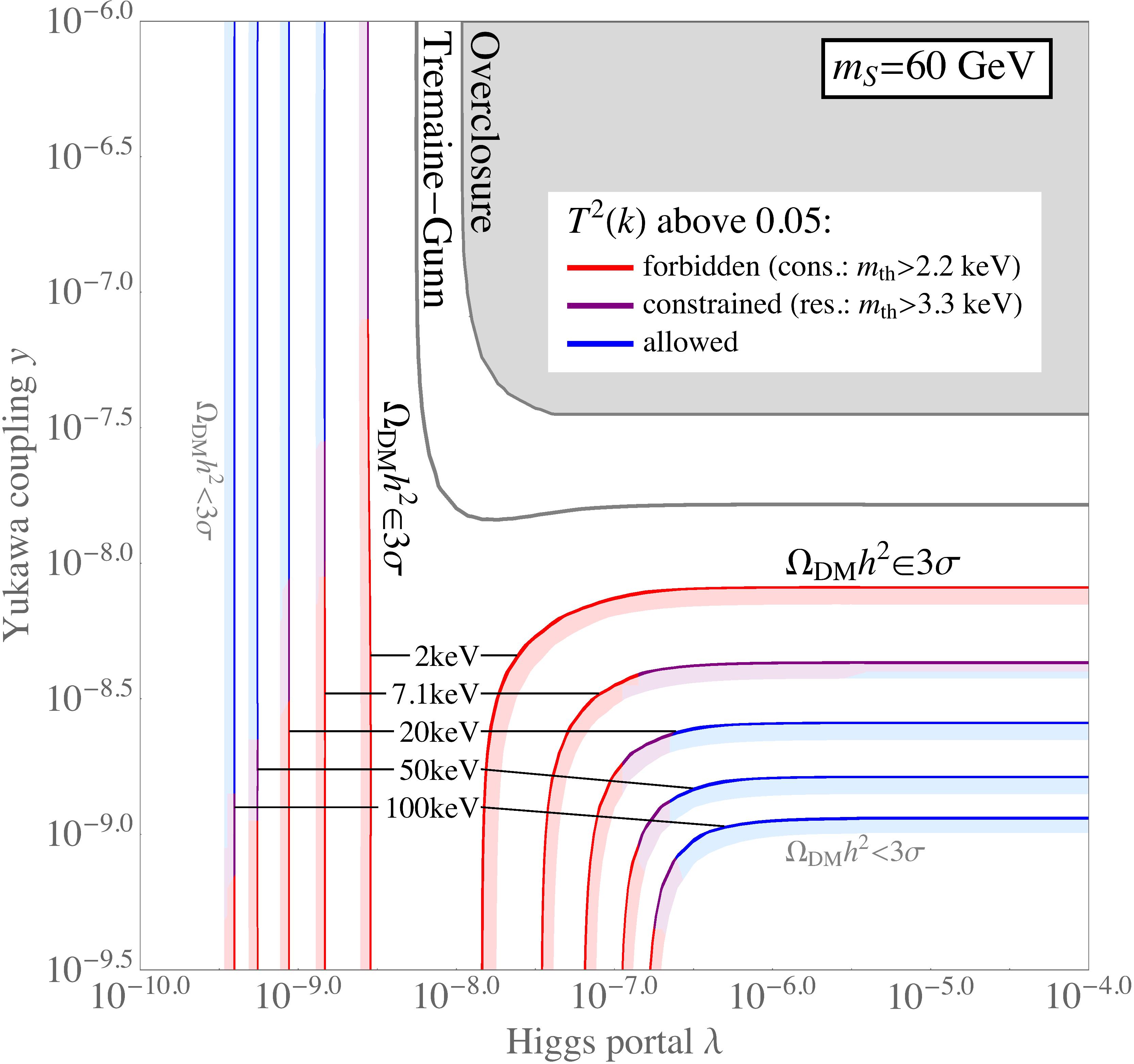}
\end{tabular}
\caption{\label{fig:HalfmodeThreshold}Compatability of the scalar decay model with $m_S=\unit{60}{GeV}$ for different threshold wave numbers. The \emph{left} panel contains the analysis with $k_x=k_{1/2}$ (as used throughout Sec.~\ref{sec:Results}; the plot displayed here is basically identical to the right Fig.~\ref{fig:verysmall_masses}, except for the model-dependent bounds not being displayed to enable a better comparison), while the \emph{right} panel displays $k_x= k_{0.05}$. The comparison shows that the changes are minor, but they will be finally specified in an upcoming work of us aiming to compared several advances methods to derive bounds from structure formation.}
\end{figure}

At first sight, the definition of the halfmode in \equref{eq:Def:Half-mode} might seem just as arbitrary as the free-streaming boundary between ``cold'' and ``warm'' (``warm'' and ``hot'') at $\unit{0.01}{Mpc}$ ($\unit{0.1}{Mpc}$). Even though the transfer function falls off steeply around $k_{1/2}$, it is a priori unclear at which value of the squared transfer function the discrimination becomes indeed negligible. While a more fundamental analysis of structure formation (e.g.~by rederiving Lyman-$\alpha$ bounds for non-thermal spectra) is beyond the scope of this paper and is projected for future work, we can nonetheless test the robustness of our analysis against changes in the threshold given in \equref{eq:Def:Half-mode}. More specifically, we can make our analysis more restrictive by comparing not only wavenumbers smaller than $k_{1/2}$ but wavenumbers smaller than some $k_x$ with $x < 0.5$, which translates to $k_{1/2} < k_x$. In order to demsonstrate that the analysis is rather robust even against large changes in this threshold, we have reanalysed the case of a scalar with $m_S = \unit{60}{GeV}$ with a $k_x=k_{0.05}$. \Figref{fig:HalfmodeThreshold} shows that even this rather drastic change in the threshold power (by one whole order of magnitude) inflicts rather mild changes on the results. This would be dramatically different when changing the values put in as boundaries in a free-streaming analysis (cf.~\Figref{fig:FS-comparison}). 

Even though being approximate itself, our method of incorporating bounds from structure formation, using Lyman-$\alpha$ data derived for thermal spectra, clearly proves more reliable than the free-streaming approach.
\end{appendices}

\bibliographystyle{./JHEP}
\bibliography{LighterScalars}

\end{document}